\documentclass[]{interact}

\usepackage{natbib}
\bibpunct[, ]{(}{)}{;}{a}{}{,}
\renewcommand\bibfont{\fontsize{10}{12}\selectfont}

\usepackage{hyperref}

\usepackage{setspace}
\usepackage{array}
\usepackage{epsfig}
\usepackage{rotating}
\usepackage{epstopdf}
\usepackage{titlesec}

\usepackage{amsmath}
\usepackage{amssymb}
\usepackage{amsfonts}
\usepackage{amsthm}
\usepackage{bm}
\usepackage{bbm}

\numberwithin{equation}{section}

\usepackage{graphicx}
\usepackage{booktabs}
\usepackage{threeparttable}
\usepackage{multirow}
\usepackage{longtable}
\usepackage{siunitx}
\usepackage{caption}
\usepackage{subcaption}
\usepackage{float}
\usepackage{placeins}

\renewcommand{\arraystretch}{1}
\usepackage{etoolbox}
\usepackage{enumitem}
\usepackage{xcolor}
\usepackage{circuitikz}

\newcommand{\rmsecell}[2]{%
  \shortstack[c]{#1\\[-1pt]{\scriptsize (#2)}}%
}

\usepackage[framemethod=TikZ]{mdframed}

\newmdenv[
  backgroundcolor=red!5,
  linecolor=blue,
  linewidth=1pt,
  skipabove=10pt,
  skipbelow=10pt,
  nobreak=true
]{ReorderedBox}

\newtheoremstyle{mytheorem}
  {6pt}
  {6pt}
  {\itshape}
  {}
  {\normalfont\bfseries\normalsize}
  {.}
  {0.5em}
  {}

\theoremstyle{mytheorem}
\newtheorem{theorem}{Theorem}
\newtheorem{proposition}{Proposition}
\newtheorem{lemma}{Lemma}

\theoremstyle{definition}

\theoremstyle{remark}
\newtheorem{remark}{Remark}

\makeatletter
\renewenvironment{proof}[1][Proof]{%
  \par\pushQED{\qed}%
  \normalfont\normalsize\textbf{#1. }\normalfont
}{\popQED}
\makeatother

\usepackage{tocloft}

\begin{document}
\newgeometry{
    margin=1in
}

\title{Functional Regression with Interacting Latent Mean and Variance Processes}

\author{
\name{Ching-Chia Chou\textsuperscript{a} and Jeng-Min Chiou\textsuperscript{a,b}}
\affil{\textsuperscript{a}Institute of Statistics and Data Science, National Taiwan University, Taipei, Taiwan, R.O.C.; \textsuperscript{b}Institute of Statistical Science, Academia Sinica, Taipei, Taiwan, R.O.C.}
}

\maketitle

\begin{abstract}
In many applications, variability itself carries important predictive information beyond the mean, and its association with an outcome may further depend on the underlying mean trajectory. However, existing functional regression approaches primarily rely on mean-based representations of functional predictors, while variability is often ignored or reduced to scalar summaries. We propose a functional regression framework that incorporates latent mean and variance processes as predictors, together with their mean–variance interaction, allowing the predictive contribution of variability to depend on the underlying mean trajectory. We establish consistency and convergence rates for the estimated coefficient functions and interaction surface, accounting for reconstruction errors, increasing observation frequency, and diverging truncation dimensions without requiring independence between the latent processes. Applications to real data demonstrate substantial gains in predictive accuracy from incorporating the variance process, with the proposed approach outperforming competing functional regression methods in the Tecator application and revealing localized mean–variance interaction patterns.
\end{abstract}



\begingroup
\singlespacing
\tableofcontents
\endgroup

\section{Introduction}

In many scientific applications, variability itself carries important information beyond the mean. In biomedical studies, for example, heart rate variability (HRV) has been shown to be a stronger predictor of cardiovascular risk and exercise capacity than mean heart rate, reflecting underlying physiological regulation that is not captured by average levels alone. Similar considerations arise in spectroscopic data, where measurements across wavelengths may exhibit subject-specific variation around an underlying spectral profile, and such variation may contain information about material composition beyond the average spectrum. However, most existing modeling approaches either summarize variability using scalar measures or ignore it altogether, potentially overlooking important patterns in how variability evolves over time or wavelength.

Functional data analysis provides a natural framework for preserving such temporal structure. In conventional functional linear regression, however, a repeatedly observed functional predictor is typically represented through its latent mean trajectory \citep{cardot2003spline, hall2007methodology, zhou2023functional}, while subject-specific variability around that trajectory is treated primarily as noise. Nevertheless, subjects with similar underlying mean trajectories may exhibit markedly different patterns of variability, and these differences may carry additional information about the response. We therefore view a noisy functional predictor as potentially containing two distinct sources of predictive information: its latent mean dynamics and its latent variance dynamics.

In many existing regression frameworks, variability is incorporated through scalar summaries such as variance or interquartile range. While convenient, such summaries cannot distinguish subjects with comparable overall variability but different patterns of fluctuation across the domain, even when variability concentrated in particular regions may be especially relevant to the response. Representing variability as a functional process instead preserves information about when and how variability occurs across the domain. The notion of a functional variance process has been introduced to characterize such heterogeneity in functional data \citep{muller2006functional}, but its role as an explicit source of predictive information in regression modeling has received limited attention.

Beyond incorporating the variance process as a functional predictor, an additional question is whether its predictive contribution depends on the underlying mean trajectory. This motivates consideration of interactions between the mean and variance processes. Functional interactions have been considered in several forms. \cite{yao2010functional} introduced functional quadratic regression, in which nonlinear effects arise through products of a functional predictor evaluated at two points in its domain. For multiple functional predictors, \cite{fuchs2015penalized} and \cite{usset2016interaction} considered interaction terms between distinct functional covariates in scalar-on-function regression, with related extensions subsequently developed for variable selection and function-on-function settings \citep{luo2019interaction, feng2022variable, jin2023optimal}. These formulations primarily describe either nonlinear self-interactions within a functional predictor or interactions between distinct functional covariates. Less attention has been given to interactions between different sources of predictive information arising from the same observed functional trajectory.

Motivated by these gaps, we propose a functional regression framework incorporating the latent mean and variance processes reconstructed from the same functional measurements, together with their interaction. This formulation allows us to assess the additional predictive contribution of subject-specific variability and how it depends on the underlying mean trajectory. Its practical utility is demonstrated in the real-data analyses, where incorporating the variance process substantially improves prediction. In the Tecator application, the proposed approach achieves substantially lower out-of-sample prediction errors than competing methods previously applied to these data, including functional quadratic modeling and partial least squares. The estimated interaction surface further reveals localized regions in which the contribution of variability depends on features of the underlying mean trajectory.

Beyond the proposed modeling framework, we develop asymptotic theory that addresses several nonstandard features of the present setting. The latent nature of the predictors is the first such challenge. In particular, reconstructing the variance process requires a logarithmic transformation of residuals computed using preliminary estimates of the subject-specific mean trajectories obtained by smoothing the observed measurements. Since this transformation has a singularity at zero, controlling the resulting reconstruction errors requires particular care. Under distributional conditions on the idiosyncratic errors, we establish second- and fourth-moment bounds for the discrepancy between the feasible transformed residuals and their oracle counterparts.

To translate these preliminary error bounds into convergence results for the reconstructed functional predictors, we develop an \(L^2\)-based perturbation analysis tailored to the integrated error structure of functional regression. In contrast to the uniform-norm analysis employed by~\cite{muller2006functional}, our treatment yields sharper bounds for reconstruction-induced stochastic errors, thereby avoiding the propagation of the slower uniform-norm reconstruction rate into subsequent error bounds for FPC score estimation and regression coefficient estimation. While their work establishes uniform-in-component consistency for estimated FPC scores, we obtain explicit convergence rates for the corresponding score estimation errors, together with empirical \(L^2\)-type bounds for score reconstruction errors uniformly over retained components and stochastic bounds on the magnitudes of estimated scores uniformly over subjects and retained components. These results accommodate diverging truncation dimensions and provide the theoretical foundation for the subsequent regression analysis.

The interaction term introduces a further layer of difficulty. Existing asymptotic analyses of functional interaction models commonly formulate the interaction through products of FPC scores and impose regularity conditions on the resulting score products \citep{yao2010functional, feng2022variable}. In the function-on-function setting, \cite{jin2023optimal} develop interaction theory under independence between the two functional predictors, thereby simplifying the treatment of cross-process dependence. Some developments consider settings in which the functional predictors are fully observed without measurement error \citep{feng2022variable}, whereas incomplete or noisy observation requires the resulting estimation errors to be carefully propagated through the analysis. Our setting is further complicated by the fact that both interacting predictors are latent processes extracted from the same noisy measurements, with reconstruction of the variance process additionally relying on a preliminary smoothed estimate of the subject-specific mean trajectory. We derive bounds for the resulting products of estimated scores and establish stability of the full joint empirical Gram matrix without assuming independence or cross-process moment factorization between the latent processes. This ultimately yields convergence results for the estimated main-effect functions and interaction surface while accounting jointly for score reconstruction, growing dimensionality, and cross-process dependence.

The remainder of this paper is organized as follows. Section~\ref{sec: model setting} introduces the model and functional predictors, and Section~\ref{sec: Estimation} describes the estimation and bootstrap inference procedures. Section~\ref{sec: consistency properties} establishes consistency and convergence rates for the proposed estimators. Sections~\ref{sec: simulation} and~\ref{sec: data example} report simulation studies and real-data applications, respectively. Section~\ref{sec: conclusion} concludes the paper. Additional estimation details are provided in the Appendix~\ref{appen A}, while technical assumptions, proofs, and supplementary materials for Sections~\ref{sec: simulation} and~\ref{sec: data example} are collected in the online Supplementary Material.

\section{Model setting}
\label{sec: model setting}

\subsection{Regression model with functional predictors}
\label{subsec: regression model with functional predictors}

To address the limitation of mean-based representations in functional regression, 
we consider a model in which functional predictors encode both central tendency 
and variability. Specifically, we focus on a functional regression framework where 
a scalar response $Y$ is related to latent functional covariates defined on a compact 
domain $\mathcal{T}\subset\mathbb{R}$.

In this work, we consider two latent functional predictors underlying the observed subject-specific trajectories: a mean process capturing the underlying signal and a variance process characterizing time-varying variability. We represent these predictors by square-integrable stochastic processes \(S(t)\) and \(V(t)\), respectively, defined on the compact domain \(\mathcal{T}\subset\mathbb{R}\). Neither process is assumed to be directly observed.

Let \(\mu_S(t)=E[S(t)]\) and \(\mu_V(t)=E[V(t)],
\) and define the centered processes
\(
S^c(t)=S(t)-\mu_S(t),
\) and \(V^c(t)=V(t)-\mu_V(t).
\) To accommodate not only the individual contributions of the mean and variance 
processes but also their joint effect on the response, we consider the model
\begin{equation}
Y=\alpha
+
\int_{\mathcal{T}}
\beta_1(s)S^c(s)\,ds
+
\int_{\mathcal{T}}
\beta_2(t)V^c(t)\,dt+
\int_{\mathcal{T}}
\int_{\mathcal{T}}
\gamma(s,t)S^c(s)V^c(t)\,ds\,dt
+
\varepsilon,
\label{eq:functional_interaction_model}
\end{equation}
where \(\varepsilon\) is a regression error with zero mean and finite variance, $\beta_1$ and $\beta_2$ are square-integrable coefficient 
functions associated with the main effects of the mean and variance processes, 
respectively, and $\gamma$ is a square-integrable bivariate coefficient surface 
characterizing their interaction.

The interaction surface \(\gamma(s,t)\) characterizes how deviations of the mean process at \(s\) and the variance process at \(t\) jointly contribute to the response. Let \(C_{SV}(s,t)=E[S^c(s)V^c(t)]\) denote their cross-covariance function. Taking expectations in~\eqref{eq:functional_interaction_model} yields \(\alpha=\mu_Y-\iint_{\mathcal T^2}\gamma(s,t)C_{SV}(s,t)\,ds\,dt\), where \(\mu_Y=E(Y)\).

As mentioned above, the underlying processes $S(t)$ and $V(t)$ are usually 
unobservable in practice. Hence, we use a two-step estimation procedure to construct 
estimators of the coefficient functions and the interaction surface. For this purpose, 
we first describe the structure of the observed data and the underlying mean and 
variance processes.

\subsection{Structure of predictors}
\label{subsec: structure of predictors}

To facilitate the estimation step introduced in Section \ref{sec: Estimation}, we now describe the structure of the observed functional data and its relationship to the latent mean and variance processes introduced in Section~\ref{subsec: regression model with functional predictors}. 
We consider the following decomposition of the observed trajectories $X(t)$, $t \in \mathcal{T}$.
\begin{equation}
\label{equa: data structure}
    X(t)=S(t)+R(t), \; t\in\mathcal{T},
\end{equation}
where $S(t)$ is a square-integrable stochastic process defined on a compact interval $\mathcal{T}$, while $R(t)$ is a measurement error process with zero mean. This decomposition is commonly encountered in the literature on functional data \citep{muller2006functional, zhang2007statistical, wang2016functional}, where observed trajectories consist of an underlying smooth signal and a noise component, and serves as a basis for separating mean and variability information. As in the traditional simple regression context, we expect that squared residuals would encode rich information of data. In this study, we consider the setting proposed in \cite{muller2006functional}, given by
\begin{equation}
\label{equa: square error}    R^2(t)=\exp(V(t)+W),
\end{equation}
where $V(t)$ is a square-integrable process, $W$ is a white noise with zero mean and finite constant variance. We adopt a specific model which integrates \eqref{equa: data structure} and \eqref{equa: square error} as a unified expression of observed data.
\begin{equation}
    \label{equa: unified expression of observed data}
    X(t)=S(t)+\text{sign}(\epsilon)\{\text{exp}(V(t)+W)\}^{{1}/{2}},
\end{equation}
where $\epsilon$ is a random variable independent of all other random variables. Moreover, it is assumed to be symmetrically distributed, that is, $P(\epsilon>0)=P(\epsilon<0)=1/2$.

In \eqref{equa: data structure}, 
for the mean process $S(t)$, we assume that it is a smooth and square-integrable stochastic process defined on a compact interval $\mathcal{T}\subset\mathbb{R}$, with the smooth mean function $\mu_S(t)=E[S(t)]$ and the covariance function $C_S(t,s)=\operatorname{Cov}(S(t), S(s))$. For the measurement error process \(R(t)\), we assume a zero mean and a diagonal covariance structure, with \(\operatorname{Cov}\{R(t),R(s)\}=\sigma_R^2(t)\mathbf 1_{\{t=s\}}\), where \(\mathbf 1_{\{\cdot\}}\) denotes the indicator function.
Consequently, from \eqref{equa: data structure}, we get \(E[X(t)]=E[S(t)]=\mu_S(t)
\)
and
\[
\operatorname{Cov}(X(t), X(s))
=C_S(t,s)+\sigma_R^2(t)\mathbf 1_{\{s=t\}}+\operatorname{Cov}(S(t), R(s))+\operatorname{Cov}(S(s), R(t)).\]

Model setting of observed data \eqref{equa: unified expression of observed data} that we adopted aligns with all the assumptions above. Specifically speaking, we have
\begin{equation}
    \label{equa: expec and cov of X}
    E[X(t)] = \mu_S(t), \; \operatorname{Cov}(X(t), X(s))=C_S(t,s)+E[\exp(V(t)+W)]\mathbf1_{\{t=s\}}.
\end{equation}

Under model~\eqref{equa: unified expression of observed data}, the above moment structure follows directly from the independence assumptions and the symmetry of $\epsilon$. Note that \eqref{equa: expec and cov of X} enable us to estimate the mean and covariance function of the mean process $S(t)$. To facilitate the least squares estimation of the coefficient functions and interaction surface in Section~\ref{sec: Estimation}, we consider the functional principal component (FPC) representations of both latent processes. For \(P\in\{S,V\}\), let \(\mu_P(t)=E[P(t)]\) and \(C_P(t,s)=\operatorname{Cov}(P(t),P(s))\) denote the mean and covariance functions, respectively. By Mercer's theorem, the covariance function admits the spectral decomposition \(
C_P(t,s)=\sum_{k=1}^{\infty}\rho_k^P\varphi_k^P(t)\varphi_k^P(s),
\) where \(\rho_1^P\geq\rho_2^P\geq\cdots\geq0\) are the eigenvalues and \(\{\varphi_k^P\}_{k\geq1}\) are the corresponding orthonormal eigenfunctions. The Karhunen–Loève expansion of \(P(t)\) is then given by \(
P(t)=\mu_P(t)+\sum_{k=1}^{\infty}\chi_k^P\varphi_k^P(t),
\) where the FPC scores are defined as
\begin{equation}
\label{eq:score}
\chi_k^P=\int_{\mathcal T}
\{P(t)-\mu_P(t)\}\varphi_k^P(t)\,dt,
\qquad k\geq1,
\end{equation}
with \(E[\chi_k^P]=0\) and \(\operatorname{Var}(\chi_k^P)=\rho_k^P\).

\section{Estimation and inference}
\label{sec: Estimation}

Let \((S_i,V_i,\varepsilon_i)\), \(i=1,\ldots,n\), be i.i.d. copies of \((S,V,\varepsilon)\) defined in Section~\ref{sec: model setting}, with corresponding responses \(Y_i\) following the regression model~\eqref{eq:functional_interaction_model}. For each subject \(i\), the latent processes are observed indirectly through measurements \(X_i(t_{ij})\), \(j=1,\ldots,m\), at design points \(t_{ij}\in\mathcal T\), according to the observation model~\eqref{equa: unified expression of observed data}, with subject-specific errors \(\varepsilon_{ij}\) and \(W_{ij}\). The design points form a regular dense grid on \(\mathcal T\), with precise conditions given in the Supplementary Material.

\subsection{Estimation of predictors – FPC scores for latent processes}
\label{subsec: Estimating mean and variance}

We estimate the FPC scores associated with the two latent processes using an FPCA-based estimation framework that combines local linear smoothing and FPCA \citep{yao2006penalized, muller2006functional}. This framework yields the associated principal component representations. All estimation details are deferred to the Section~\ref{appen A} of the Appendix.

Although the mean and variance processes play different modeling roles, they share the same estimation structure under FPCA. The only distinction is in the input data: the raw trajectories \(X_i(t_{ij})\) are used for the mean process, whereas the variance process is inferred from log-transformed squared residuals. Specifically, for each subject \(i\), we obtain a preliminary smooth estimate \(\tilde S_i(t)\) of its latent trajectory via local linear smoothing of \(\{(t_{ij}, X_i(t_{ij}))\}\). 
We then form
\begin{equation}
\label{equa: TR}
    \widehat{Z}_i(t_{ij}) = \log\left\{(X_i(t_{ij}) - \widetilde{S}_i(t_{ij}))^2\right\},
\end{equation}
which serves as a proxy observation for the latent variance trajectory \(V_i(t)\).

For each process, we estimate the population mean function by pooled smoothing, obtain the covariance surface through two-dimensional local smoothing of off-diagonal empirical covariances, extract eigenvalues and eigenfunctions by eigendecomposition, and compute subject-specific FPC scores by numerical integration.

The estimated FPC scores are computed as follows, by plugging the estimated mean functions and eigenfunctions obtained in the previous steps into the theoretical score definitions and approximating the integrals \eqref{eq:score} using Riemann sums:
\begin{align*}
\widehat{\chi}_{ik}^S &= \sum_{j=2}^{m} \bigl( X_i(t_{ij}) - \widehat{\mu}_S(t_{ij}) \bigr) \widehat{\varphi}_k^S(t_{ij}) (t_{ij} - t_{i,j-1}),
\quad i=1,\ldots,n;\; k=1,\ldots,K_1, \\[-0.25em]
\widehat{\chi}_{ik}^V &= \sum_{j=2}^{m} \bigl( \widehat{Z}_i(t_{ij}) - \widehat{\mu}_V(t_{ij}) \bigr) \widehat{\varphi}_k^V(t_{ij}) (t_{ij} - t_{i,j-1}),
\quad i=1,\ldots,n;\; k=1,\ldots,K_2.
\end{align*}
The numbers of retained components \(K_1=K_S\) and \(K_2=K_V\) are chosen using the fraction-of-variance-explained (FVE) criterion, with each chosen as the smallest number of components for which the cumulative proportion of variation explained by the corresponding estimated eigenvalues reaches a prespecified threshold. The set of notations $(K_1, K_2)$ and $(K_S,K_V)$ are used interchangeably throughout the article.


\subsection{Estimation of coefficient functions and interaction surface}
\label{subsec:parameter estimation}

In this subsection, we derive the estimators of the coefficient functions and
the interaction surface. Using the notation introduced in Section~\ref{sec: model setting}, the subject-specific
Karhunen--Lo\`eve expansions of the centered latent processes are \(S_i^c(s)
=
\sum_{k=1}^{\infty}
\chi_{ik}^S\varphi_k^S(s)\) and \(V_i^c(t)
=
\sum_{\ell=1}^{\infty}
\chi_{i\ell}^V\varphi_\ell^V(t).
\) The FPC scores satisfy \(E[\chi_{ik}^S]=0\), \(\operatorname{Var}(\chi_{ik}^S)=\rho_k^S\) and \(E[\chi_{ik}^S\chi_{ik'}^S]=0 \text{ for }k\ne k',
\)
and analogous for FPC scores of variance process. Note that $E[\chi_{ik}^S\chi_{i\ell}^V]$ is generally allowed to be
nonzero. 

We expand the two main-effect coefficient functions in their respective
eigenfunction bases as \(
\beta_1(s)
=
\sum_{k=1}^{\infty}
b_k\varphi_k^S(s)\) and \(\beta_2(t)
=
\sum_{\ell=1}^{\infty}
c_\ell\varphi_\ell^V(t),
\) and represent the interaction surface using the corresponding tensor-product
basis \(
\gamma(s,t)
=
\sum_{k=1}^{\infty}
\sum_{\ell=1}^{\infty}
d_{k\ell}
\varphi_k^S(s)\varphi_\ell^V(t).
\)
Substituting these expansions into \eqref{eq:functional_interaction_model} gives \(
Y_i
=
\alpha
+
\sum_{k=1}^{\infty}b_k\chi_{ik}^S
+
\sum_{\ell=1}^{\infty}c_\ell\chi_{i\ell}^V
+
\sum_{k=1}^{\infty}\sum_{\ell=1}^{\infty}
d_{k\ell}\chi_{ik}^S\chi_{i\ell}^V
+
\varepsilon_i.
\label{eq:score_regression}
\)

For estimation, we truncate the FPC expansions at $K_1$ and $K_2$ for the
mean and variance processes, respectively, and replace the latent scores by
their estimators obtained in Section~\ref{subsec: Estimating mean and variance}. This leads to the finite-dimensional
least-squares criterion
\begin{equation}
(\widehat{\alpha},\widehat{\boldsymbol b},
 \widehat{\boldsymbol c},\widehat{\boldsymbol D})
=
\arg\min_{\alpha,\boldsymbol b,\boldsymbol c,\boldsymbol D}
\sum_{i=1}^n
\Bigg[
Y_i-\alpha
-
\sum_{k=1}^{K_1}
b_k\widehat{\chi}_{ik}^S
-
\sum_{\ell=1}^{K_2}
c_\ell\widehat{\chi}_{i\ell}^V-
\sum_{k=1}^{K_1}\sum_{\ell=1}^{K_2}
d_{k\ell}
\widehat{\chi}_{ik}^S
\widehat{\chi}_{i\ell}^V
\Bigg]^2,
\label{eq:interaction_ls}
\end{equation}
where \(
\boldsymbol b=(b_1,\ldots,b_{K_1})^\top\), \(\boldsymbol c=(c_1,\ldots,c_{K_2})^\top\), \(
\boldsymbol D=(d_{k\ell})_
{1\le k\le K_1,\;1\le\ell\le K_2},
\)
and $\widehat{\chi}_{ik}^S$ and $\widehat{\chi}_{i\ell}^V$,
$i=1,\ldots,n$, $k=1,\ldots,K_1$, and $\ell=1,\ldots,K_2$, are the
estimated FPC scores obtained in Section~\ref{subsec: Estimating mean and variance}. The truncation levels
$K_1$ and $K_2$ may be selected by the fraction of variance explained
or specified in advance, as described in Section~\ref{sec: simulation} and~\ref{sec: data example}.

For a matrix representation of the least-squares estimator, define
$\widehat{\boldsymbol X}^{S}
=(\widehat{\chi}_{ik}^{S})_{n\times K_1}$
and
$\widehat{\boldsymbol X}^{V}
=(\widehat{\chi}_{i\ell}^{V})_{n\times K_2}$,
and let the interaction design matrix
$\widehat{\boldsymbol X}^{SV}\in
\mathbb R^{n\times K_1K_2}$ have $i$th row
$(
\widehat{\chi}_{i1}^{S}\widehat{\chi}_{i1}^{V},
\ldots,
\widehat{\chi}_{i1}^{S}\widehat{\chi}_{iK_2}^{V},
\ldots,
\widehat{\chi}_{iK_1}^{S}\widehat{\chi}_{i1}^{V},
\ldots,
\widehat{\chi}_{iK_1}^{S}\widehat{\chi}_{iK_2}^{V}
)$.
Define
$\widehat{\boldsymbol Z}
=[\boldsymbol 1_n,
\widehat{\boldsymbol X}^{S},
\widehat{\boldsymbol X}^{V},
\widehat{\boldsymbol X}^{SV}]$
and
$\boldsymbol\theta
=(\alpha,\boldsymbol b^\top,\boldsymbol c^\top,
\operatorname{vec}(\boldsymbol D^\top)^\top)^\top$,
where $\boldsymbol 1_n$ is the $n$-dimensional vector
of ones and $\operatorname{vec}(\cdot)$ denotes
column-wise vectorization.
Let
$\boldsymbol Y=(Y_1,\ldots,Y_n)^\top$,
provided that
$\widehat{\boldsymbol Z}^\top\widehat{\boldsymbol Z}$
is nonsingular, then the least-squares estimator is
$\widehat{\boldsymbol\theta}
=(\widehat{\boldsymbol Z}^\top
\widehat{\boldsymbol Z})^{-1}
\widehat{\boldsymbol Z}^\top\boldsymbol Y$.

The resulting estimators of the coefficient functions and interaction
surface are \(
\widehat{\beta}_1(s)
=
\sum_{k=1}^{K_1}
\widehat b_k\widehat{\varphi}_k^S(s)\), \(\widehat{\beta}_2(t)
=
\sum_{\ell=1}^{K_2}
\widehat c_\ell\widehat{\varphi}_\ell^V(t)\) and \(
\widehat{\gamma}(s,t)
=
\sum_{k=1}^{K_1}
\sum_{\ell=1}^{K_2}
\widehat d_{k\ell}
\widehat{\varphi}_k^S(s)
\widehat{\varphi}_\ell^V(t),\)
where $\widehat{\varphi}_k^S(t)$, $\widehat{\varphi}_k^V(t)$ are the eigenfunctions estimated from the observed data $X_i(t_{ij})$ as in the Section \ref{subsec: Estimating mean and variance}.

\subsection{Bootstrap-based simultaneous band}

Based on the estimators obtained in Section~\ref{subsec:parameter estimation}, we employ a pairs bootstrap procedure to construct simultaneous bands for the slope and surface estimators as a practical tool for uncertainty assessment. Although confidence bands have been studied extensively in finite-dimensional or well-specified nonparametric settings \citep{claeskens2003bootstrap}, relatively few works have addressed their construction in the functional linear regression (FLR) framework. In particular, simultaneous confidence bands for response trajectories in FLR models with multiple functional predictors have been studied \citep{chiou2016multivariate}. Pointwise confidence intervals for conditional mean responses in scalar-on-function linear regression have also been considered \citep{gonzalez2011bootstrap, yeon2023bootstrap, yeon2024bootstrap}. However, extending these approaches to the present functional setting is nontrivial due to the additional randomness introduced by data-driven basis representations.

Constructing asymptotically valid simultaneous confidence bands in FPCA-based functional regression is challenging because estimated eigenfunctions introduce additional uncertainty. Bootstrap methods provide a flexible alternative for approximating the sampling variability of such estimators \citep{efron1979bootstrap}. Existing approaches include residual and wild bootstrap procedures for pointwise inference \citep{gonzalez2011bootstrap,yeon2023bootstrap}, paired bootstrap methods for scalar-on-function regression \citep{yeon2024bootstrap}, and subject-level resampling of entire observed curves with re-estimation of functional quantities \citep{rathnayake2016tolerance}.

We adapt a pairs bootstrap procedure to construct simultaneous confidence bands for the slope functions and interaction surface. While establishing the theoretical validity of these bands in the present functional setting remains an open problem, the proposed procedure provides a practical means of assessing estimation uncertainty. Simulation studies demonstrate satisfactory empirical coverage across various data-generating scenarios. The pairs bootstrap procedure is summarized below. 

\vspace{1em}

\noindent\textbf{Pairs bootstrap procedure for constructing simultaneous bands}

\begin{enumerate}

\item \textbf{Initial estimation.}
Using the original sample, estimate the mean and variance processes
$S_i$ and $V_i$ and their corresponding FPC scores
$\widehat{\chi}_{ik}^{S}$ and $\widehat{\chi}_{i\ell}^{V}$.
Fit the regression model to obtain $\widehat{\beta}_1(s)$, $\widehat{\beta}_2(t)$ and $\widehat{\gamma}(s,t)$ as described in Section~\ref{subsec:parameter estimation}.

\item \textbf{Pairs bootstrap resampling.}
For $b=1,\ldots,B$:
\begin{enumerate}

\item Sample $n$ subject indices \(I_1^{*(b)},\ldots,I_n^{*(b)}\)
independently with replacement from $\{1,\ldots,n\}$. For each selected
index, retain the subject's entire set of functional observations together
with the corresponding scalar response, thereby forming the bootstrap sample \(
\{
(
\{X_{I_i^{*(b)}}(t_{I_i^{*(b)}j})\}_{j=1}^{m},
Y_{I_i^{*(b)}}
)
\}_{i=1}^{n}.
\)

\item From the bootstrap sample, perform FPCA, and obtain the corresponding
bootstrap FPC scores
$\widehat{\chi}_{ik}^{S,*(b)}$ and
$\widehat{\chi}_{i\ell}^{V,*(b)}$.

\item Refit the regression model using the bootstrap estimates to obtain
$\widehat{\beta}_1^{*(b)}(s)$,
$\widehat{\beta}_2^{*(b)}(t)$, and
$\widehat{\gamma}^{*(b)}(s,t)$.

\end{enumerate}

\item \textbf{Band construction.}
For $b=1,\ldots,B$, compute the supremum deviations \(
M_1^{*(b)}
=
\sup_{s\in\mathcal{T}}
|
\widehat{\beta}_1^{*(b)}(s)-\widehat{\beta}_1(s)
|\), \(
M_2^{*(b)}
=
\sup_{t\in\mathcal{T}}
|
\widehat{\beta}_2^{*(b)}(t)-\widehat{\beta}_2(t)
|\) and \(M_{\gamma}^{*(b)}
=
\sup_{(s,t)\in\mathcal{T}^2}
|
\widehat{\gamma}^{*(b)}(s,t)\)
\(-\widehat{\gamma}(s,t)
|.
\)
Let $c_{1,\alpha}$, $c_{2,\alpha}$, and $c_{\gamma,\alpha}$ denote the
empirical $(1-\alpha)$ quantiles of
$\{M_1^{*(b)}\}_{b=1}^{B}$,
$\{M_2^{*(b)}\}_{b=1}^{B}$, and
$\{M_{\gamma}^{*(b)}\}_{b=1}^{B}$, respectively.
The simultaneous confidence bands for the coefficient functions are
then given by
\[
\mathcal{C}_{1,\alpha}(s)
=
\widehat{\beta}_1(s)\pm c_{1,\alpha},
\quad s\in\mathcal{T}, \quad \text{ and }\quad\mathcal{C}_{2,\alpha}(t)
=
\widehat{\beta}_2(t)\pm c_{2,\alpha},
\quad t\in\mathcal{T}.
\]

Similarly, the simultaneous confidence band for the interaction surface is
\[
\mathcal{C}_{\gamma,\alpha}(s,t)
=
\widehat{\gamma}(s,t)\pm c_{\gamma,\alpha},
\quad (s,t)\in\mathcal{T}^2.
\]
\end{enumerate}

Steps 1 and 2 in the pairs bootstrap procedure follow the pairs bootstrap strategy \citep{freedman1981bootstrapping}, which has also been employed in functional data analysis \citep{rathnayake2016tolerance, yeon2024bootstrap}. Step 3, which constructs equal-width bands by computing the pointwise supremum over $\mathcal{T}$ and then taking the empirical $(1 - \alpha)$ quantile of the resulting maxima, is inspired by the methodology of \cite{claeskens2003bootstrap}, whose theoretical framework, while developed under different settings, provides useful guidance for constructing simultaneous bands.

\section{Consistency properties}
\label{sec: consistency properties}

We now investigate the asymptotic properties of the proposed estimators under the observation setting described in Section~\ref{sec: model setting} and~\ref{sec: Estimation}. All proofs are deferred to the Supporting Material available online. We first introduce the asymptotic notation and rate quantities used in stating the theoretical results.

\noindent\textbf{Asymptotic notation.}
For two positive deterministic sequences $a_n$ and $b_n$, we write
$a_n\lesssim b_n$ if $a_n\leq Cb_n$ for some constant $C>0$
independent of $n$ and all sufficiently large $n$, and write
$a_n\asymp b_n$ if both $a_n\lesssim b_n$ and $b_n\lesssim a_n$
hold. We use $a_n=o(b_n)$ to mean $a_n/b_n\to0$. For random sequences, \(X_n=O_p(1)\) means that \(X_n\) is bounded in probability. More generally, \(X_n=O_p(a_n)\) means \(X_n/a_n=O_p(1)\).
We also write $a\vee b=\max(a,b)$. For square-integrable functions on \(\mathcal T\) or \(\mathcal T^2\), \(\|\cdot\|_2\) denotes the corresponding \(L^2\) norm. For a symmetric matrix $A$, $\lambda_{\min}(A)$
denotes its smallest eigenvalue. All asymptotic statements are
understood as $n,m\to\infty$, with $K_1$ and $K_2$ allowed to
diverge according to the growth conditions specified below.

\noindent\textbf{Rate notation.} For the asymptotic results below, let \(h_{C_S}\) and \(h_{C_V}\) denote the covariance-smoothing bandwidths for the mean and variance processes, respectively, and let \(b_S\) denote the common-order bandwidth sequence associated with the subject-specific smoothing of the mean trajectories, and define $
\theta_{m,2}^2 = b_S^4 + (mb_S)^{-1}.$

\subsection{Error bounds for functional predictor reconstruction}
\label{subsec:error bounds for functional predictor reconstruction}

We first establish the score-level bounds needed for the subsequent analysis of the interaction design. In addition to uniform convergence over the retained FPC components, we derive empirical mean-squared reconstruction error bounds across subjects, together with uniform bounds over subjects and retained FPC components. These results allow us to control the empirical Gram matrix directly through the Euclidean norms of the regression design vectors.

We consider the regular dense design specified in Assumption~(A3)
of Section~\ref{appen S1} of the Supporting Material, where each of the $n$ subjects is
observed at $m$ time points, and assume the asymptotic regime $n\asymp m^2$
for simplicity of the convergence-rate expressions. To establish the
required bounds for $\widehat{\chi}_{ik}^S$ and
$\widehat{\chi}_{ik}^V$, we impose the following assumptions.

\noindent
(R1) The white noise $W$ admits a density $f_W$
satisfying \(
\sup_{w\in\mathbb R} f_W(w)<\infty.
\) Also, there exist constants $\delta>0$, $C_W>0$,
and $w_0\in\mathbb R$ such that \(
f_W(w)\le C_W e^{(2+\delta)w}, \text{ for all } w\le w_0.\)

\noindent (R2) For $P\in\{S,V\}$, the eigenvalues are simple and satisfy \(
\rho_k^S\lesssim k^{-\alpha_1}\), 
\(\rho_k^V\lesssim k^{-\alpha_2},
\) and \(
\rho_k^S-\rho_{k+1}^S\gtrsim k^{-\alpha_1-1}\),
\(\rho_k^V-\rho_{k+1}^V\gtrsim k^{-\alpha_2-1},
\)
uniformly in $k\ge1$, where $\alpha_j>1$, $j=1,2$.

The estimation of the variance process involves a logarithmic transformation of residuals obtained after presmoothing the mean trajectory. Since residuals close to zero may amplify the discrepancy between the feasible and oracle transformed residuals, we impose a condition on the left-tail behavior of the white noise $W$. This controls the behavior of the residual distribution near zero. (R2) is a standard eigenvalue decay condition in FPCA which imposes polynomial upper and lower bounds on eigenvalues and their eigenvalue decay rates, respectively. In addition to Assumption (R1) and (R2), we impose technical conditions adapted from~\cite{yao2006penalized} and \cite{muller2006functional}. These are based on Assumptions (a)–(b) of~\cite{yao2006penalized}, (A1)–(A4), (A6) and (B2.1)–(B2.2) of~\cite{muller2006functional}, with their bandwidth conditions consolidated into (A2.1)–(A2.6). All conditions, including our modifications, are stated in the Supplementary Material, whose numbering is used throughout this paper.

\begin{proposition}
\label{prop:transformed_residual}
Under Assumptions (R1), (A1), (A2.3), (A2.4), (A3) and (B2.1),
\[
\sup_{1\le i\le n}\sup_{1\le j\le m}
E\left[
\{\widehat Z_i(t_{ij})-Z_i(t_{ij})\}^q
\right]
=
O(\theta_{m,2}^q) 
\quad\text{for}\quad
q = 2,4,
\quad\theta_{m,2}^2
=
b_S^4+\frac{1}{mb_S}.
\]
where $Z_i(t_{ij})=\log(R_i^2(t_{ij}))$ and $\widehat Z_i(t_{ij})$ is defined in \eqref{equa: TR}. Consequently,
\[
\frac{1}{n}\sum_{i=1}^n
\sum_{j=2}^m|\widehat Z_{ij}-Z_{ij}|^q\Delta_{ij}=
O_p(\theta_{m,2}^q) 
\quad\text{for}\quad q=2,4,
\quad\Delta_{ij} = t_{ij} - t_{i,j-1}. 
\]

\end{proposition}

With this mean-square bound established, we now present the convergence rates and stochastic bounds for the estimated FPC scores \(\widehat\chi_{ik}^S\) and \(\widehat\chi_{ik}^V\) under the proposed model.

\begin{theorem}[Convergence and stochastic bounds for the estimated FPC scores]
\label{thm:score-convergence}
\noindent
\begin{enumerate}[
    label=(\alph*),
    leftmargin=1.5em,
    itemsep=1em
]
\item {Variance-process scores.}
Define \(
\Delta_{ik}^V=\widehat{\chi}_{ik}^V-\chi_{ik}^V,
\) then under Assumptions (R1), (R2), (A1), (A2.3)--(A2.6), (A3)--(A5),
and (B2.1)--(B2.2), 
\begin{enumerate}[
    label=(\roman*),
    leftmargin=2em,
    itemsep=0.6em
]
\item\textit{Uniform estimation errors.}
Let $K_2=K_2(n)\to\infty$. Define \(r_{V,n,K_2}
=
K_2^{\alpha_2+1}\theta_{m,2}.
\)
Then, for each fixed $i$,
\[
\max_{1\leq k\leq K_2}
|\Delta_{ik}^V|
=
O_p(r_{V,n,K_2}).
\]

\item\textit{Empirical second-moment bounds.}
We have
\[
\max_{1\leq k\leq K_2}
\frac{1}{n}\sum_{i=1}^n|\Delta_{ik}^V|^2
=
O_p(r^2_{V,n,K_2}), \quad \max_{1\le k\le K_2}
\frac{1}{n}\sum_{i=1}^n|\widehat{\chi}_{ik}^{V}|^2
=
O_p(1).
\]

\end{enumerate}

\item\textit{Mean-process scores.}
Define \(
\Delta_{ik}^S
=
\widehat{\chi}_{ik}^S-\chi_{ik}^S,
\) then under Assumptions (R2), (A1), (A2.1)--(A2.2), (A3)--(A5), and
(B2.1)--(B2.2),

\begin{enumerate}[
    label=(\roman*),
    leftmargin=2em,
    itemsep=0.6em
]
\item\textit{Uniform estimation errors.}
Let $K_1=K_1(n)\to\infty$. Define \(
r_{S,n,K_1} = 
K_1^{\alpha_1+1}h_{C_S}^2\vee m^{-1/2}\) and \(r_{S,n,K_1}^\mathrm{ind}
=
K_1^{\alpha_1+1}h_{C_S}^2
\vee
K_1^{1/8}m^{-1/2}.
\)
Then, for each fixed $i$,
\[
\max_{1\le k\le K_1}
|\Delta_{ik}^S|
=
O_p(r_{S,n,K_1}^\mathrm{ind}).
\]
\item\textit{Empirical second-moment bounds.}
We have
\[
\max_{1\le k\le K_1}
\frac1n\sum_{i=1}^n
|\Delta_{ik}^S|^2
=
O_p(r_{S,n,K_1}^2),\quad
\max_{1\le k\le K_1}
\frac1n\sum_{i=1}^n
|\widehat{\chi}_{ik}^S|^2
=
O_p(1).
\]
\item \textit{Uniform stochastic magnitude bound.}
The estimated mean-process scores satisfy
\[
\max_{\substack{1\le i\le n\\1\le k\le K_1}}
|\widehat{\chi}_{ik}^S|
=
O_p(1).
\]

\end{enumerate}

\end{enumerate}

\end{theorem}

\subsection{Consistency of \texorpdfstring{$\widehat{\beta}_j(t)$}{beta j hat}}
\label{subsubsec:covergence rate of coefficient functions}
Having established in Theorem~\ref{thm:score-convergence} that the estimated FPC scores converge at appropriate rates, we now proceed to analyze the consistency of the regression coefficient estimators $\widehat{\beta}_1(s)$, $\widehat \beta_2(t)$ and interaction surface estimator $\widehat{\gamma}(s,t)$. Here we introduce the necessary assumptions used in the subsequent theoretical developments. Let $\boldsymbol{\Lambda}_K$ be the population counterpart of the empirical Gram matrix $n^{-1}\widehat{\boldsymbol{Z}}^\top\widehat{\boldsymbol{Z}}$, See Section~\ref{subsec:parameter estimation} for the definition of $\widehat{\boldsymbol{Z}}$.

\noindent (S1) $E[(\chi_k^{S})^4] \lesssim (\rho_k^{S})^2$ and $E[(\chi_k^{V})^4] \lesssim (\rho_k^{V})^2$ uniformly in $k\ge 1$.

\noindent (S2) $|b_k|\lesssim k^{-\eta_1}$ and $|c_k| \lesssim k^{-\eta_2}$ uniformly in $k \ge 1$, for some constants $\eta_j > 1$, $j=1,2$.

\noindent (S3) $|d_{k\ell}| \lesssim k^{-\eta_3}\ell^{-\eta_4}$ uniformly in $k,\ell\ge1$, for some constants
$\eta_j>1$, $j=3,4$.

\noindent (S4) For some deterministic sequence $\underline{\lambda}_K>0$, $
\lambda_{\min}(\boldsymbol\Lambda_K)
\gtrsim \underline{\lambda}_K,
$
and
\begin{align*}
&\sqrt{K_1K_2}
\left(
r_{S,n,K_1}\vee r_{V,n,K_2}
\right)=o(\underline{\lambda}_K),
\\
&K_1^{\{(1-\eta_3)\vee (1/2-\eta_1)\}-\alpha_1/2}
\vee
K_2^{\{(1-\eta_4)\vee (1/2-\eta_2)\}-\alpha_2/2}
=
o(\underline{\lambda}_K^{1/2}).
\end{align*}

(S1) is a standard moment condition in functional regression and FPCA \citep{hall2007methodology, yuan2010reproducing, zhou2023functional}. Specifically, it controls the fourth moments of FPC scores relative to squared eigenvalues. (S2) imposes standard smoothness conditions on the slope coefficients \citep{hall2007methodology, zhou2023functional}, while (S3) extends these conditions to the interaction surface through polynomial decay of its tensor-product coefficients. Although the polynomial decay conditions in (S2)–(S3) appear weaker than those commonly imposed in functional linear regression, additional restrictions on the coefficient decay rates are implicitly imposed by (S4). These restrictions and their implications are discussed in Remark~\ref{rmk:decay_rate_restriction_of_S2S3} below. (S4) controls the stability of the regression design by allowing the minimum eigenvalue of the population Gram matrix to decrease with \(K_1\) and \(K_2\). It requires the score-estimation perturbations to be sufficiently small relative to \(\underline\lambda_K\), ensuring asymptotic invertibility of the empirical Gram matrix, while imposing a weaker restriction on the truncation errors to establish consistency of the regression estimators.

\begin{remark}
\label{rmk:first_assertion_of_S4}
The first condition in Assumption (S4) imposes restrictions on the growth
of the truncation levels. To illustrate these restrictions, consider
$K_1\asymp K_2\asymp K$,
$\alpha_1=\alpha_2=\alpha>1$, and
$\underline\lambda_K\asymp K^{-\beta}$ for some $\beta>0$.
For illustration, let $m\asymp n^\delta$,
$b_S\asymp m^{-1/5}$, and
$h_{C_S}\asymp h_{C_V}\asymp n^{-1/8}$, and define
\(
q=\min\left\{2\delta/5,1/4\right\}.
\) Then \(
r_{S,n,K_1}\vee r_{V,n,K_2}
=O(K^{\alpha+1}n^{-q}).
\) Consequently, the first condition in (S4) is satisfied
provided that \(
K=o\left(n^{q/(\alpha+\beta+2)}\right).
\) Thus, faster eigenvalue decay (larger \(\alpha\)) or a more rapidly vanishing minimum eigenvalue of the population Gram matrix (larger \(\beta\)) imposes a more restrictive growth rate on \(K\).

The exponent \(q\) reflects the combined effects of observation density and smoothing. In particular, estimation of the latent variance process requires an additional reconstruction step based on the estimated mean trajectories, introducing an error of order \(\theta_{m,2}=b_S^2+(mb_S)^{-1/2}\). With \(b_S\asymp m^{-1/5}\), this error is of order \(m^{-2/5}\asymp n^{-2\delta/5}\), whereas the covariance smoothing errors are of order \(n^{-1/4}\) under the specified bandwidth choices. Consequently, the relative growth rates of \(n\) and \(m\) play an important role in determining \(q\). For example, \(q=1/5\) when \(\delta=1/2\), while \(q=1/4\) when \(\delta\ge5/8\). Beyond this threshold, further increases in observation density do not improve \(q\) under the fixed covariance bandwidth order \(n^{-1/8}\).

More generally, the choice of smoothing bandwidths also affects the attainable rate. For covariance bandwidths of order \(n^{-a}\), the corresponding bias contribution is of order \(n^{-2a}\), so smaller bandwidths may permit a larger \(q\), provided that the remaining stochastic errors and bandwidth conditions are appropriately controlled. Thus, the restriction on the growth of \(K\) reflects not only the eigenvalue decay and the conditioning of the population Gram matrix, but also the observation density and the smoothing procedures involved in reconstructing the latent functional predictors.
\end{remark}

\begin{remark}
\label{rmk:decay_rate_restriction_of_S2S3}
Using the same example as in remark~\ref{rmk:first_assertion_of_S4}, the second condition in (S4) requires
\(
\eta_1,\eta_2>(1+\beta-\alpha)/2
\) and \(\eta_3,\eta_4>(1+\beta-\alpha)/2,
\) in addition to $\eta_j>1$, $j=1,\ldots,4$.
Thus, (S4) implicitly imposes smoothness requirements
on the regression coefficients relative to the eigenvalue
decay and the conditioning of the population Gram matrix.

As a special case, suppose that $S$ and $V$ are independent.
Then the population Gram matrix is block diagonal.
Under the eigenvalue decay and eigengap conditions in
(R2), its smallest eigenvalue satisfies \(
\lambda_{\min}(\boldsymbol\Lambda_K)\asymp K^{-2\alpha},
\) so that $\beta=2\alpha$ can be taken. The preceding
conditions reduce to \(
\eta_1,\eta_2>(\alpha+1)/2\) and \(\eta_3,\eta_4>1+\alpha/2\). These coefficient decay requirements are comparable to those commonly imposed in functional linear regression. In particular, in this special case, the requirement on the interaction coefficients coincides with the smoothness condition imposed on the slope coefficients by \cite{hall2007methodology}, while the condition on the main-effect coefficients is slightly weaker. Slightly stronger smoothness conditions are imposed by \cite{zhou2023functional} for functional regression with discretely observed predictors.
\end{remark}

\begin{theorem}[Convergence of the coefficient functions and interaction surface]
Under the assumptions of Theorem~\ref{thm:score-convergence},
together with Assumptions (S1)--(S4),
\begin{align*}
\|\widehat{\beta}_1-\beta_1\|_2^2
&= O_p\left(A_{n,K}^2 \vee K_1^{1-2\eta_1}\right),\\
\|\widehat{\beta}_2-\beta_2\|_2^2
&= O_p\left(A_{n,K}^2 \vee K_2^{1-2\eta_2}\right),\\
\|\widehat{\gamma}-\gamma\|_2^2
&= O_p\left(
A_{n,K}^2
\vee K_1^{1-2\eta_3}
\vee K_2^{1-2\eta_4}
\right),
\end{align*}
where
\[
A_{n,K}
=
\underline{\lambda}_K^{-1/2}
\left\{
r_{S,n,K_1}
\vee r_{V,n,K_2}
\vee
K_1^{\{(1-\eta_3)\vee(1/2-\eta_1)\}-\alpha_1/2}
\vee
K_2^{\{(1-\eta_4)\vee(1/2-\eta_2)\}-\alpha_2/2}
\right\}.
\]

\end{theorem}

\section{Simulation}
\label{sec: simulation}
\subsection{Simulation setting}
\label{subsec: simulation setting}

We evaluate the finite-sample performance of the proposed method under varying strengths of the mean, variance, and interaction effects. The functional predictors are generated as 
\begin{equation}
\label{simu equa:KL expansion of mean and var process}
S_i(t)=\sum_{k=1}^{50} \chi_{ik}^S\varphi_k^S(t),
\;\;
V_i(t)=\sum_{k=1}^{50} \chi_{ik}^V\varphi_k^V(t), \quad t\in\mathcal T=[0,1]
\end{equation}
where the FPC scores are independently generated across subjects and component indices, with
\[
\begin{pmatrix}
\chi_{ik}^S\\
\chi_{ik}^V
\end{pmatrix}
\sim
N\left[
\begin{pmatrix}
0\\
0
\end{pmatrix},
\begin{pmatrix}
\rho_k^S &
\rho_{SV}\sqrt{\rho_k^S\rho_k^V}\\
\rho_{SV}\sqrt{\rho_k^S\rho_k^V} &
\rho_k^V
\end{pmatrix}
\right].
\]
Thus, $\rho_{SV}$ controls the dependence between the mean- and
variance-process scores corresponding to the same component index.
The main simulation experiments use $\rho_{SV}=0$, while an additional
simulation with $\rho_{SV}=0.5$ is reported in the Supplementary Material
to assess robustness to dependence between the two latent processes.

We generate the observed data as in \eqref{equa: unified expression of observed data}, where the specific values of $(n,m)$ are specified in the simulation
scenarios below. The processes $S_i$ and $V_i$ are generated as in
\eqref{simu equa:KL expansion of mean and var process}, and
$W_{ij}\overset{\mathrm{i.i.d.}}{\sim}N(0,0.01)$ and
$\epsilon_{ij}\overset{\mathrm{i.i.d.}}{\sim}\mathrm{Uniform}(-1,1)$
for all $i$ and $j$. For the main simulation experiments, the
observation times are generated from a jittered regular design,
\(
t_{ij}
=
({j-0.5})/{m}+\zeta_{ij}\), where \(\zeta_{ij}\overset{\mathrm{i.i.d.}}{\sim}
\mathrm{Uniform}(-0.25m^{-1},0.25m^{-1})\), \( j=1,\ldots,m.
\) As a robustness analysis, we also consider a fully random observation
design in which \(t_{ij}\overset{\mathrm{i.i.d.}}{\sim}\mathrm{Uniform}(0,1).
\) 

Except for the large-sample study, we used fixed bandwidth settings throughout the simulations. The preliminary subject-specific smoothing bandwidth was set to 5\% of the observation-domain length, while FPCA was implemented using the default bandwidth settings in the \texttt{fdapace} package, corresponding to 5\% and 10\% of the domain length for mean and covariance smoothing, respectively. Bandwidth sequences chosen to satisfy the asymptotic conditions underlying the theoretical results are used in the analysis of large sample properties, as specified therein.

We use \(L=200\) Monte Carlo replicates for all simulation studies, except for the bootstrap confidence band analysis, where \(L=100\) replicates are used. The response $Y_i$ is generated by \eqref{eq:functional_interaction_model}, where we set the intercept term $\alpha=0$.

We next describe the parameter settings used in the simulation.
The eigenfunctions and eigenvalues for the mean and variance processes
are specified as follows. We set \(
\rho_k^S=\rho_k^V=k^{-1.5},
\) and choose the eigenfunctions according to two configurations:
\begin{itemize}

\item \textbf{Basis setting 1:}\hspace{0.5em} $
\varphi_k^S(t)=\varphi_k^V(t)=\sqrt{2}\sin(k\pi t),
\text{ } t\in[0,1].
$

\item \textbf{Basis setting 2:}\hspace{0.5em} $
\varphi_k^S(t)=\sqrt{2}\sin(k\pi t);\text{ }
\varphi_1^V(t)=1,\text{ }
\varphi_{k+1}^V(t)=\sqrt{2}\cos(k\pi t),\text{ } k\ge1.
$

\end{itemize}
For the main-effect coefficient functions, we let \(
\beta_1(s)=\sum_{k=1}^{50} b_{1k}\varphi_k^S(s)\) and \(\beta_2(t)=\sum_{k=1}^{50}\) \( b_{2k}\varphi_k^V(t),
\)
where \(
b_{1k}=b_{2k}=4k^{-3}, \text{  }k\ge 1, \) with the leading coefficients modified as $b_{11}=b_{21}=1$. The interaction coefficient surface is represented as
\(
\gamma(s,t)
=
\sum_{k=1}^{50}\sum_{\ell=1}^{50}
d_{k\ell}\varphi_k^S(s)\varphi_\ell^V(t).
\)
For the baseline interaction surface, the coefficients are specified by
\(
d_{k\ell}
=
k^{-3}\ell^{-3}
\{
1+0.5\sin (0.2\pi k\ell)
\},
\) except for the leading $3\times3$ block, which is set to
\[
(d_{k\ell})_{k,\ell=1}^3
=
\begin{pmatrix}
0.80 & 0.50 & -0.30\\
-0.45 & 0.40 & 0.25\\
0.25 & -0.20 & 0.15\\
\end{pmatrix}.
\]

To systematically vary the contributions of the mean main effect,
variance main effect, and mean--variance interaction, we introduce
three scaling parameters, $\delta_S$, $\delta_V$, and $\delta_I$,
respectively. Specifically, the response is generated according to \(
Y_i
=
\delta_S\sum_{k=1}^{50} b_{1k}\chi_{ik}^S
+
\delta_V\sum_{\ell=1}^{50} b_{2\ell}\chi_{i\ell}^V
+
\delta_I
\sum_{k=1}^{50}\sum_{\ell=1}^{50}
d_{k\ell}\chi_{ik}^S\chi_{i\ell}^V
+
\varepsilon_i.
\)

For the main simulation scenarios, we set \(\delta_S=1\), \(\delta_V\in\{0,0.25,1\}\), and \(\delta_I\in\{0,1\}\). We additionally consider \((\delta_S,\delta_V,\delta_I)=(0.25,1,1)\) to examine the effect of a weak mean main effect. Throughout the simulations, \(\varepsilon_i\sim N(0,1)\), and numerical integrations are approximated using the trapezoidal rule.

\subsection{Simulation results}
\noindent\textbf{Effects of the mean, variance, and interaction components}

We first examine the finite-sample performance of the proposed model under different configuration. Specifically, we vary the relative strengths of these three components through \((\delta_S,\delta_V,\delta_I)\), allowing us to assess model performance when the variance and interaction effects are present, weak, or absent. We evaluate both the estimation accuracy of the coefficient functions and interaction surface and the fitted-response accuracy of models with different predictor specifications. For the latter, we use the relative root mean squared error (relRMSE), defined as
\[
\operatorname{relRMSE}
=
\left[
\frac{1}{L}
\sum_{l=1}^{L}
\frac{
\sum_{i=1}^{n}
\left\{
Y_i^{(l)}-\widehat{Y}_i^{(l)}
\right\}^{2}
}{
\sum_{i=1}^{n}
\left\{
Y_i^{(l)}-\overline{Y}^{(l)}
\right\}^{2}
}
\right]^{1/2},
\]
where the superscript \((l)\) indexes the Monte Carlo replications, \(L\) is the total number of replications, and \(\overline{Y}^{(l)}=n^{-1}\sum_{i=1}^{n}Y_i^{(l)}\).

Coefficient estimation accuracy is evaluated using the relative root mean integrated squared error (relRMISE). Let \( \beta_{1,\delta}=\delta_S\beta_1,\text{ } \beta_{2,\delta}=\delta_V\beta_2, \text{ and } \gamma_\delta=\delta_I\gamma,
\)
denote the true coefficient functions and interaction surface under a given simulation scenario. For \(j=1,2\), when \(\|\beta_{j,\delta}\|_2>0\), we define
\[
\operatorname{relRMISE}(\widehat\beta_j) = \frac{ \left[ L^{-1}\sum_{l=1}^L \int_{\mathcal T} \{\widehat\beta_j^{(l)}(t)-\beta_{j,\delta}(t)\}^2dt \right]^{1/2} }{ \left[\int_{\mathcal T}\beta_{j,\delta}^2(t)dt\right]^{1/2} },
\]
when $\|\gamma_\delta\|_2>0$, the relRMISE of
$\widehat{\gamma}$ is defined analogously,
replacing $\beta_{j,\delta}$ and
$\widehat{\beta}_j^{(l)}$ with $\gamma_\delta$
and $\widehat{\gamma}^{(l)}$, respectively,
and integrating over $\mathcal T^2$. 

The numerical results in Tables~\ref{tab:response_relRMSE} is computed under the scenario \((n,m)=(200,100)\). It reports the fitted-response relRMSEs for the mean-only, additive, and full models under different configurations of the mean, variance, and interaction effects. As expected, incorporating a component that is present in the data-generating model generally improves fitted-response accuracy. For examples, When all three components are present, as in \((\delta_S,\delta_V,\delta_I)=(1,1,1)\), the additive model substantially improves upon the mean-only model, while the full model yields a further reduction in relRMSE by accounting for the mean–variance interaction.


\begin{table}[!t]
\centering
\caption{relRMSEs under two basis settings and three FVE thresholds.}
\label{tab:response_relRMSE}

\small
\setlength{\tabcolsep}{5.5pt}
\renewcommand{\arraystretch}{1}

\begin{tabular}{@{}
c l c c c c
@{\hspace{25pt}}
c c c c
@{}}

\toprule

$(\delta_S,\delta_V,\delta_I)$
& Model
& Basis
& 80\%
& 85\%
& 90\%
&
$(\delta_S,\delta_V,\delta_I)$
& 80\%
& 85\%
& 90\% \\

\cmidrule(r){1-6}
\cmidrule(l){7-10}


\multirow{6}{*}{$(1,1,1)$}
& \multirow{2}{*}{Mean-only}
& 1 & 0.82 & 0.81 & 0.81
&
\multirow{6}{*}{$(1,0.25,1)$}
& 0.71 & 0.71 & 0.71 \\

&
& 2 & 0.81 & 0.81 & 0.81
&
& 0.71 & 0.71 & 0.70 \\

\cmidrule(r){2-6}

&
\multirow{2}{*}{Additive}
& 1 & 0.59 & 0.59 & 0.59
&
& 0.69 & 0.68 & 0.68 \\

&
& 2 & 0.59 & 0.59 & 0.58
&
& 0.68 & 0.68 & 0.68 \\

\cmidrule(r){2-6}

&
\multirow{2}{*}{Full}
& 1 & 0.40 & 0.39 & 0.37
&
& 0.43 & 0.42 & 0.40 \\

&
& 2 & 0.37 & 0.36 & 0.35
&
& 0.39 & 0.38 & 0.37 \\


\midrule

\multirow{6}{*}{$(1,0,1)$}
& \multirow{2}{*}{Mean-only}
& 1 & 0.70 & 0.70 & 0.70
&
\multirow{6}{*}{$(0.25,1,1)$}
& 0.98 & 0.98 & 0.98 \\

&
& 2 & 0.70 & 0.70 & 0.69
&
& 0.98 & 0.98 & 0.98 \\

\cmidrule(r){2-6}

&
\multirow{2}{*}{Additive}
& 1 & 0.69 & 0.69 & 0.69
&
& 0.71 & 0.70 & 0.70 \\

&
& 2 & 0.69 & 0.69 & 0.68
&
& 0.70 & 0.70 & 0.70 \\

\cmidrule(r){2-6}

&
\multirow{2}{*}{Full}
& 1 & 0.43 & 0.42 & 0.40
&
& 0.44 & 0.43 & 0.41 \\

&
& 2 & 0.39 & 0.38 & 0.37
&
& 0.41 & 0.40 & 0.39 \\


\midrule

\multirow{6}{*}{$(1,0,0)$}
& \multirow{2}{*}{Mean-only}
& 1 & 0.36 & 0.36 & 0.35
&
\multirow{6}{*}{$(1,1,0)$}
& 0.74 & 0.74 & 0.73 \\

&
& 2 & 0.34 & 0.34 & 0.34
&
& 0.73 & 0.73 & 0.73 \\

\cmidrule(r){2-6}

&
\multirow{2}{*}{Additive}
& 1 & 0.35 & 0.35 & 0.35
&
& 0.34 & 0.34 & 0.33 \\

&
& 2 & 0.34 & 0.34 & 0.33
&
& 0.32 & 0.32 & 0.32 \\

\cmidrule(r){2-6}

&
\multirow{2}{*}{Full}
& 1 & 0.35 & 0.34 & 0.33
&
& 0.33 & 0.33 & 0.32 \\

&
& 2 & 0.33 & 0.33 & 0.32
&
& 0.32 & 0.32 & 0.31 \\

\bottomrule

\end{tabular}
\end{table}

The results also illustrate the behavior of the models when a component is absent. Under \((1,0,1)\), adding the variance main effect alone provides little improvement over the mean-only model, whereas including the interaction substantially reduces relRMSE. In contrast, under \((1,0,0)\), where neither the variance main effect nor the interaction is present, the three models perform similarly. These patterns are consistent across the two basis settings and the FVE thresholds considered, suggesting that the additional model components provide substantial gains when they carry meaningful response information, while incurring relatively little loss when their contribution is weak or absent.

We also assess the estimation accuracy of the coefficient functions and interaction surface using relRMISEs, with detailed results reported in Table~\ref{tab:simulation} of the Supplementary Material. The main-effect coefficients generally exhibit smaller estimation errors than the interaction surface, reflecting the greater complexity of estimating a bivariate coefficient surface. The results also reveal a trade-off in FVE threshold selection, with no single threshold uniformly minimizing estimation errors across the settings considered.
\vspace{0.5em}

\noindent\textbf{Finite-sample performance under increasing sample size}

To examine the estimation behavior under the increasing-dimensional
asymptotic framework, we consider a sequence of configurations in
which $n$ and $m$ increase jointly, with $m\asymp n^{1/2}$,
while the truncation levels $K_1$ and $K_2$ increase gradually.
The smoothing bandwidths are selected in accordance with the
asymptotic conditions in Section~\ref{appen S1} of the Supplementary Material.
The resulting relative RMISEs of the estimated main-effect
coefficient functions and interaction surface decrease steadily
along this sequence, providing numerical evidence consistent
with the large-sample results in
Section~\ref{sec: consistency properties}.
The detailed simulation settings and results are provided
in the Supplementary Material.

\vspace{0.5em}

\noindent\textbf{Bootstrap-based simultaneous bands}

We evaluate the finite-sample performance of the bootstrap-based
simultaneous confidence bands under the full-signal scenario
$(\delta_S,\delta_V,\delta_I)=(1,1,1)$ and Basis setting~2,
using the jittered regular design described in
Section~\ref{subsec: simulation setting}.
We consider representative configurations of $(n,m)$ and FVE thresholds,
with $L=100$ Monte Carlo datasets and $B=1,000$ subject-level pairs
bootstrap replicates per dataset.
Performance is assessed using empirical simultaneous coverage
probabilities (SCP) and median normalized half-widths, with detailed
definitions and numerical results provided in the Supplementary Material.

Overall, the 95\% simultaneous confidence bands achieve coverage
reasonably close to the nominal level in many settings, although
their performance depends on the sampling configuration and FVE
threshold. Increasing the FVE threshold generally improves coverage
by reducing truncation bias, at the expense of wider confidence bands.
The interaction surface exhibits substantially larger normalized
band widths than the main-effect coefficient functions, reflecting
the greater uncertainty associated with estimating a two-dimensional
effect.

\section{Real data application}
\label{sec: data example}

In this section, we present two data examples to illustrate the empirical performance of the proposed method, focusing on prediction accuracy, reported in terms of the RMSE and MAE of \(\widehat{Y}_i\), and on the interpretation of the estimated coefficient functions.

\subsection{Tecator meat data}
\label{subsec:tecator}

We apply the proposed method to the Tecator meat data, comprising 215 samples with near-infrared absorbance spectra measured at 100 wavelengths from 850 to 1050 nm and chemically determined fat, protein, and moisture contents. After removing one observation from each of 22 pairs of exact duplicates in both spectra and chemical compositions, 193 distinct samples remain.

\begin{figure}[!t]
    \centering
    \includegraphics[width=0.5\textwidth]{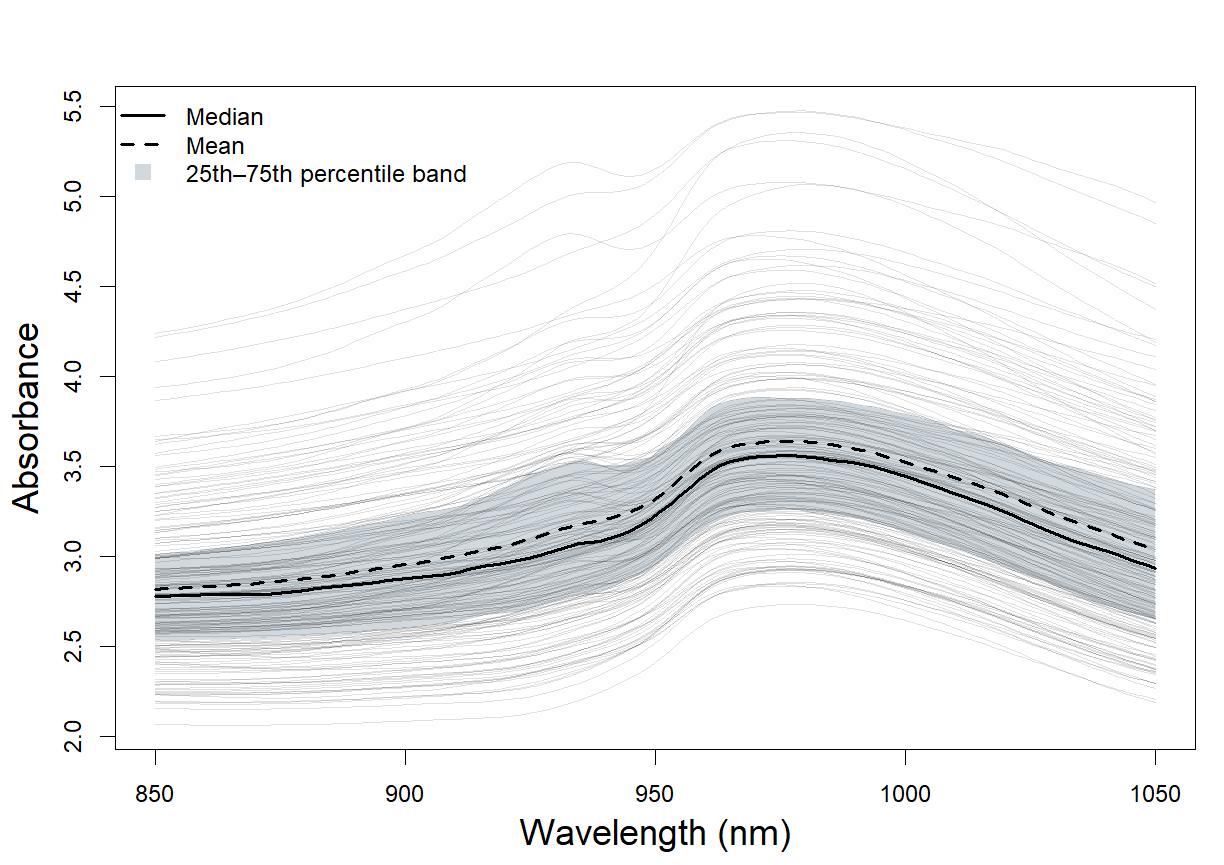}
    \caption{Near-infrared absorbance spectra for the Tecator meat data.}
    \label{fig:tecator_raw_spectra}
\end{figure}

Figure~\ref{fig:tecator_raw_spectra} displays the observed spectra with their pointwise mean, median, and interquartile band. Using fat content as the response, we examine the predictive contributions of the latent mean and variance processes and their interaction.

\begin{figure}[!t]
    \centering

    \begin{subfigure}[t]{0.38\textwidth}
        \centering
        \includegraphics[
        width=\textwidth,
        trim=0 0 10 0,
        clip
    ]{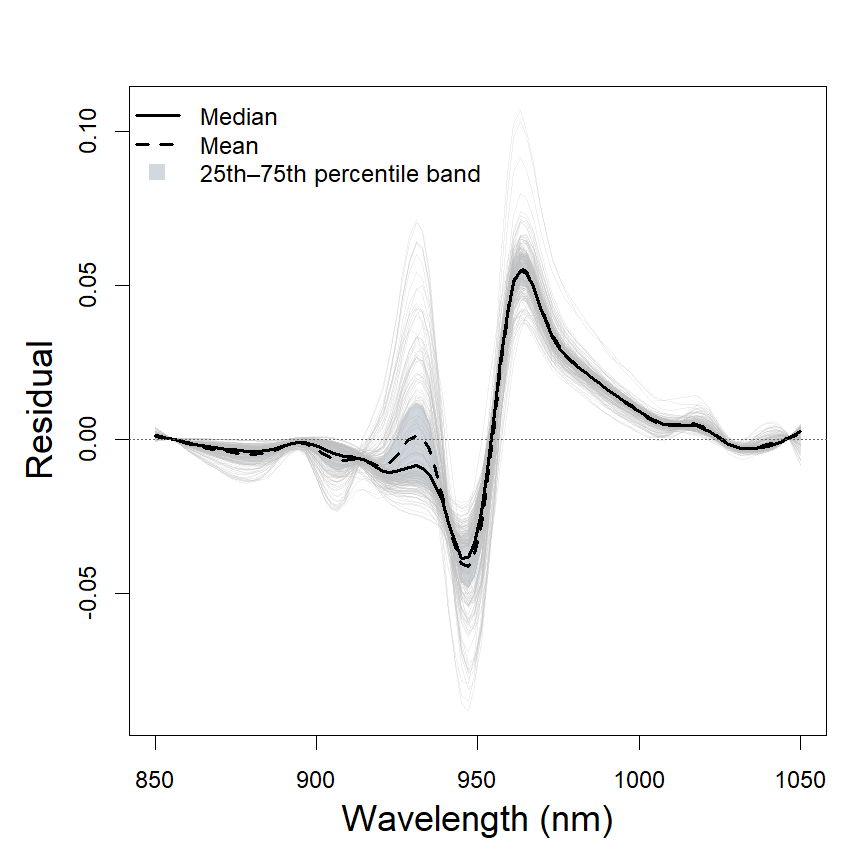}
        \caption{}
        \label{fig:tecator_residual}
    \end{subfigure}
    \hspace{3em}
    \begin{subfigure}[t]{0.38\textwidth}
        \centering
        \includegraphics[
        width=\textwidth,
        trim=10 0 0 0,
        clip
    ]{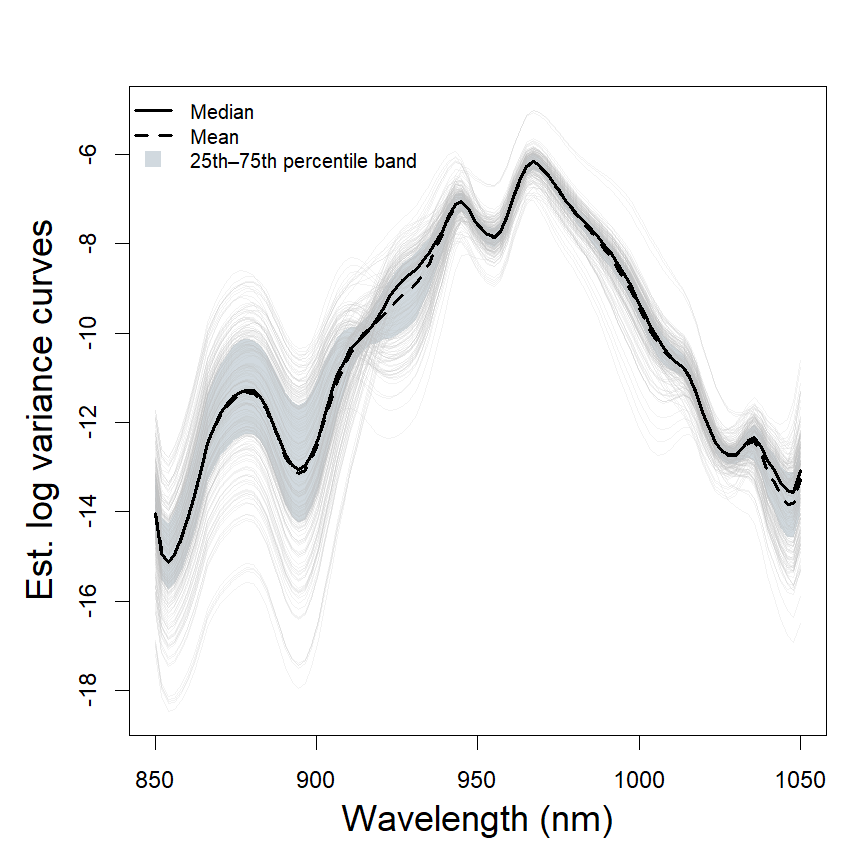}
        \caption{}
        \label{fig:tecator_variance}
    \end{subfigure}

    \caption{(a) Residual curves obtained after preliminary smoothing of the individual spectra. (b) Estimated
    subject-specific variance curves.}
    \label{fig:tecator_variance_construction}
\end{figure}

Figure~\ref{fig:tecator_variance_construction} displays the residual curves obtained from preliminary local linear smoothing and the corresponding estimated variance processes. To handle near-zero residuals, we use the stabilized transformation \(\log\{(X_i(t_{ij})-\widetilde S_i(t_{ij}))^2+10^{-8}\}\), following \citet{muller2006functional}. Both panels reveal pronounced wavelength-dependent variation, particularly around 930–970 nm. Previous spectroscopic studies have also identified a fat-associated absorption feature near 930 nm and substantial variation in spectral shape around 970 nm \citep{giglio2026demystifying}. These features motivate examining whether the variance process and its interaction with the mean process provide additional predictive information.

We compare the proposed method with partial least squares (PLS) regression \citep{wold2001pls} and functional quadratic regression (FQM) \citep{yao2010functional}. Prediction accuracy is evaluated using leave-one-out cross-validation (LOOCV) for all models. For PLS, the number of components is selected by five-fold cross-validation within each LOOCV training sample, using the \texttt{plsr} function in the R package \texttt{pls} \citep{mevik2007pls}. FQM is fitted by ordinary least squares using the mean-process FPC scores and their quadratic terms, with FPCA performed separately within each training sample. Following the default convention in \texttt{fdapace}, bandwidths for one- and two-dimensional smoothing were set to 5\% and 10\% of the spectral-domain range, respectively. For each combination of \(K_S\), \(K_V\), and model specification, the preliminary smoothing bandwidth \(b_S\) was selected from \(\{4,6,8,10,12,14,16\}\) nm by five-fold cross-validation within each LOOCV training sample, minimizing prediction RMSE.

We considered \(K_S=1,\ldots,10\), matching the range of mean-process truncation levels examined in the functional quadratic regression analysis of the Tecator data by \cite{yao2010functional}, and used the same range for the proposed models. For the variance process, FVE thresholds of 85\%, 90\%, 95\%, and 99\% yielded \(K_V=2,2,3,\) and \(5\), respectively. Accordingly, we considered \(K_V=2,\ldots,5\), covering the range of variance-process truncation levels implied by these FVE thresholds.

\begin{figure}[!t]
    \centering

    \begin{subfigure}{0.4\textwidth}
        \centering
        \includegraphics[width=\textwidth]
        {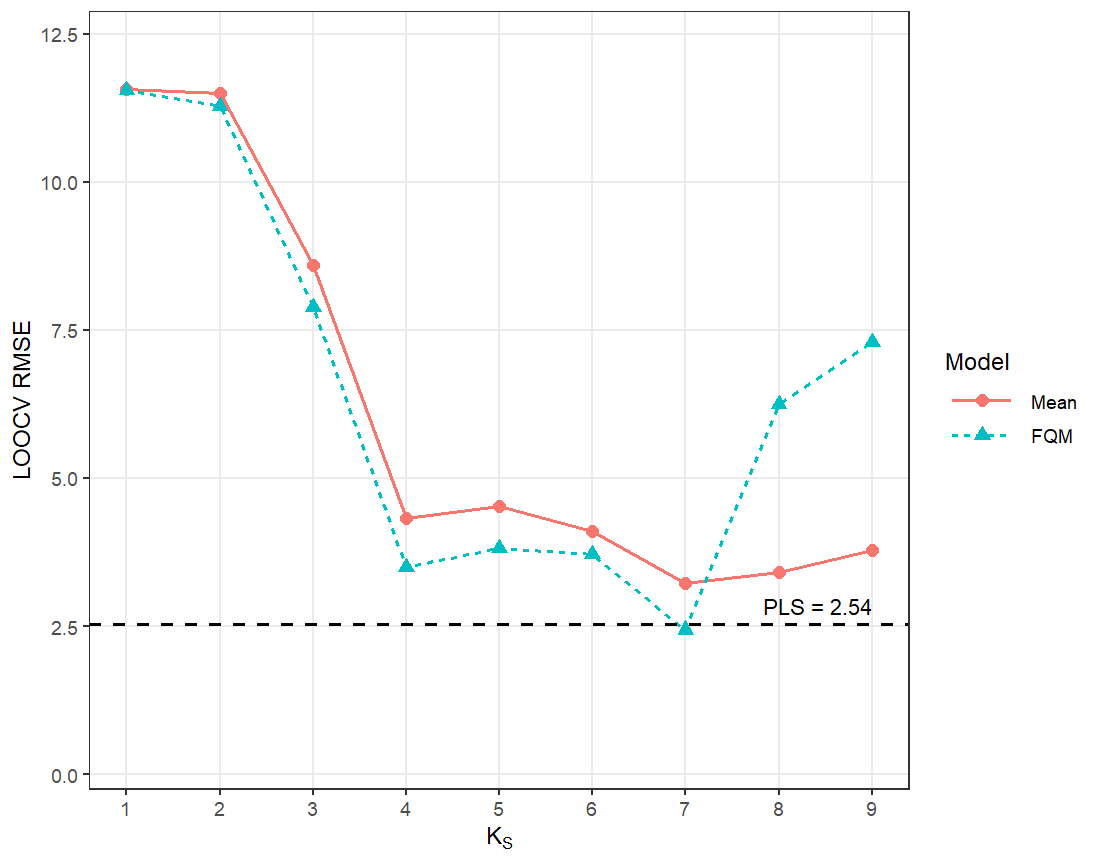}
        \caption{}
        \label{fig:tecator_fat_baseline}
    \end{subfigure}
    \hspace{1.5em}
    \begin{subfigure}{0.4\textwidth}
        \centering
        \includegraphics[width=\textwidth]
        {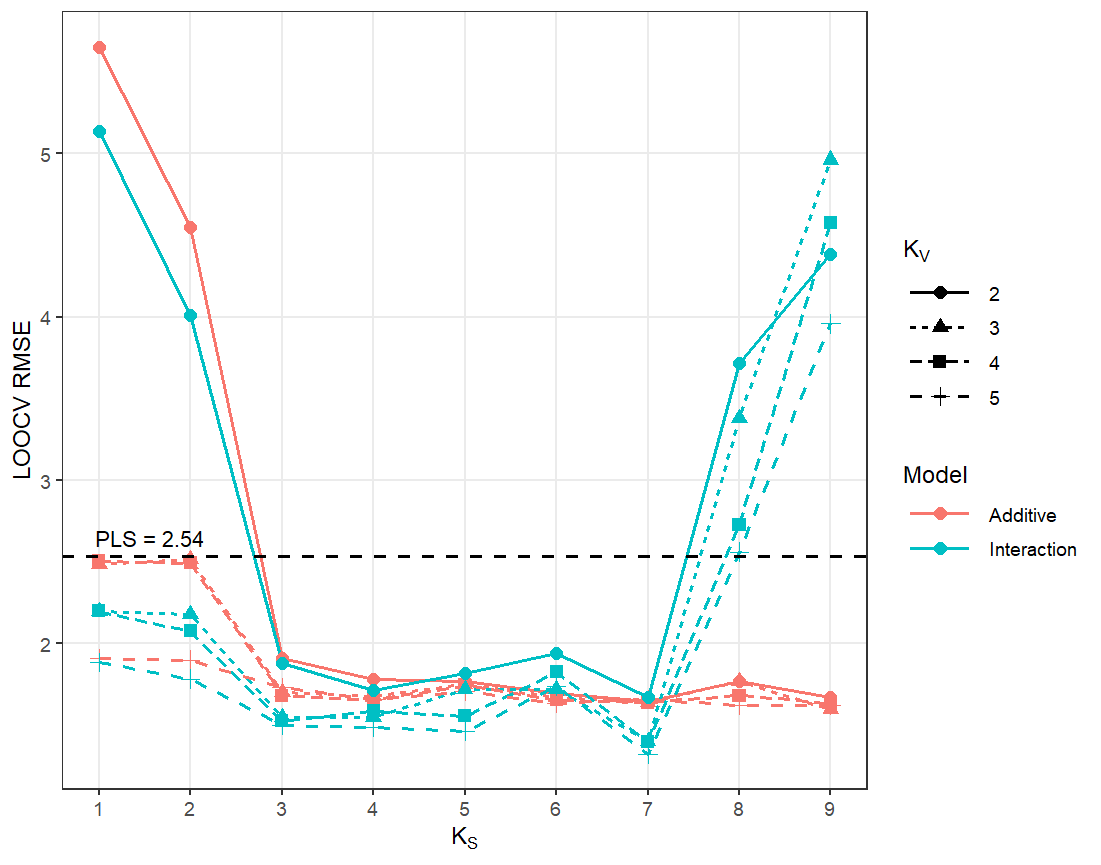}
        \caption{}
        \label{fig:tecator_fat_proposed}
    \end{subfigure}

    \caption{Leave-one-out prediction RMSE for the Tecator data
(fat content): (a) Mean and FQM models;
(b) Additive and Interaction models with $K_V=2,\ldots,5$.
The dashed horizontal line indicates the PLS benchmark.
The panels use different vertical scales.}
    \label{fig:tecator_fat_prediction}
\end{figure}

Prediction accuracy was evaluated using the root mean squared error,
$\mathrm{RMSE}=\{n^{-1}$ $\sum_{i=1}^n(Y_i-\widehat{Y}_{i,-i})^2\}^{1/2}$,
and the mean absolute error,
$\mathrm{MAE}=n^{-1}\sum_{i=1}^n|Y_i-\widehat{Y}_{i,-i}|$,
where $n=193$ and $\widehat{Y}_{i,-i}$ denotes the prediction for
subject $i$ obtained from the model fitted without that subject.
Incorporating the latent variance process substantially improved
prediction accuracy over mean-only models, incorporating
interactions further improved prediction accuracy in 22 of
the 28 configurations with $K_S\leq 7$, although performance
deteriorated at larger truncation levels. For example, at
$(K_S,K_V)=(7,5)$, the RMSE decreased from 1.66 for the
additive model to 1.32 for the interaction model, a 20.5\%
reduction. Across all configurations, the lowest RMSEs were
1.60 for the additive model and 1.32 for the interaction
model, corresponding to a 17.5\% reduction and the lowest RMSE of the interaction model
represents a 48\% reduction relative to the PLS benchmark (2.54). Figure~\ref{fig:tecator_fat_prediction} illustrates RMSE results
for $K_S\leq 9$, excluding $K_S=10$ to avoid excessive expansion
of the vertical scale. Detailed LOOCV prediction results are
reported in Table~\ref{tab:tecator_fat_fixedK} of the Supplementary Material.

\begin{figure}[!t]
    \centering


    \begin{subfigure}[t]{0.4\textwidth}
        \centering
        \includegraphics[
        width=\textwidth,
        trim=0 20 0 10,
        clip
    ]{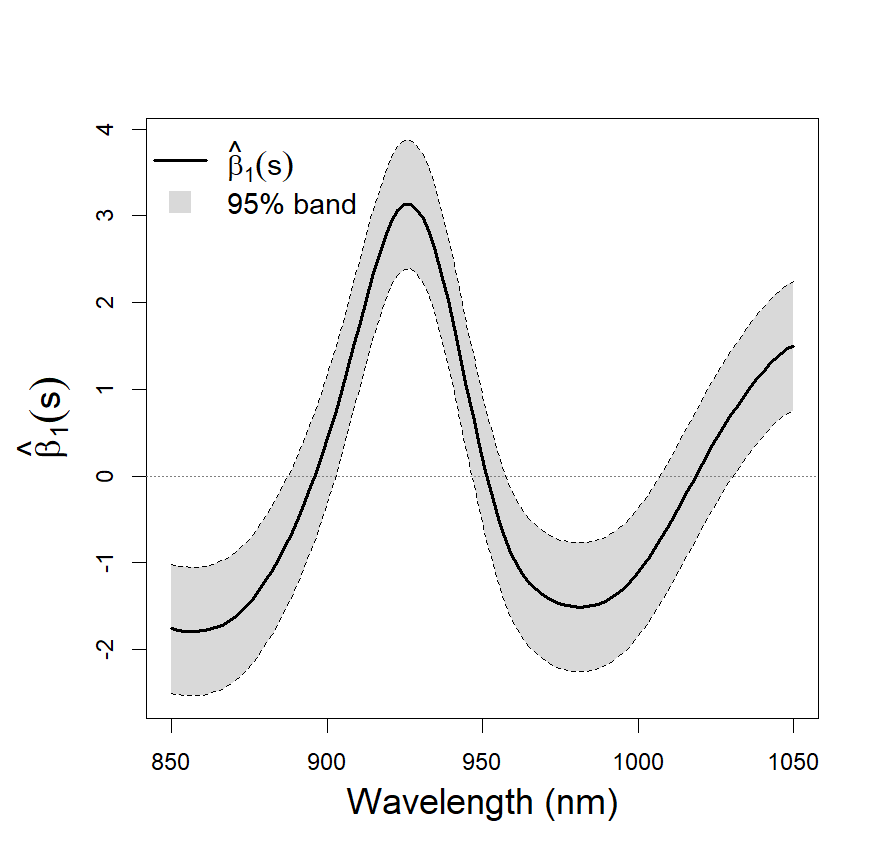}
        \caption{$\widehat{\beta}_1(s)$.}
        \label{fig:tecator_fat_beta1}
    \end{subfigure}
    \hspace{0.005\textwidth}
    \begin{subfigure}[t]{0.4\textwidth}
        \centering
        \includegraphics[
        width=\textwidth,
        trim=0 20 0 10,
        clip
    ]{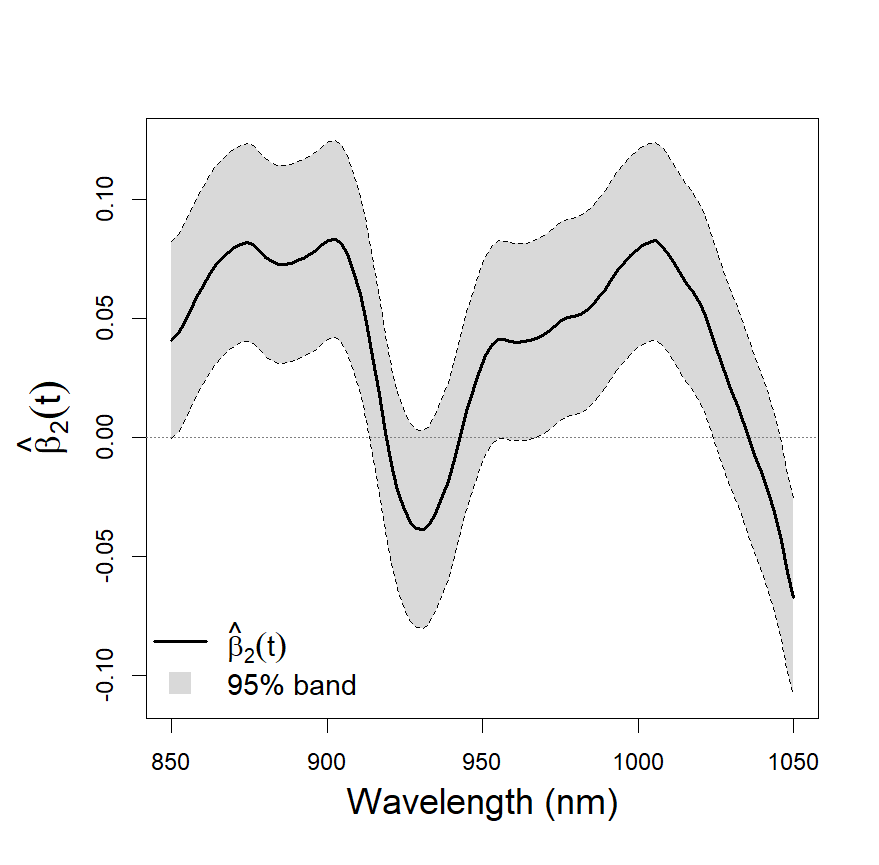}
        \caption{$\widehat{\beta}_2(t)$.}
        \label{fig:tecator_fat_beta2}
    \end{subfigure}


    \begin{subfigure}[t]{0.4\textwidth}
        \centering
        \includegraphics[
        width=\textwidth,
        trim=0 10 0 10,
        clip
    ]{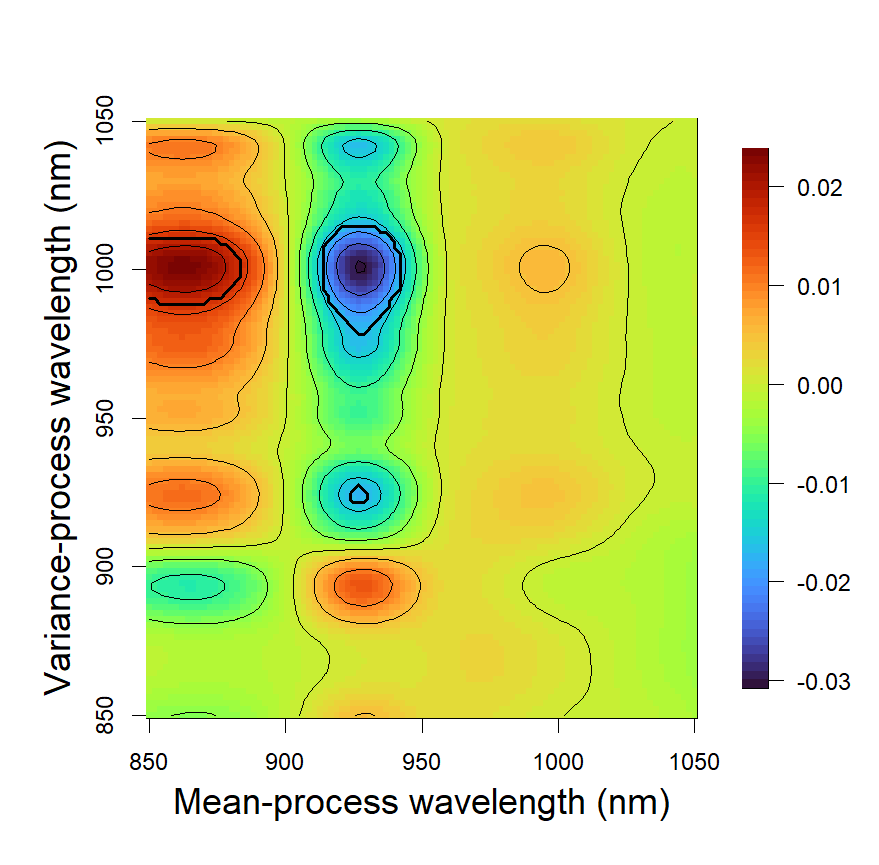}
        \caption{$\widehat{\gamma}(s,t)$.}
        \label{fig:tecator_fat_gamma}
    \end{subfigure}
    \hspace{0.005\textwidth}
    \begin{subfigure}[t]{0.4\textwidth}
        \centering
        \includegraphics[
        width=\textwidth,
        trim=0 10 0 10,
        clip
    ]{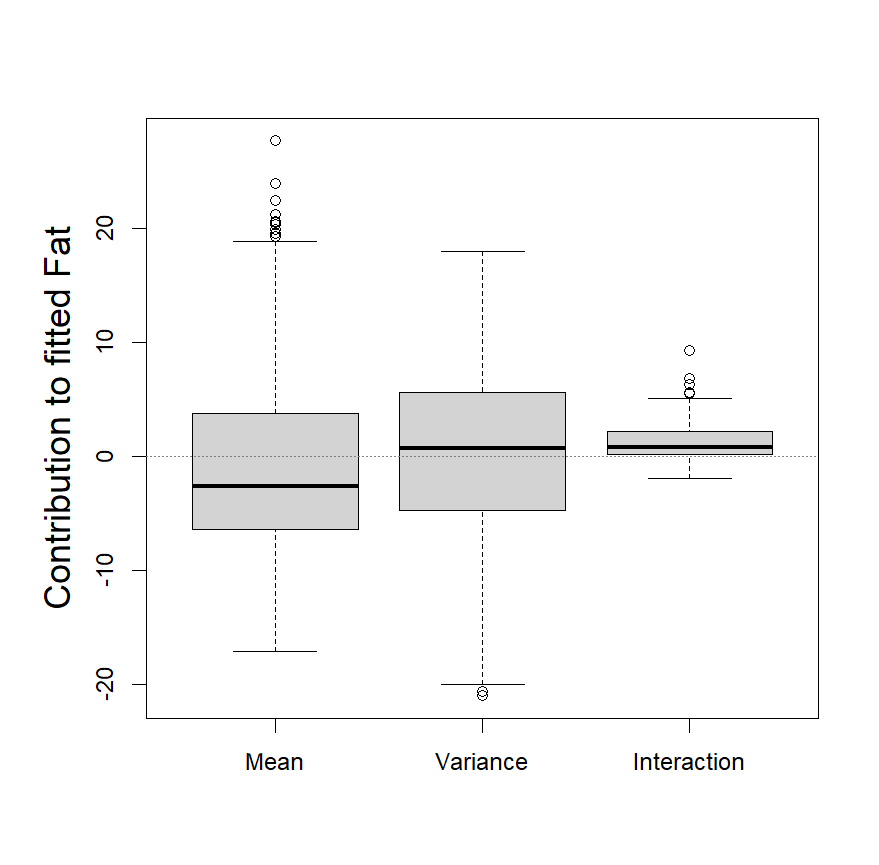}
        \caption{Subject-specific fitted contributions.}
        \label{fig:tecator_fat_contributions}
    \end{subfigure}

    \vspace{-0.15em}

    \caption{
    Estimated effects for the Tecator data with fat content as the response,
    using $(K_S,K_V)=(4,3)$.
    Panels (a) and (b) show the estimated mean- and variance-process
    coefficient functions, respectively, together with their 95\%
    bootstrap confidence bands.
    Panel (c) shows the estimated interaction surface; thick black contours
    delineate regions where the 95\% bootstrap confidence band excludes
    zero, while thin black contours represent levels of the estimated
    interaction surface.
    Panel (d) shows the distributions across subjects of the fitted
    contributions from each effect.
    }
    \label{fig:tecator_fat_effects}

\end{figure}

To further examine how the mean and variance processes contribute to the response, Figure~\ref{fig:tecator_fat_effects} shows the estimated coefficient functions and their 95\% simultaneous confidence bands. Although the interaction model attained its lowest prediction error at \((K_S,K_V)=(7,5)\), this larger truncation resulted in substantially greater estimation variability in the coefficient functions. We therefore report the coefficient estimates under the more parsimonious specification \((K_S,K_V)=(4,3)\), which involves 20 rather than 48 regression parameters. This choice also retains the fourth mean-process component, which appears to contain substantial predictive information: increasing \(K_S\) from 3 to 4 reduced the RMSE from 8.60 to 4.33 for the mean-process model and from 7.89 to 3.49 for FQM. The \((4,3)\) specification also retained strong predictive performance for the proposed interaction model.

The estimated mean-process coefficient $\widehat{\beta}_1(s)$ exhibits positive effects around 900--950 nm and negative effects immediately above 950 nm, with its 95\% simultaneous confidence band excluding zero over portions of these regions. The variance-process coefficient $\widehat{\beta}_2(t)$ is predominantly positive, although its confidence band includes zero over several wavelength ranges. The estimated interaction surface $\widehat{\gamma}(s,t)$ exhibits localized significant effects around variance-process wavelengths of 920 and 1000 nm, with their magnitude and direction varying across the mean-process wavelength. These findings suggest that the contribution of spectral variability depends on the underlying mean spectral pattern rather than acting purely additively.

To assess the magnitude of these effects on the response scale, we also examined the subject-specific fitted contributions of the three model components, shown in Figure~\ref{fig:tecator_fat_contributions}. Specifically, for subject \(i\), the mean-process, variance-process, and interaction contributions were defined as \(\widehat C_i^S=\sum_{k=1}^{4}\widehat b_{k}\widehat\chi_{ik}^S\), \(\widehat C_i^V=\sum_{\ell=1}^{3}\widehat c_{\ell}\widehat\chi_{i\ell}^V\) and \(\widehat C_i^I=\sum_{k=1}^{4}\sum_{\ell=1}^{3}\widehat d_{k\ell}\widehat\chi_{ik}^S\widehat\chi_{i\ell}^V
\), respectively, so that \(\widehat Y_i-\widehat\alpha=\widehat C_i^S+\widehat C_i^V+\widehat C_i^I\). The variance-process contribution was comparable in magnitude to that of the mean process: the standard deviations of the fitted contributions were 8.53 and 7.33 for the mean and variance terms, respectively, while their mean absolute contributions were 6.68 and 5.97. The interaction contribution was smaller overall, with a standard deviation of 1.73 and a mean absolute contribution of 1.52, but was not uniformly negligible across subjects, ranging from \(-1.91\) to \(9.34\). These results indicate that subject-specific spectral variability contributes materially to the fitted response, while the interaction term provides a smaller but non-negligible additional contribution for some subjects.

\subsection{R-peak-to-R-peak interval data}

We apply our method to the open dataset \emph{Cardiorespiratory measurement from graded cycloergometer exercise testing} \citep{chabert2022cardiorespiratory, mongin2022decrease}, comprising 18 adolescent athletes (10 males and 8 females; mean age $15.2 \pm 2$ years). Beat-to-beat RR intervals, defined as the time between consecutive R-wave peaks on an electrocardiogram (ECG), were recorded around exercise onset ($t=0$). We analyze the window from 35 seconds before to 140 seconds after onset, with subject-specific, irregularly spaced observation times. The scalar response is the power at the first ventilatory threshold, $P_{\mathrm{vt1}}$, and the functional predictors are the estimated mean and variance processes of the RR trajectories.

Given the small sample size, we focus on the additive mean--variance model, as preliminary analyses across several truncation levels showed no consistent predictive advantage from including all interactions, with substantial deterioration in some higher-dimensional specifications. We also omit bootstrap simultaneous confidence bands, which were relatively wide and provided limited information. Instead, the estimated coefficient functions are presented descriptively to illustrate the temporal patterns captured by the fitted model.

\begin{figure}[!t]
\centering

\includegraphics[
    width=0.5\textwidth,
    trim=25 20 25 15,
    clip
]{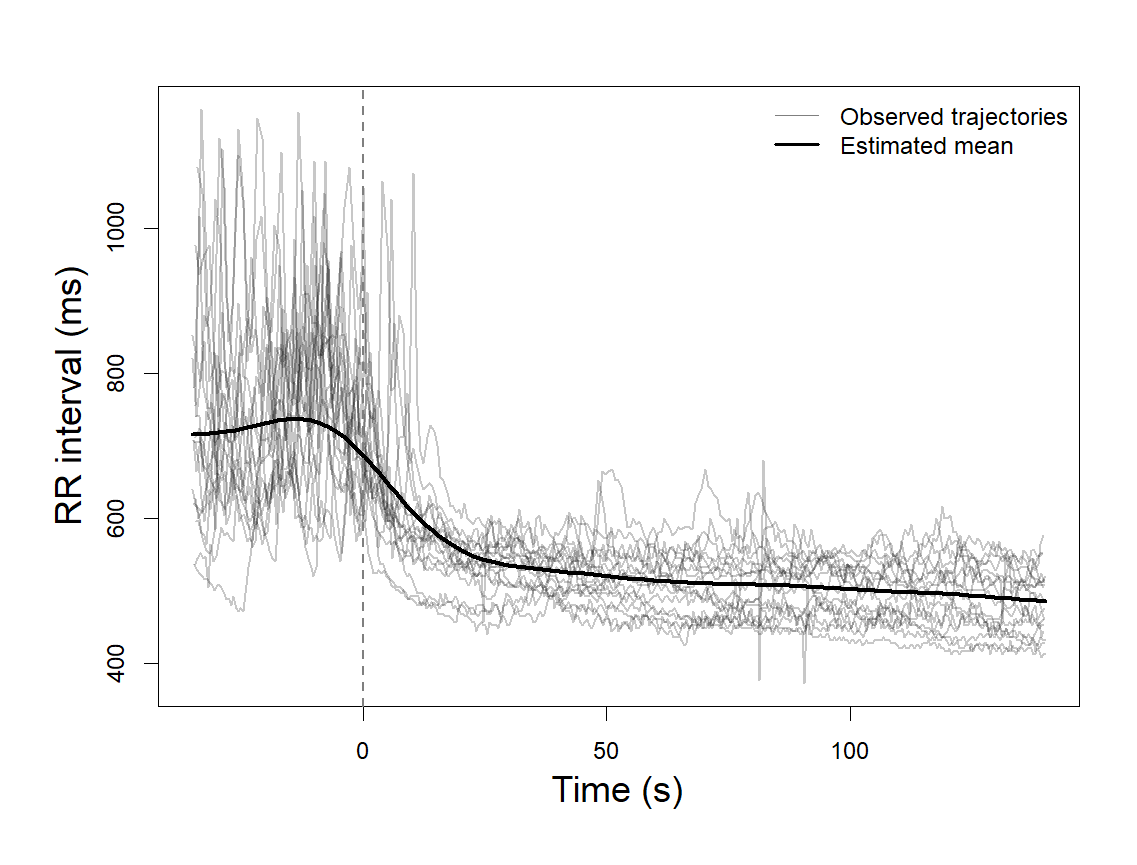}

\caption{
Observed RR interval trajectories aligned at the onset of exercise. The black curve represents the estimated population mean function obtained from FPCA of the raw RR trajectories.
}
\label{fig:raw_rr_curves}

\end{figure}

\begin{figure}[!t]
\centering

\begin{subfigure}{0.325\textwidth}
    \centering
    \includegraphics[
        width=\textwidth,
        trim=25 20 25 15,
        clip
    ]{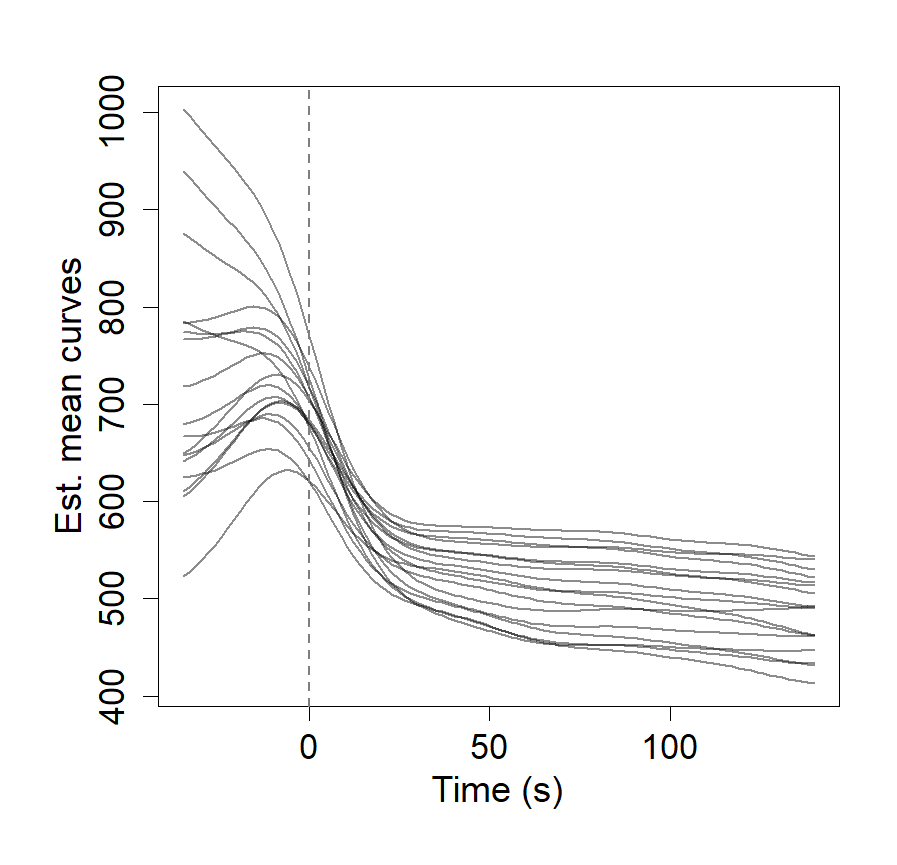}
    \caption{}
    \label{fig:rr_mean_processes}
\end{subfigure}
\hspace{2em}
\begin{subfigure}{0.325\textwidth}
    \centering
    \includegraphics[
        width=\textwidth,
        trim=25 20 25 15,
        clip
    ]{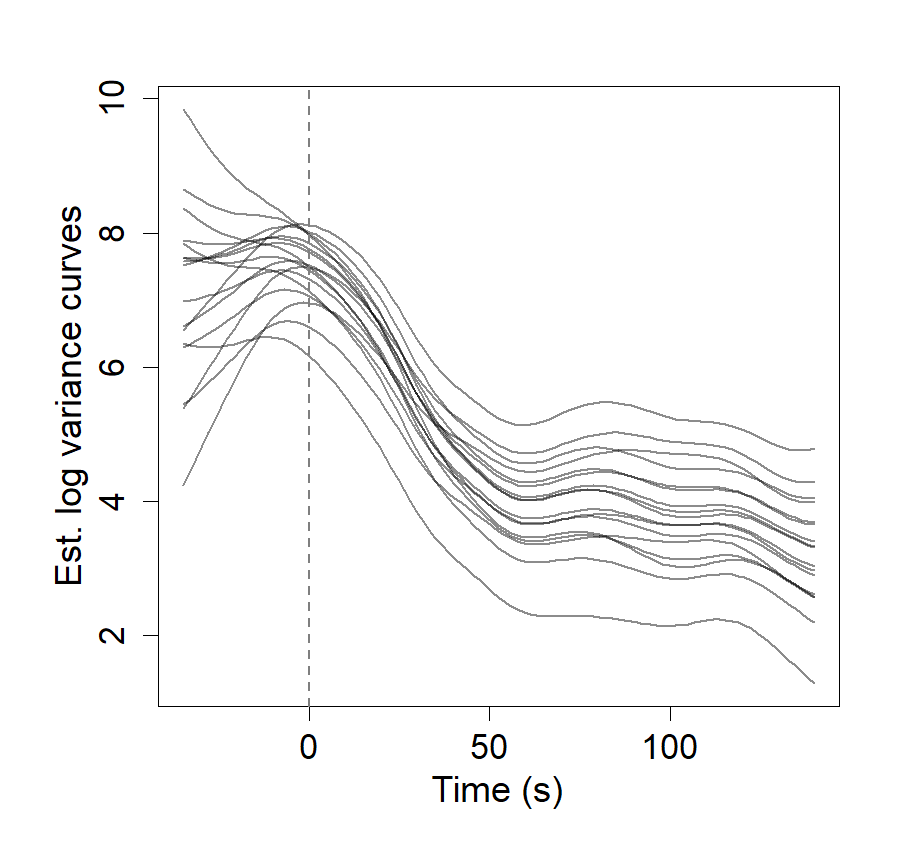}
    \caption{}
    \label{fig:rr_variance_processes}
\end{subfigure}

\caption{
Estimated functional predictors for the RR interval data.
Panels (a) and (b) display the estimated mean and variance processes, respectively.
}
\label{fig:rr_functional_predictors}

\end{figure}

The raw RR trajectories in Figure~\ref{fig:raw_rr_curves} exhibit substantial variability across individuals, particularly before the onset of exercise. A pronounced decline is observed shortly after $t = 0$, followed by a gradual stabilization phase, reflecting the expected acceleration of heart rate and subsequent recovery. The estimated mean and variance processes in Figures~\ref{fig:rr_mean_processes} and \ref{fig:rr_variance_processes} reveal clear temporal patterns in both the first- and second-order structures of the RR trajectories. In particular, the mean curves capture the overall decreasing trend after exercise onset, while the variance curves indicate time-varying variability, with elevated fluctuations around the onset of exercise followed by stabilization.

To assess predictive performance, we conduct a leave-one-subject-out cross-validation (LOOCV) analysis following the same nested procedure as in the preceding application. For the preliminary smoothing step, we set \(b_{S,i}=c|\mathcal{T}|\), where \(|\mathcal{T}|=175\) seconds denotes the length of the observation domain, and select \(c\) from \(\{0.05,0.075,0.10,0.125,0.15,0.175,0.20\}\) using inner LOOCV.

Table~\ref{tab:rr_loocv} presents the LOOCV prediction errors across four FVE thresholds. The additive mean--variance model consistently achieves the lowest RMSE and MAE, outperforming both the mean-only model and a model incorporating a scalar summary of detrended variability (see Supplementary Material for the description of this model). These results suggest that the predictive contribution of subject-specific variability is robust across the FVE thresholds considered and that modeling variability as a functional process provides additional information beyond a scalar summary.

\begin{table}[!t]
\centering
\caption{LOOCV prediction errors for the additive functional regression model and benchmark models
for the RR interval data under different FVE thresholds.
Each entry reports RMSE, with MAE in parentheses.}
\label{tab:rr_loocv}
\small
\setlength{\tabcolsep}{7pt}
\begin{tabular}{lcccc}
\toprule
& \multicolumn{4}{c}{FVE threshold} \\
\cmidrule(lr){2-5}
Model
& 80\%
& 85\%
& 90\%
& 95\% \\
\midrule
Mean
& 40.90 (30.47)
& 40.90 (30.47)
& 43.89 (33.13)
& 38.17 (29.18) \\

Mean + SDRR
& 43.49 (34.09)
& 43.49 (34.09)
& 42.18 (33.33)
& 33.76 (29.03) \\

\midrule
Mean + Variance
& \textbf{28.47 (22.23)}
& \textbf{29.03 (22.48)}
& \textbf{33.91 (28.32)}
& \textbf{31.35 (26.60)} \\
\bottomrule
\end{tabular}
\end{table}

\begin{figure}[!t]
\centering
\begin{minipage}{0.4\textwidth}
\centering
\includegraphics[width=\textwidth]{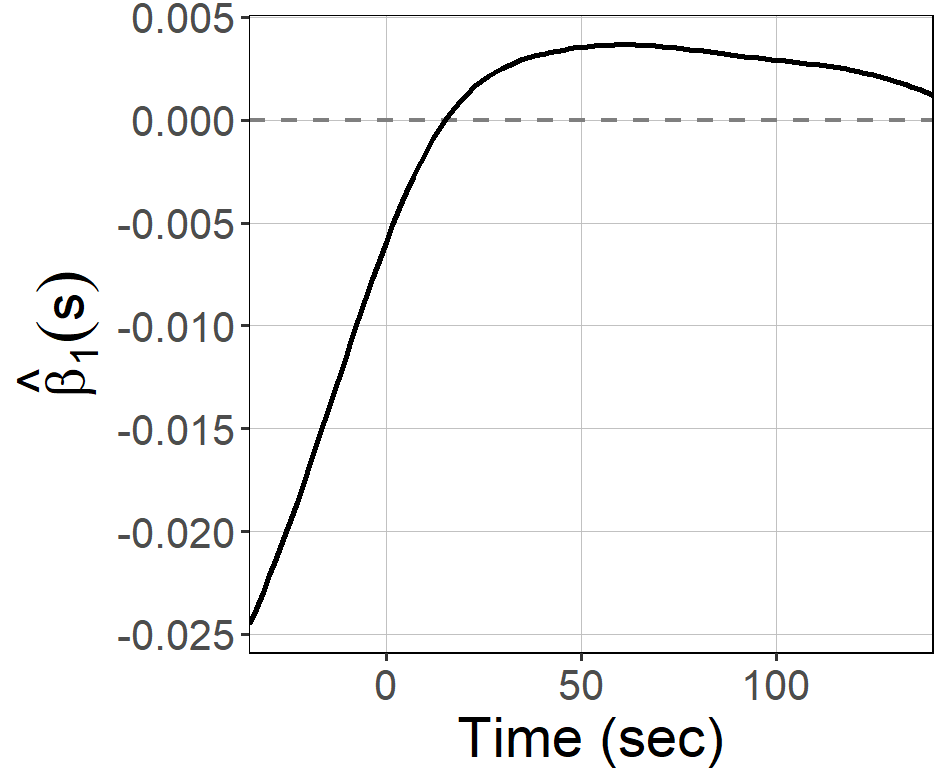}\
(a)
\end{minipage}
\hspace{0.055\textwidth}
\begin{minipage}{0.4\textwidth}
\centering
\includegraphics[width=\textwidth]{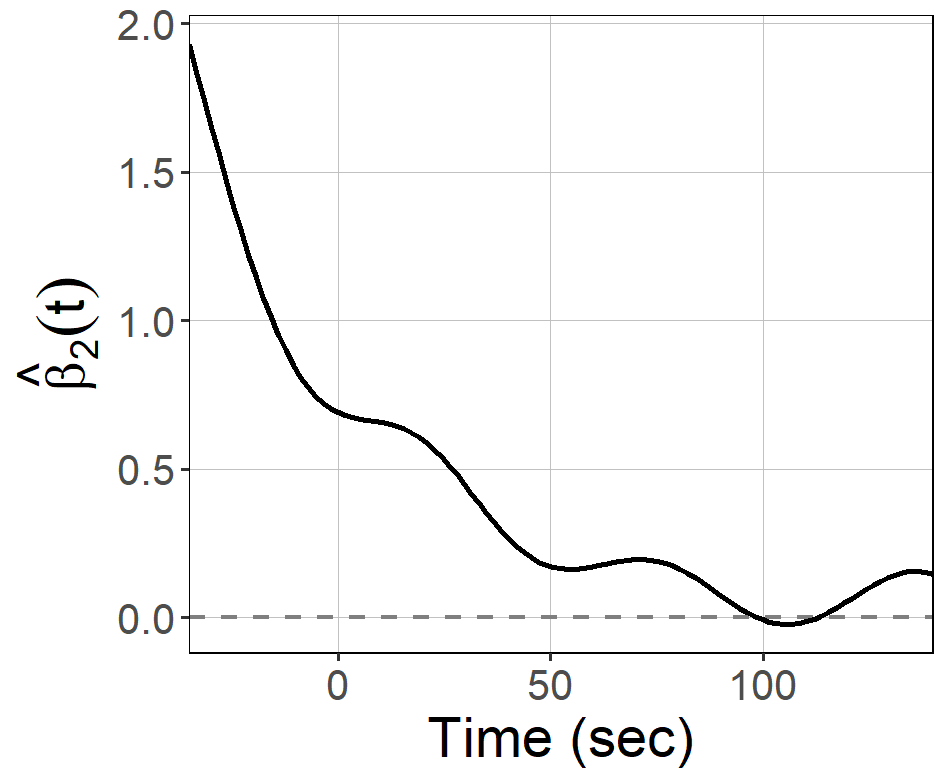}
(b)
\end{minipage}
\caption{Estimated coefficient functions (a) $\widehat{\beta}_1(s)$ and (b) $\widehat{\beta}_2(t)$ for the RR interval data.}
\label{fig:rr beta}
\end{figure}

Figure~\ref{fig:rr beta} presents the estimated coefficient functions from the additive mean--variance model at an FVE threshold of 85\% for both processes. The estimated mean coefficient $\widehat{\beta}_1(s)$ is negative near exercise onset and becomes positive afterward, while the variance coefficient $\widehat{\beta}_2(t)$ exhibits a pronounced positive peak around $t=0$. This finding is of interest in light of \cite{mongin2022decrease}, who reported an inverse association between the rate of decline in heart rate variability (HRV) and $P_{\mathrm{vt1}}$. Although interpreted descriptively given the small sample size, these estimates illustrate how modeling variability as a functional process allows us to examine when and how it is associated with the response, beyond what a scalar summary can reveal.

On the response scale, the root-mean-square fitted contributions across subjects, based on the subject-specific contributions defined in Section~\ref{subsec:tecator}, were 53.8 W and 43.6 W for the mean and variance processes, respectively, indicating contributions of comparable magnitude.

Although the full interaction model did not consistently improve
prediction for the RR interval data, this does not preclude the
predictive relevance of particular mean--variance interactions.
To explore this possibility, we conducted an exhaustive search
over subsets of FPC scores and their pairwise interactions.
The best-performing interaction specification achieved a
substantially lower LOOCV RMSE than the best additive specification
(9.26 versus 25.08), and several interaction terms appeared
consistently among the top-ranked models.
These findings suggest that selected mean--variance interactions
may provide additional predictive information.
However, given the small sample size and extensive model search,
the results should be interpreted as exploratory.
Details are provided in the Supplementary Material.

\section{Conclusion}
\label{sec: conclusion}

We proposed a functional regression framework incorporating latent
mean and variance processes and their interaction as predictors
of a scalar response. Estimation method combines FPCA and least squares and we established
estimation convergence results and constructed bootstrap-based
simultaneous confidence bands.

Simulation studies supported the finite-sample performance of the
proposed methods. In the Tecator application, incorporating
functional variability and its interaction with the mean improved
prediction over functional quadratic regression and partial least
squares. The RR interval analysis demonstrated the predictive value
of functional variability beyond scalar summaries, with exploratory
results suggesting that selected interactions may be informative
even when their corresponding main effects are weak.

Future work includes establishing the theoretical validity of the
bootstrap confidence bands, developing regularized estimation and
variable selection for interaction effects without restrictive
heredity constraints, and extending the framework to generalized
responses and other functional data settings.

\appendix

\section*{Appendix}
\section{Estimation details}
\setcounter{equation}{0}
\renewcommand{\theequation}{A.\arabic{equation}}
\label{appen A}
\noindent
Let $\kappa_1(\cdot)$ and $\kappa_2(\cdot,\cdot)$ denote nonnegative univariate and bivariate kernel functions satisfying assumptions (B2.1)–(B2.2) stated in Section~\ref{appen S1} of the Supplementary Material. These serve as weighting functions in the local linear smoothing procedures. The preliminary subject-specific smoothing bandwidth is denoted by \(b_{S,i}\), while the bandwidths used for estimating the mean and covariance functions of the mean and variance processes are denoted by \(h_{\mu_S}, h_{C_S}, h_{\mu_V}\), and \(h_{C_V}\), respectively. In finite-sample implementation, the bandwidths may be specified according to the scale of the observation domain or selected using data-driven criteria, depending on the smoothing task and the objective of the analysis. In the asymptotic development, they are assumed to satisfy the bandwidth sequences and rate conditions specified in the corresponding assumptions of the Supplementary Material. 

\vspace{0.5em}
\noindent\textbf{(1) Estimation of subject-specific trajectories.}
For each subject $i$, we first obtain a smoothed trajectory $\widetilde S_i(t)$ from noisy observations $(t_{ij}, X_i(t_{ij}))$ by local linear scatterplot smoothing \citep{fan1996local}, that is, by minimizing
\begin{equation*}
\sum_{j=1}^{m} \kappa_1\!\left( \frac{t_{ij}-t}{b_{S,i}} \right)\!\left\{X_i(t_{ij})-\beta_{i,0}-\beta_{i,1}(t-t_{ij})\right\}^2,
\end{equation*}
with respect to $(\beta_{i,0},\beta_{i,1})$. The estimated trajectory is then $\widetilde S_i(t)=\widehat{\beta}_{i,0}(t)$.

\vspace{0.5em}
\noindent\textbf{(2) Estimation of mean functions.}
The mean functions $\mu_S(t)$ and $\mu_V(t)$ of the mean and variance processes are estimated via local linear smoothing, where $\widehat\beta_0^{S}$ and $\widehat\beta_1^{S}$ are defined as the minimizers of the respective weighted least squares criteria. The estimate is then $\widehat\mu_S(t)=\widehat\beta_0^S(t)$, and similar for $\widehat\mu_V(t)$. $X_i(t_{ij})$ stand for observed trajectories, and $\widehat Z_i(t_{ij})$ are the transformed residuals as defined in Section~\ref{sec: Estimation}.
\begin{align*}
\sum_{i=1}^n \sum_{j=1}^m \kappa_1 \!\left(\frac{t_{ij}-t}{h_{\mu_S}}\right)\!\left\{X_i(t_{ij})-\beta_0^S-\beta_1^S(t-t_{ij})\right\}^2, \\
\sum_{i=1}^n \sum_{j=1}^m \kappa_1 \!\left(\frac{t_{ij}-t}{h_{\mu_V}}\right)\!\left\{\widehat Z_i(t_{ij})-\beta_0^V-\beta_1^V(t-t_{ij})\right\}^2.
\end{align*}

\vspace{0.5em}
\noindent\textbf{(3) Estimation of covariance functions.}
The covariance surfaces $C_S(s,t)$ and $C_V(s,t)$ are estimated by local linear surface smoothing \citep{fan1996local}, using the empirical covariances 
$C_i^S(t_{ij_1},t_{ij_2})=(X_i(t_{ij_1})-\widehat\mu_S(t_{ij_1}))(X_i(t_{ij_2})-\widehat\mu_S(t_{ij_2}))$
and 
$C_i^V(t_{ij_1},t_{ij_2})=(\widehat Z_i(t_{ij_1})-\widehat\mu_V(t_{ij_1}))(\widehat Z_i(t_{ij_2})-\widehat\mu_V(t_{ij_2}))$,
excluding diagonal pairs $t_{ij_1}=t_{ij_2}$. Each surface is defined by the minimizer of
\begin{equation*}
\sum_{i=1}^n \sum_{1 \le j_1 \ne j_2 \le m} 
\kappa_2\!\left(\frac{t_{ij_1}-s}{h_{C_S}}, \frac{t_{ij_2}-t}{h_{C_S}}\right)
\!\left\{ C_i^S(t_{ij_1},t_{ij_2}) - f(\boldsymbol\beta,(s,t),(t_{ij_1},t_{ij_2})) \right\}^2,
\end{equation*}
where
\begin{equation*}
f(\boldsymbol\beta,(s,t),(t_{ij_1},t_{ij_2}))=\beta_0+\beta_{11}(s-t_{ij_1})+\beta_{12}(t-t_{ij_2}).
\end{equation*}
The estimate is thus $\widehat C_S(s,t)=\widehat\beta_0(s,t)$, and $\widehat C_V(s,t)$ is obtained analogously by repeating the same step with $h_{C_V}$ and $C_i^V(t_{ij_1}, t_{ij_2})$.

\vspace{0.5em}
\noindent\textbf{(4) Estimation of eigenfunctions and eigenvalues.}
The eigenfunctions $\widehat\varphi_k^S$ and $\widehat\varphi_k^V$ for $S$ and $V$ are obtained from the smoothed covariance surfaces $\widehat C_S$ and $\widehat C_V$ by solving
\begin{align*}
\int_{\mathcal T}\widehat C_S(s,t)\widehat\varphi_k^S(s)\,ds=\widehat\rho_k^S\,\widehat\varphi_k^S(t), \;\;
\int_{\mathcal T}\widehat C_V(s,t)\widehat\varphi_k^V(s)\,ds=\widehat\rho_k^V\,\widehat\varphi_k^V(t),
\end{align*}
subject to standard orthonormality constraints. In practice, these are solved on a discrete grid via numerical eigen-decomposition of the smoothed covariance matrices.

\bibliographystyle{chicago}      
\bibliography{references}   

\clearpage

\counterwithout{figure}{section}
\counterwithout{table}{section}

\setcounter{section}{0}
\setcounter{equation}{0}
\setcounter{table}{0}
\setcounter{figure}{0}

\renewcommand{\thesection}{S\arabic{section}}
\renewcommand{\theequation}{S\arabic{section}.\arabic{equation}}
\renewcommand{\thetable}{S\arabic{table}}
\renewcommand{\thefigure}{S\arabic{figure}}

\titleformat{\section}
  {\normalfont\bfseries\large}
  {\thesection.}
  {0.5em}
  {}

\titleformat{\subsection}
  {\normalfont\bfseries\itshape\normalsize}
  {\thesubsection.}
  {0.5em}
  {}

\centerline{\large\bf Supplementary Material for Functional Regression with}
\centerline{\large\bf Interacting Latent Mean and Variance Processes}

\setcounter{section}{0}
\setcounter{equation}{0}
\def\theequation{S\arabic{section}.\arabic{equation}}
\def\thesection{S\arabic{section}}

\fontsize{12}{14pt plus.8pt minus .6pt}\selectfont

\section{Auxiliary assumptions and notations}
\label{appen S1}
\setcounter{equation}{0}
\noindent
The following assumptions are adapted from \cite{yao2006penalized}
and \cite{muller2006functional}. Assumptions (A2.1)--(A2.6)
consolidate the bandwidth conditions from both sources.
The remaining assumptions follow the original numbering
in \cite{muller2006functional}, except that (A5)
corresponds to the original (A6), while (B1) and (B2)
correspond to the original (B2.1) and (B2.2), respectively.

We modify (A1) by requiring the first derivative of the
variance process to have a finite second moment in $L^2$,
which facilitates the control of Riemann-sum approximation
errors. In addition, we assume uniformly bounded second
derivatives of the mean functions and second-order partial
derivatives of the covariance functions for both the mean
and variance processes, ensuring the required smoothness
for local linear estimation and eigenfunction analysis. Unlike the common observation grid adopted in
\cite{muller2006functional}, (A3) allows subject-specific
observation times under uniform spacing conditions.
We also strengthen (A4) by assuming a uniformly bounded
eighth moment of $X(t)$, rather than the fourth moment
required in \cite{muller2006functional}, to establish
stochastic bounds for the estimated mean-process FPC
scores uniformly over subjects and components via a
union bound.

Finally, we modify the kernel conditions (B1)--(B2)
by imposing continuity, uniform Lipschitz continuity
of the bivariate kernel in its first argument, and
positivity at the origin. The Lipschitz condition
facilitates the control of weak derivatives of the
smoothing weights and estimated covariance surfaces,
which is needed to transfer continuous $L^2$ convergence
rates to discrete observation grids. The positivity
condition ensures uniform nonsingularity of the local
linear design matrices, thereby supporting the required
smoothing-weight bounds. The absolute-integrability
conditions on the Fourier transforms imposed in
\cite{muller2006functional} are omitted, as they are
not needed in our direct $L^2$-based analysis.

\begin{itemize}
  \item[(A1)] Processes $S$ and $V$ have twice and once
continuously differentiable sample paths, respectively.
Moreover, there exists $C>0$ such that
\[
\sup_{t\in\mathcal T}|S^{(\nu)}(t)|<C,
\quad \nu=0,1,2,
\qquad
\sup_{t\in\mathcal T}|V(t)|<C
\]
almost surely. In addition, we assume that
\(
E[\int_{\mathcal T}|V_1'(t)|^2\,dt]<\infty.
\) For $P\in\{S,V\}$, the mean function $\mu_P$ has a
uniformly bounded second derivative on $\mathcal T$,
and the covariance function $C_P$ has uniformly
bounded second-order partial derivatives on
$\mathcal T^2$, including
$\partial_{11}C_P$, $\partial_{12}C_P$,
and $\partial_{22}C_P$.
\end{itemize}
Let \(h_{\mu_S},h_{C_S},h_{\mu_V},h_{C_V},b_{S,i}\) be the bandwidths used for estimating \(\widehat\mu_S\), \(\widehat C_S\), \(\widehat\mu_V\), \(\widehat C_V\), and $\widetilde S_i$ respectively.  
As \(n,m\to\infty\), the following conditions are assumed:

\begin{itemize}[leftmargin=3.1em,labelsep=1.2em,align=parleft,itemsep=0.6ex]
  \item[(A2.1)] 
  $h_{\mu_S}\to0,\: nh_{\mu_S}^4\to\infty,\: nh_{\mu_S}^6<\infty.$
  \item[(A2.2)] \(h_{C_S}\to0,\ nh_{C_S}^6\to\infty,\ nh_{C_S}^8<\infty.\)
  \item[(A2.3)] There exists a common bandwidth sequence \(b_S\) such that \(0<c_1<\inf_i(b_{S,i}/b_S)\le\sup_i(b_{S,i}/b_S)<c_2<\infty.\)
  \item[(A2.4)] \(m\to\infty,\ b_S\to0,\ mb_S^2\to\infty.\)
  \item[(A2.5)] \(h_{\mu_V}\to0,\ n h_{\mu_V}^4\to\infty,\ \limsup_n n h_{\mu_V}^6<\infty.\)
  \item[(A2.6)] \(h_{C_V}\to0,\ n h_{C_V}^6\to\infty,\ \limsup_n n h_{C_V}^8<\infty.\)
\end{itemize}

\begin{itemize}
  \item[(A3)]
Let $\mathcal T=[a,b]$, and suppose that
$a\le t_{i1}<\cdots<t_{im}\le b$.
Define $t_{i0}=a$ and $t_{i,m+1}=b$. There exist constants $c_1,c_2>0$ such that,
for all sufficiently large $m$,

\[
\min_{\substack{1\le i\le n\\2\le j\le m}}
(t_{ij}-t_{i,j-1})
\ge \frac{c_1}{m},
\qquad
\max_{\substack{1\le i\le n\\1\le j\le m+1}}
(t_{ij}-t_{i,j-1})
\le \frac{c_2}{m}.
\]
  \item[(A4)] The eighth moment of $X(t)$ and fourth moment of $Z(t)$ are uniformly bounded:
\[
\sup_{t\in\mathcal T} E\{X(t)^8\}<\infty,
\qquad
\sup_{t\in\mathcal T} E\{Z(t)^4\}<\infty.
\]
  \item[(A5)]
For $P\in\{S,V\}$,
\[
\sum_{j=1}^{K_P}
\|\varphi_j^P\|_\infty
=
o\!\left(
\sqrt m
\right)
\quad\text{ and }\quad
\sum_{j=1}^{K_P}
\|\varphi_j^P\|_\infty
\|(\varphi_j^P)'\|_\infty
=
o(m).
\]
\end{itemize}

\noindent\textbf{Kernel conditions.}
The kernels \(\kappa_1(\cdot)\) and \(\kappa_2(\cdot, \cdot)\) satisfy the following:

\begin{itemize}
    \item[(B1)] $\kappa_1$ is a symmetric, continuous,
    compactly supported density function.

    \item[(B2)] $\kappa_2$ is a continuous, compactly
    supported density function, symmetric in each argument.
    Moreover, $\kappa_2$ is Lipschitz continuous in its
    first argument uniformly in its second argument, and satisfies $\kappa_2(0,0)>0$.
\end{itemize}

\noindent\textbf{Additional Notation Used in the Supplementary Material.}
\begin{itemize}
    \item For a matrix $A\in\mathbb{R}^{m\times n}$, let
\[
\|A\|_{\mathrm{op}}
=
\sup_{\|x\|_2=1}\|Ax\|_2\quad\text{and}\quad \|A\|_{\mathrm F}
=\left(\sum_{i=1}^{n}\sum_{j=1}^{m}|a_{ij}|^2\right)^{1/2}
\]
denote the operator norm induced by the Euclidean norm, and the Frobenius norm of a matrix, respectively.
\item 
For a Lebesgue-measurable set $D\subseteq\mathbb{R}^d$
and a measurable function $f:D\to\mathbb{R}$, define
\[
\operatorname*{ess\,sup}_{x\in D}|f(x)|
=
\inf\left\{
M\ge 0:
|f(x)|\le M
\text{ for almost every }x\in D
\right\},
\]
where the essential supremum is taken with respect to
the $d$-dimensional Lebesgue measure.

\item For two functions $f$ and $g$, their tensor product
$f\otimes g$ is defined by
\[
(f\otimes g)(s,t)=f(s)g(t).
\]
In particular, $\varphi_k^S\otimes\varphi_\ell^V$
denotes the product basis function
$(s,t)\mapsto\varphi_k^S(s)\varphi_\ell^V(t)$.
\item For a random sequence $X_n$ and a positive deterministic sequence $a_n$, we say that $X_n = \Omega_p(a_n)$ if for every $\epsilon > 0$, there exists a positive constant $M_\epsilon$ such that
\[
P\left(|X_n| \le {a_n}M_{\epsilon}\right)\le \epsilon,
\]
for all sufficiently large $n$.
\end{itemize}

\section{Proofs}
\label{appen B}

\begin{proof}[Proof of proposition 1]

\vspace{0.5em}

\noindent\textbf{Step 1. Mean-square error of the leave-one-out estimator.}
\vspace{0.5em}

For brevity, we write $X_{ij}=X_i(t_{ij})$,
$R_{ij}=R_i(t_{ij})$, $Z_{ij}=Z_i(t_{ij})$, $\widehat Z_{ij} = \widehat Z_i(t_{ij})$
throughout this proof. Let \(
e_{ij}^{(-j)}
=
\widetilde S_{i,-j}(t_{ij})-S_i(t_{ij})\) and \(u_{i\ell}(t)=(t_{i\ell}-t)/b_{S_i}.
\) The leave-one-out local linear estimator admits the representation \(
\widetilde S_{i,-j}(t)
=
\sum_{\ell\ne j}w_{i\ell}^{(-j)}(t)X_{i\ell}\), where
\[
w_{i\ell}^{(-j)}(t)
=
\frac{
K(u_{i\ell}(t))
\left[
M_{i,2}^{(-j)}(t)
-u_{i\ell}(t)M_{i,1}^{(-j)}(t)
\right]
}{
M_{i,0}^{(-j)}(t)M_{i,2}^{(-j)}(t)
-\left[M_{i,1}^{(-j)}(t)\right]^2
},
\]
$K$ is a univariate kernel function that satisfies (B2.1) and \(
M_{i,r}^{(-j)}(t)
=
\sum_{\ell\ne j}
K(u_{i\ell}(t))u_{i\ell}(t)^r\), \(r=0,1,2.
\)

By (A3), compact support
of $K$, and the positivity of $K$ on a nondegenerate
interval, standard local linear arguments yield,
uniformly in $i,j$ and $t\in\mathcal T$,
\[
\max_{\ell\ne j}|w_{i\ell}^{(-j)}(t)|
=O\left(\frac{1}{mb_S}\right),
\qquad
\sum_{\ell\ne j}|w_{i\ell}^{(-j)}(t)|=O(1),
\]
and
\[
\sum_{\ell\ne j}\left[w_{i\ell}^{(-j)}(t)\right]^2
=O\left(\frac{1}{mb_S}\right).
\]

Write $X_{i\ell}=S_i(t_{i\ell})+R_{i\ell}$.
Then
\[
e_{ij}^{(-j)}
=
\underbrace{
\sum_{\ell\ne j}
w_{i\ell}^{(-j)}(t_{ij})S_i(t_{i\ell})
-S_i(t_{ij})
}_{B_{ij}^{(-j)}}
+
\underbrace{
\sum_{\ell\ne j}
w_{i\ell}^{(-j)}(t_{ij})R_{i\ell}
}_{U_{ij}^{(-j)}}.
\]

By Taylor's expansion, the local linear reproduction
properties, and the boundedness of $S_i''$ given by (A1), \(
|B_{ij}^{(-j)}|
\le
C b_{S_i}^2
\sum_{\ell\ne j}|w_{i\ell}^{(-j)}(t_{ij})|
=O(b_S^2).
\)

Let \(
\mathcal F_i
=
\sigma(S_i,V_i)
\) denote the sigma-field generated by the latent trajectories
$S_i$, $V_i$.
Conditional on $\mathcal F_i$, the errors $R_{i\ell}$
are independent across $\ell$ with mean zero and uniformly bounded
second moments. Hence
\begin{align*}
&E\left[\left(U_{ij}^{(-j)}\right)^4
\mid \mathcal F_i\right] \\
&\quad =
\sum_{\ell\ne j}
\left[w_{i\ell}^{(-j)}(t_{ij})\right]^4
E\left[R_{i\ell}^4\mid\mathcal F_i\right] \\
&\qquad\quad +
6\sum_{\substack{\ell<k\\ \ell,k\ne j}}
\left[w_{i\ell}^{(-j)}(t_{ij})\right]^2
\left[w_{ik}^{(-j)}(t_{ij})\right]^2
E\left[R_{i\ell}^2\mid\mathcal F_i\right]
E\left[R_{ik}^2\mid\mathcal F_i\right] \\
&\quad \le
C\sum_{\ell\ne j}
\left[w_{i\ell}^{(-j)}(t_{ij})\right]^4
+
C\sum_{\substack{\ell<k\\ \ell,k\ne j}}
\left[w_{i\ell}^{(-j)}(t_{ij})\right]^2
\left[w_{ik}^{(-j)}(t_{ij})\right]^2 \\
&\quad \le
C\left(
\sum_{\ell\ne j}
\left[w_{i\ell}^{(-j)}(t_{ij})\right]^2
\right)^2=
O\left(\frac{1}{m^2b_S^2}\right).
\end{align*}

Consequently, using $|a+b|^4\le 8(|a|^4+|b|^4)$,
\begin{align*}
E\left[\left|e_{ij}^{(-j)}\right|^4
\mid\mathcal F_i\right]
&\le
8\left|B_{ij}^{(-j)}\right|^4
+
8E\left[\left(U_{ij}^{(-j)}\right)^4
\mid\mathcal F_i\right]=
O\left(b_S^8+\frac{1}{m^2b_S^2}\right).
\end{align*}
Taking expectations and using the uniformity of
the bounds gives
\[
\sup_{1\le i\le n}\sup_{1\le j\le m}
E\left[(e_{ij}^{(-j)})^4\right]
=
O(\theta_{m,2}^4),
\quad
\theta_{m,2}
=
b_S^2+\frac{1}{\sqrt {mb_S}}.
\]

\noindent\textbf{Step 2. Mean-square error of the leave-one-out transformed residuals.}

Because \(
X_{ij}
=
S_i(t_{ij})+R_{ij},
\) we have \(X_{ij}-\widetilde S_{i,-j}(t_{ij})
=
R_{ij}-e_{ij}^{(-j)}.
\) Therefore,
\[
\begin{aligned}
\widehat Z_{ij}^{(-j)}-Z_{ij}
=
2\log\left|
1-\frac{e_{ij}^{(-j)}}{R_{ij}}
\right|,
\end{aligned}
\]
where $\widehat Z_{ij}^{(-j)}$ denotes the transformed residual with $\widetilde S_i(t_{ij})$ replaced by $\widetilde S_{i, -j}(t_{ij})$. Conditional on $\mathcal F_i$, since $\widetilde S_{i,-j}(t_{ij})$ does not use $X_{ij}$,
the error $e_{ij}^{(-j)}$ depends only on $\epsilon_{i\ell}$ and $W_{i\ell}$ with $\ell \ne j$.
On the other hand, $R_{ij}$
depends only on $\epsilon_{ij}$ and $W_{ij}$. Hence \(
e_{ij}^{(-j)}
\perp R_{ij}
\mid \mathcal F_i.
\) 

Fix $\mathcal F_i$ and for brevity, define \(
R=R_{ij}\), \(V=V_i(t_{ij})\),
\(W = W_{ij},
\)
and define
\[
g_v(e)
=
E\left[
\left.
\log^4\left|
1-\frac{e}{R}
\right|
\right|
V=v
\right].
\]
We show that, uniformly in $|v|\le M_V$ for some constant $M_V > 0$ and $e\in\mathbb R$, \(
g_v(e)\le C e^2
\)
for some constant $C<\infty$.

Let \(
\mathcal S(e)
=
\left\{
1/2
<
e/R
<
3/2
\right\}
\) denote a fixed neighborhood of the logarithmic singularity. On $\mathcal S(e)^c$, there exists a constant $C_1<\infty$
such that \(
|\log|1-x||
\le
C_1|x|\), \(
x\in
(-\infty,1/2]\cup[3/2,\infty).
\) It follows that
\[
\begin{aligned}
E\left[
\left.
\log^4\left|
1-\frac{e}{R}
\right|
\mathbf 1_{\mathcal S(e)^c}
\right|
V=v
\right]\le
C_1^2 e^2 E(R^{-4}\mid V=v).
\end{aligned}
\]
Since \(R^4=e^{2v+2W},\) we have \(
E(R^{-4}\mid V=v)
=
e^{-2v}E(e^{-2W}).
\) (R1) implies that \(
E(e^{-2W})<\infty.
\)
 By (A1), there exists $M_V>0$ such that $|v| \le M_V$. Since $|2v|\le 2M_V$, \(
E(R^{-4}\mid V=v)
\le
e^{2M_V}E(e^{-2W})
<\infty.
\)
Thus
\[
E\left[
\left.
\log^4\left|
1-\frac{e}{R}
\right|
\mathbf 1_{\mathcal S(e)^c}
\right|
V=v
\right]
\le
C e^4.
\]

It remains to consider $\mathcal S(e)$. Put
\(
a=|e|\) and \(Q=|R|.\) The event $\mathcal S(e)$ requires $e$ and $R$ to have the
same sign and satisfies \(
1/2
<
|e|/|R|
<
3/2.
\) Hence \(
\mathcal S(e)
\subseteq
\left\{
2a/3
<
Q
<
2a
\right\}.
\) Consequently, 
\[
\begin{aligned}
E\left[
\left.
\log^4\left|
1-\frac{e}{R}
\right|
\mathbf 1_{\mathcal S(e)}
\right|
V=v
\right]\le
I_v(a),
\end{aligned}
\]
where
\[
I_v(a)
=
\int_{2a/3}^{2a}
\log^4\left|
1-\frac{a}{q}
\right|
f_{Q\mid V}(q\mid v)\,dq.
\]

Conditional on $V=v$, \(
Q=e^{(v+W)/2}. \) Therefore,
\(
f_{Q\mid V}(q\mid v)
=
2f_W(2\log q-v)/q\), \(q>0.
\) We first establish a near-zero bound for this density. Set \(
q_0
=
\min\left[
1,\exp\left\{(w_0-M_V){/2}\right\}
\right].
\) For $0<q\le q_0$ and $|v|\le M_V$, \(
2\log q-v\le w_0.
\) Hence
\begin{align*}
f_{Q\mid V}(q\mid v)
&=
\frac{2}{q}f_W(2\log q-v)\\
&\le
\frac{2C_W}{q}
\exp\{(2+\delta)(2\log q-v)\}\\
&=
2C_W e^{-(2+\delta)v}q^{3+2\delta}\le C_Qq^3,
\end{align*}
where the last inequality follows from $q\le 1$ and
$|v|\le M_V$.

Suppose first that $0<a\le q_0/2$. Then the whole interval
$(2a/3,2a)$ lies in $(0,q_0]$. Using the change of variables
\(
x=a/q\), \(q=a/x\), \(dq=(-a/x^2)\,dx,
\) we obtain
\[
I_{v,4}(a)
=
a\int_{1/2}^{3/2}
\log^4|1-x|\,
f_{Q\mid V}(a/x\mid v)\,
\frac{dx}{x^2}.
\]

The near-zero density bound gives
\(
f_{Q\mid V}(a/x\mid v)
\le C_Q\left(a/x\right)^3,
\) and therefore
\[
I_{v,4}(a)
\le
C_Qa^4
\int_{1/2}^{3/2}
\frac{\log^4|1-x|}{x^5}\,dx
\le Ca^4,
\]
because
\[
\int_{1/2}^{3/2}
\frac{\log^4|1-x|}{x^5}\,dx<\infty.
\]

Now suppose that $a>q_0/2$. By (R1), let
$M_f=\sup_{w\in\mathbb R}f_W(w)<\infty$, then
\(
f_{Q\mid V}(q\mid v)\le{2M_f}/{q}.
\) Using the same change of variables,
\[
I_{v,4}(a)
\le
2M_f\int_{1/2}^{3/2}
\frac{\log^4|1-x|}{x}\,dx
\le C.
\]
Since $a>q_0/2$, \(
1\le16a^4/q_0^4,
\) and therefore \(
I_{v,4}(a)\le Ca^4.
\) Combining the two cases yields \(
I_{v,4}(a)\le Ca^4
\)
uniformly in $a>0$ and $|v|\le M_V$.
Together with the bound on $S(e)^c$, we conclude that \(
g_{v,4}(e)\le Ce^4.
\) Returning to the leave-one-out smoother error and using
conditional independence,
\[
\begin{aligned}
&E\left[
\left.
\left(
\widehat Z_{ij}^{(-j)}-Z_{ij}
\right)^4
\right|
\mathcal F_i,e_{ij}^{(-j)}
\right]
\\
&\qquad=
16E\left[
\left.
\log^4\left|
1-\frac{e_{ij}^{(-j)}}{R_{ij}}
\right|
\right|
\mathcal F_i,e_{ij}^{(-j)}
\right]
\le
C\left(e_{ij}^{(-j)}\right)^4.
\end{aligned}
\]
Taking expectations gives \(
E\left[
\widehat Z_{ij}^{(-j)}-Z_{ij}
\right]^4
\le
C
E\left\{e_{ij}^{(-j)}\right\}^4.
\)
Hence
\[
\sup_{1\le j\le m}
E\left[
\widehat Z_{ij}^{(-j)}-Z_{ij}
\right]^4
=
O\left(\theta_{m,2}^4\right).
\]

Finally,
\[
\begin{aligned}
E\left[\sum_{j=1}^m
\left(
\widehat Z_{ij}^{(-j)}-Z_{ij}
\right)^4
\Delta_{ij}\right] 
=
O\left(\theta_{m,2}^4\right),
\end{aligned}
\]
uniformly in $i$. Therefore,
\[
E\left[
\frac{1}{n}
\sum_{i=1}^n
\sum_{j=1}^m
\left(
\widehat Z_{ij}^{(-j)}-Z_{ij}
\right)^4
\Delta_{ij}
\right]
=
O\left(\theta_{m,2}^4\right).
\]
An application of Markov's inequality yields
\[
\frac{1}{n}
\sum_{i=1}^n
\sum_{j=1}^m
\left(
\widehat Z_{ij}^{(-j)}-Z_{ij}
\right)^4
\Delta_{ij}
=
O_p\left(\theta_{m,2}^4\right),
\]

\noindent\textbf{Step 3. Recovery of the full-sample transformed residuals.}

Let $H_{ij}$ denote the diagonal element of the
full-sample local linear smoothing matrix. In particular,
\[
H_{ij}
=
\frac{
K(0)M_{i,2}(t_{ij})
}{
M_{i,0}(t_{ij})M_{i,2}(t_{ij})
-\left[M_{i,1}(t_{ij})\right]^2
},
\]
where \(
M_{i,r}(t)
=
\sum_{\ell=1}^m
K(u_{i\ell}(t))u_{i\ell}(t)^r\), \(r=0,1,2
\). By the weight bounds established in Step 1, which also
hold for the full-sample local linear estimator,
\[
\sup_{i,j}|H_{ij}|
=
O\left(\frac{1}{mb_S}\right).
\]
Since $mb_S\to\infty$, we have
$\sup_{i,j}|H_{ij}|\le 1/2$ for sufficiently large $m$. The standard leave-one-out identity for linear smoothers
gives \(
X_{ij}-\widetilde S_i(t_{ij})
=
(1-H_{ij})
\left(
X_{ij}-\widetilde S_{i,-j}(t_{ij})
\right).
\) Consequently, \(
\widehat Z_{ij}-\widehat Z_{ij}^{(-j)}
=
2\log|1-H_{ij}|.
\)
Using $|\log(1-x)|\le 2|x|$ for $|x|\le 1/2$,
we obtain, uniformly in $i,j$,
\[
\begin{aligned}
\left(
\widehat Z_{ij}-\widehat Z_{ij}^{(-j)}
\right)^4\le 32H_{ij}^4=O\left(\frac{1}{m^4b_S^4}\right)
=o(\theta_{m,2}^4).
\end{aligned}
\]

Combining this with the leave-one-out bound
established in Step 2 yields
\[
\begin{aligned}
E\left[
\left(\widehat Z_{ij}-Z_{ij}\right)^4
\right]
&\le
2E\left[
\left(\widehat Z_{ij}^{(-j)}-Z_{ij}\right)^4
\right]+
2E\left[
\left(\widehat Z_{ij}-\widehat Z_{ij}^{(-j)}\right)^4
\right]=O(\theta_{m,2}^4),
\end{aligned}
\]
uniformly in $i,j$. Therefore,
\[
\sup_{1\le i\le n}\sup_{1\le j\le m}
E\left[
(\widehat Z_{ij}-Z_{ij})^4
\right]
=
O(\theta_{m,2}^4).
\]

Finally, since
$\max_i\sum_{j=2}^m\Delta_{ij}=O(1)$,
\[
\begin{aligned}
E\left[
\frac1n\sum_{i=1}^n
\|\widehat Z_i-Z_i\|_{m,2}^4
\right]
&=
\frac1n\sum_{i=1}^n\sum_{j=2}^m
\Delta_{ij}
E\left[
(\widehat Z_{ij}-Z_{ij})^4
\right]=O(\theta_{m,2}^4).
\end{aligned}
\]
Hence, by Markov's inequality,
\[
\frac1n\sum_{i=1}^n
\sum_{j=1}^m
\left\{
\widehat Z_{ij}^{(-j)}-Z_{ij}
\right\}^4
\Delta_{ij}
=
O_p(\theta_{m,2}^4).
\]
This completes the proof.
\end{proof}
\vspace{0.5em}

Before proceeding to the main theoretical results, we state the following lemma, which provides the estimation error bounds for the oracle and feasible estimators. Its proof is deferred to the end of this supplementary material.

\begin{lemma}
\label{lem:results_of_L2_norm}
Suppose that the assumptions of Proposition~1 hold, together with Assumptions (R2), (A2.1)--(A2.2), (A2.5)--(A2.6),
(A4) and (B2.2). For any function $f:\mathcal T\to\mathbb R$, define
the discrete weighted $L^2$ norm \(
\|f\|_{i,m}
=
\left\{
\sum_{j=2}^{m}|f(t_{ij})|^2\Delta_{ij}
\right\}^{1/2}\), 
\(\Delta_{ij}=t_{ij}-t_{i,j-1}.
\)
Let $\widetilde{\mu}_V$ and $\widetilde{G}_V$
denote the oracle estimators constructed from $Z_{ij}$,
and let $\widehat{\mu}_V$ and $\widehat{G}_V$
denote their feasible counterparts constructed from
$\widehat{Z}_{ij}$.
Let $\varphi_k^V$, $\widetilde{\varphi}_k^V$,
and $\widehat{\varphi}_k^V$ denote the corresponding
true, oracle, and feasible eigenfunctions, respectively,
with their signs appropriately aligned.
Define \(
\theta_{m,2}=b_S^2+(mb_S)^{-1/2}.
\) Then, as $n,m\to\infty$, the following results hold.

\noindent
\textup{(i) Non-oracle perturbation.}
\[
\max_{1\le i\le n}
\|\widehat{\mu}_V-\widetilde{\mu}_V\|_{i,m}^2
=O_p(\theta_{m,2}^2),
\qquad
\|\widehat{C}_V-\widetilde{C}_V\|_2
=O_p(\theta_{m,2}).
\]

\noindent
\textup{(ii) Oracle estimation error.}
\[
\max_{1\le i\le n}
E\left[\|\widetilde{\mu}_V-\mu_V\|^2_{i,m}\right]
=
O\left(h_{\mu_V}^4+
n^{-1}\right),
\quad
\|\widetilde{C}_V-C_V\|_2
=
O_p\left({h_{\mu_V}^2}+{h_{C_V}^2}+n^{-1/2}\right),
\]
\[
\max_{1\le i\le n}
E\left[\|\widehat{\mu}_S-\mu_S\|_{i,m}^2\right]
=O(h_{\mu_S}^4+n^{-1}),
\qquad
\|\widehat{C}_S-C_S\|_2
=O_p(h_{\mu_S}^2+h_{C_S}^2+n^{-1/2}).
\]

\noindent
\textup{(iii) Eigenfunction estimation error.}
For eigenvalues of multiplicity one,
\[
\max_{\substack{1\le k\le K_2\\1\le i\le n}}
\|\widetilde{\varphi}_k^V-\varphi_k^V\|_{i,m}
=
O_p\left(
K_2^{\alpha_2+1}({h_{\mu_V}^2+h_{C_V}^2}+n^{-1/2})
\right),
\]
\[
\max_{\substack{1\le k\le K_1\\1\le i\le n}}
\|\widehat{\varphi}_k^S-\varphi_k^S\|_{i,m}
=
O_p\left(
K_1^{\alpha_1+1}
(h_{\mu_S}^2+h_{C_S}^2+n^{-1/2})
\right),
\]
and
\[
\max_{\substack{1\le k\le K_2\\1\le i\le n}}
\|\widehat{\varphi}_k^V-\widetilde{\varphi}_k^V\|_{i,m}
=
O_p\left(
K_2^{\alpha_2+1}\theta_{m,2}
\right).
\]
\end{lemma}

\vspace{1.5em}

\begin{proof}[\hspace{-1.25em}Proof of Theorem 1]
We introduce the oracle estimators based on the true transformed residuals
$Z_i(t_{ij})$, following \cite{muller2006functional}. Denote by
$\widetilde{\mu}_V$, $\widetilde{C}_V$,
$\widetilde{\varphi}_k^V$, and $\widetilde{\chi}_{ik}^V$
the oracle analogues of
$\widehat{\mu}_V$, $\widehat{C}_V$,
$\widehat{\varphi}_k^V$, and $\widehat{\chi}_{ik}^V$,
respectively.

Write
\begin{equation}
\label{eq:score_basic_decomp}
|\widehat{\chi}_{ik}^V-\chi_{ik}^V|
\le
|\widetilde{\chi}_{ik}^V-\chi_{ik}^V|
+
|\widehat{\chi}_{ik}^V-\widetilde{\chi}_{ik}^V|.
\end{equation}

\noindent\textbf{Step 1: Variance score error - oracle part.}
Let
\[
\begin{alignedat}{2}
B_{1ik}
&=
\int_{\mathcal T}
\{V_i(t)-\mu_V(t)\}\varphi_k^V(t)\,dt
=\chi_{ik}^V,
&\qquad
B_{2ik}
&=
\sum_{j=2}^m
\{V_i(t_{ij})-\mu_V(t_{ij})\}
\varphi_k^V(t_{ij})\Delta_{ij},
\\
B_{3ik}
&=
\sum_{j=2}^m
\{V_i(t_{ij})-\mu_V(t_{ij})\}
\widetilde{\varphi}_k^V(t_{ij})\Delta_{ij},
&\qquad
B_{4ik}
&=
\sum_{j=2}^m
\{V_i(t_{ij})-\widetilde{\mu}_V(t_{ij})\}
\widetilde{\varphi}_k^V(t_{ij})\Delta_{ij}.
\end{alignedat}
\]
where $\Delta_{ij}=t_{ij}-t_{i,j-1}$. Then
\begin{equation}
\label{eq:oracle_decomp}
|\chi_{ik}^V-\widetilde{\chi}_{ik}^V|
\le
|B_{1ik}-B_{2ik}|
+
|B_{2ik}-B_{3ik}|
+
|B_{3ik}-B_{4ik}|
+
|B_{4ik}-\widetilde{\chi}_{ik}^V|.
\end{equation}

For \(
g_{V,ik}(t)
=
\{V_i(t)-\mu_V(t)\}\varphi_k^V(t),
\)
Assumption (A1) gives \(
\int_\mathcal T|g'_{V,ik}(t)|dt
\le
C(1+\|V_i'\|_2+\|(\varphi_k^V)'\|_2),
\) where $C>0$ does not depend on $i$ or $k$. Hence, by (A1) and (A3), for fixed $i$, \(
|B_{1ik}-B_{2ik}|
\lesssim
m^{-1}\{1+\|(\varphi_k^V)'\|_2\},
\) almost surely. Moreover, by the eigenequation, (A1), and (R2), \(
\|(\varphi_k^V)'\|_2
\le
\|\partial_1 G_V\|_2/\rho_k^V
\lesssim k^{\alpha_2}
\), where $\|\partial_1 G_V\|_2<\infty$ follows from (A1).
Consequently, \(
\max_{1\le k\le K_2}
\|(\varphi_k^V)'\|_2
=O(K_2^{\alpha_2}),
\)
and hence, for fixed $i$,
\begin{equation}
\label{eq:B1B2_bound}
\max_{1\le k\le K_2}|B_{1ik}-B_{2ik}|
\lesssim \frac{K_2^{\alpha_2}}{m}
\end{equation}

Under (A2), the oracle mean- and eigenfunction-estimation bounds of Lemma~\ref{lem:results_of_L2_norm} yield
\begin{equation}
|B_{2ik}-B_{3ik}|
=O_p\left(
k^{\alpha_2+1}h_{C_V}^2
\right),\quad
|B_{3ik}-B_{4ik}|
=O_p\left(h_{\mu_V}^2\right),
\label{eq:rate_of_|B2ik-B3ik|_&_|B_3ik-B_4ik|}
\end{equation}
uniformly over $1\le k\le K_2$ whenever the corresponding
maximum is taken. Here we use
\begin{equation}
\label{eq:uniform_rate_of_empirical_two_norm_of_oracle_eigenfunction}
\max_{\substack{1\le i\le n\\1\le k\le K_2}}
\left(
\sum_{j=2}^{m}
|\widetilde\varphi_k^V(t_{ij})|^2
\Delta_{ij}
\right)^{1/2}
=O_p(1).
\end{equation}
Indeed, by the triangle inequality,
\[
\left(
\sum_{j=2}^{m}
|\widetilde\varphi_k^V(t_{ij})|^2
\Delta_{ij}
\right)^{1/2}
\le
\left(
\sum_{j=2}^{m}
|\varphi_k^V(t_{ij})|^2
\Delta_{ij}
\right)^{1/2}
+
\left(
\sum_{j=2}^{m}
|\widetilde\varphi_k^V(t_{ij})
-\varphi_k^V(t_{ij})|^2
\Delta_{ij}
\right)^{1/2}.
\]
The first term is uniformly bounded by (A5),
while the second is $o_p(1)$ uniformly over
$i$ and $1\le k\le K_2$ by the oracle
discrete eigenfunction bound given by Lemma~\ref{lem:results_of_L2_norm}.
This proves~\eqref{eq:uniform_rate_of_empirical_two_norm_of_oracle_eigenfunction}.

Since $Z_{ij}=V_i(t_{ij})+W_{ij}$, \(
B_{4ik}-\widetilde{\chi}_{ik}^V
=
-\sum_{j=2}^m
W_{ij}\widetilde{\varphi}_k^V(t_{ij})\Delta_{ij}.
\)
Decompose
\begin{equation*}
T_{ik}^{(0)}
=
\sum_{j=2}^m
W_{ij}\varphi_k^V(t_{ij})\Delta_{ij},
\quad
T_{ik}^{(1)}
=
\sum_{j=2}^m
W_{ij}
\{\widetilde{\varphi}_k^V(t_{ij})
-\varphi_k^V(t_{ij})\}\Delta_{ij}.
\end{equation*}
By the independence across $j$ and zero mean of $W_{ij}$,
the bounded fourth moments implied by (A1) and (A4), and the
regular design condition (A3),
\begin{equation}
\label{eq:T0_fourth}
\max_{1\le k\le K_2}
E\left[|T_{ik}^{(0)}|^4\right]
=
O(m^{-2}).
\end{equation}
Consequently,
\begin{equation}
\label{eq:rate_of_max_k_|Tik^(0)|}
\max_{1\le k\le K_2}|T_{ik}^{(0)}|
=
O_p\left(\frac{K_2^{1/4}}{\sqrt m}\right).
\end{equation}
Also, by the Cauchy--Schwarz inequality,
\[
|T_{ik}^{(1)}|
\le
\left(
\sum_{j=2}^m W_{ij}^2\Delta_{ij}
\right)^{1/2}
\left(
\sum_{j=2}^m
|\widetilde{\varphi}_k^V(t_{ij})
-\varphi_k^V(t_{ij})|^2\Delta_{ij}
\right)^{1/2}.
\]
Since the second moments of $W_{ij}$ are uniformly bounded,
\begin{equation}
\label{eq:bound_of_empirical_one_norm_of_Wij}
\sum_{j=2}^m W_{ij}^2\Delta_{ij}=O_p(1).
\end{equation}
Consequently, the oracle discrete eigenfunction bound given by Lemma~\ref{lem:results_of_L2_norm} yields
\begin{equation}
\label{eq:T1_bound}
\max_{1\le k\le K_2}|T_{ik}^{(1)}|
=
O_p\left(
K_2^{\alpha_2+1}h_{C_V}^2
\right).
\end{equation}

Combining
\eqref{eq:B1B2_bound}, \eqref{eq:rate_of_|B2ik-B3ik|_&_|B_3ik-B_4ik|}, \eqref{eq:rate_of_max_k_|Tik^(0)|} and \eqref{eq:T1_bound},
\begin{equation}
\label{eq:oracle_uniform_bound}
\max_{1\le k\le K_2}
|\chi_{ik}^{V}-\widetilde{\chi}_{ik}^{V}|
=O_p\Bigg(
\frac{K_2^{\alpha_2}}{m}
\vee
\frac{K_2^{\alpha_2+1}}{\,h_{C_V}^{-2}}
\vee h_{\mu_V}^2
\vee\frac{K_2^{1/4}}{\sqrt m}
\Bigg).
\end{equation}

\noindent\textbf{Step 2: Variance score error - non-oracle part.}
Define
\begin{align*}
C_{1ik}
&=
\sum_{j=2}^m
\{Z_i(t_{ij})-\widetilde{\mu}_V(t_{ij})\}
\widetilde{\varphi}_k^V(t_{ij})\Delta_{ij}
=
\widetilde{\chi}_{ik}^V,\\
C_{2ik}
&=
\sum_{j=2}^m
\{Z_i(t_{ij})-\widetilde{\mu}_V(t_{ij})\}
\widehat{\varphi}_k^V(t_{ij})\Delta_{ij},\\
C_{3ik}
&=
\sum_{j=2}^m
\{\widehat Z_i(t_{ij})-\widetilde{\mu}_V(t_{ij})\}
\widehat{\varphi}_k^V(t_{ij})\Delta_{ij}.
\end{align*}
Then
\begin{equation}
\label{eq:nonoracle_decomp}
|\widetilde{\chi}_{ik}^V-\widehat{\chi}_{ik}^V|
\le
|C_{1ik}-C_{2ik}|
+
|C_{2ik}-C_{3ik}|
+
|C_{3ik}-\widehat{\chi}_{ik}^V|.
\end{equation}

Combining Lemma~\ref{lem:results_of_L2_norm} with the decomposition
\begin{equation}
\label{eq:decomp}
Z_i(t_{ij})-V_i(t_{ij})+V_i(t_{ij})-\mu_V(t_{ij})+\mu_V(t_{ij})-\tilde\mu_V(t_{ij}),
\end{equation}
then \eqref{eq:bound_of_empirical_one_norm_of_Wij}, (A1) and Lemma~\ref{lem:results_of_L2_norm} yield 
\begin{equation}
\label{eq:C1C2_bound}
\max_{1\le k\le K_2}|C_{1ik}-C_{2ik}|
=
O_p\left(
K_2^{\alpha_2+1}\theta_{m,2}
\right).
\end{equation}
Furthermore,
\begin{equation}
\label{eq:discrete_phi_bound}
\max_{1\le k\le K_2}
\left(
\sum_{j=2}^m
|\widehat{\varphi}_k^V(t_{ij})|^2\Delta_{ij}
\right)^{1/2}
=
O_p(1),
\end{equation}
by Lemma~\ref{lem:results_of_L2_norm} and the similar argument used in showing \eqref{eq:uniform_rate_of_empirical_two_norm_of_oracle_eigenfunction}. 
Moreover, the proof of proposition 1 implies \(
\left(
\sum_{j=2}^m
|\widehat Z_i(t_{ij})-Z_i(t_{ij})|^2\Delta_{ij}
\right)^{1/2}
=
O_p(\theta_{m,2}),
\)
and hence \(\max_{1\le k\le K_2}\) \(|C_{2ik}-C_{3ik}|
=
O_p(\theta_{m,2})\). Similarly, by Lemma~\ref{lem:results_of_L2_norm}, \(
\max_{1\le k\le K_2}
|C_{3ik}-\widehat{\chi}_{ik}^V|
=
O_p(
\theta_{m,2}
).
\) Therefore,
\begin{equation}
\label{eq:nonoracle_uniform_bound}
\max_{1\le k\le K_2}
|\widetilde{\chi}_{ik}^V-\widehat{\chi}_{ik}^V|
=
O_p\left(K_2^{\alpha_2+1}\theta_{m,2}
\right).
\end{equation}

By (A2.5), (A2.6), and $n\asymp m^2$, \(
h_{\mu_V}^2
=o(\theta_{m,2})\) and \(h_{C_V}^2
=o(\theta_{m,2})
\). Moreover, since $\theta_{m,2}\gtrsim m^{-2/5}$, \(
K_2^{1/4}m^{-1/2}
=o\left(K_2^{\alpha_2+1}\theta_{m,2}\right)
\).
Combining
\eqref{eq:score_basic_decomp},
\eqref{eq:oracle_uniform_bound}, and
\eqref{eq:nonoracle_uniform_bound}
yields, for each fixed $i$,
\begin{equation*}
\boxed{
\max_{1\le k\le K_2}
|\widehat{\chi}_{ik}^V-\chi_{ik}^V|
=
O_p(r_{V,n,K_2}),
}
\end{equation*}
where \(
r_{V,n,K_2}=
K_2^{\alpha_2+1}
\theta_{m,2}.
\) This proves \textit{(i)} in the part (a) of Theorem 1.

\noindent\textbf{Step 3: Empirical second-moment bound of variance score.}
We next establish the empirical $L_2$ bound for the
variance-score estimation errors. By \eqref{eq:oracle_decomp} and \eqref{eq:nonoracle_decomp},
it suffices to control the squared contribution of each
term separately.

For the oracle part, \eqref{eq:rate_of_|B2ik-B3ik|_&_|B_3ik-B_4ik|}, whose upper bounds do not depend on $i$, yield
\[
\max_{1\le k\le K_2}
\frac{1}{n}\sum_{i=1}^n
|B_{2ik}-B_{3ik}|^2
=
O_p\left(K_2^{\alpha_2+1} h_{C_V}^2\right),
\]
and
\[
\max_{1\le k\le K_2}
\frac{1}{n}\sum_{i=1}^n
|B_{3ik}-B_{4ik}|^2
=
O_p\left(h_{\mu_V}^2\right).
\]

For \eqref{eq:B1B2_bound}, by (A1), we have
\[
\begin{aligned}
\max_{1\le k\le K_2}\frac1n\sum_{i=1}^n
|B_{1ik}-B_{2ik}|^2
&\lesssim \frac1{m^2}
\left(
1+\frac1n\sum_{i=1}^n\|V_i'\|_2^2
+\max_{1\le k\le K_2}\|(\varphi_k^V)'\|_2^2
\right)\\
&=O_p\left(\frac{K_2^{2\alpha_2}}{m^2}\right).
\end{aligned}
\]

For $T_{ik}^{(0)}$, the independence and moment
conditions used in deriving \eqref{eq:T0_fourth} give, uniformly in $k$, \(
E\bigl[|T_{ik}^{(0)}|^2\bigr]=O(m^{-1})\) and \(E\bigl[|T_{ik}^{(0)}|^4\bigr]=O(m^{-2}).
\) Since the subjects are independent, \(
\operatorname{Var}
\left(
n^{-1}\sum_{i=1}^n|T_{ik}^{(0)}|^2
\right)
=
O\left(n^{-1}m^{-2}\right).
\)
Hence, by Chebyshev's inequality, a union bound
over $1\le k\le K_2$, and (A5),
\[
\max_{1\le k\le K_2}
\frac{1}{n}\sum_{i=1}^n
|T_{ik}^{(0)}|^2
=
O_p(m^{-1}).
\]

Moreover,
\[
\max_{1\le k\le K_2}
\frac{1}{n}\sum_{i=1}^n
|T_{ik}^{(1)}|^2
=
O_p\left\{\left(K_2^{\alpha_2+1}{h_{C_V}^2}\right)^2\right\},
\]
since \(
n^{-1}\sum_{i=1}^n
\left(\sum_{j=2}^mW_{ij}^2\Delta_{ij}\right)^2
=
O_p(1).
\)

For the non-oracle part, same  conditions used to derive \eqref{eq:C1C2_bound} imply
\[
\max_{1\le k\le K_2}
\frac{1}{n}\sum_{i=1}^n
|C_{1ik}-C_{2ik}|^2
=
O_p\left(K_2^{2\alpha_2+2}\theta_{m,2}^2\right).
\]
Furthermore, Proposition~1
and \eqref{eq:discrete_phi_bound} yield
\[
\max_{1\le k\le K_2}
\frac{1}{n}\sum_{i=1}^n
|C_{2ik}-C_{3ik}|^2
=
O_p(\theta_{m,2}^2).
\]
Similarly, by \eqref{eq:discrete_phi_bound},
\[
\max_{1\le k\le K_2}
\frac{1}{n}\sum_{i=1}^n
|C_{3ik}-\widehat\chi_{ik}^V|^2
=
O_p(\theta_{m,2}^2).
\]

Combining the preceding bounds and the rate
conditions in (A2), (A5), we obtain
\begin{equation}
\label{eq:second moment bound of var score error}
    \boxed{\max_{1\le k\le K_2}
\frac{1}{n}\sum_{i=1}^n
|\Delta_{ik}^V|^2
=
O_p\left(
K_2^{2\alpha_2+2}
\theta_{m,2}^2
\right).}
\end{equation}

This proves the first assertion in \textit{(ii)} in the part (a) of Theorem 1. Finally, with
$\Delta_{ik}^V=\widehat{\chi}_{ik}^V-\chi_{ik}^V$,
\[
\max_{1\le k\le K_2}
\frac1n\sum_{i=1}^n
|\widehat{\chi}_{ik}^V|^2
\le
2\max_{1\le k\le K_2}
\frac1n\sum_{i=1}^n|\chi_{ik}^V|^2
+
2\max_{1\le k\le K_2}
\frac1n\sum_{i=1}^n|\Delta_{ik}^V|^2.
\]
The first term is $O(1)$ by (A1) and $\|\varphi_k^V\|_2=1$ for all $k\ge 1$, while the second term is $o_p(1)$ by \eqref{eq:second moment bound of var score error}. Hence we get
\[
\boxed{
\max_{1\le k\le K_2}\frac{1}{n}\sum_{i=1}^n|\widehat \chi^V_k|^2=O_p(1).}
\]

This proves the second assertion in \textit{(ii)} in the part (a) of Theorem 1.

\noindent\textbf{Step 4: Mean-process score error.}
We now turn to the mean-process scores. Define
\[
\begin{alignedat}{2}
A_{1ik}
&=
\int_{\mathcal T}
\{S_i(t)-\mu_S(t)\}\varphi_k^S(t)\,dt
=\chi_{ik}^S,
&\qquad
A_{2ik}
&=
\sum_{j=2}^m
\{S_i(t_{ij})-\mu_S(t_{ij})\}
\varphi_k^S(t_{ij})\Delta_{ij},
\\
A_{3ik}
&=
\sum_{j=2}^m
\{S_i(t_{ij})-\mu_S(t_{ij})\}
\widehat{\varphi}_k^S(t_{ij})\Delta_{ij},
&\qquad
A_{4ik}
&=
\sum_{j=2}^m
\{S_i(t_{ij})-\widehat{\mu}_S(t_{ij})\}
\widehat{\varphi}_k^S(t_{ij})\Delta_{ij} .
\end{alignedat}
\]
Since \(
\widehat{\chi}_{ik}^S
=
\sum_{j=2}^m
\{X_i(t_{ij})-\widehat{\mu}_S(t_{ij})\}
\widehat{\varphi}_k^S(t_{ij})\Delta_{ij},
\) with $\Delta_{ik}^S=\widehat{\chi}_{ik}^S - \chi_{ik}^S$, we have
\begin{equation}
\label{eq:expansion of error of mean scores}
|\Delta_{ik}^S|
\le
|A_{1ik}-A_{2ik}|
+
|A_{2ik}-A_{3ik}|
+
|A_{3ik}-A_{4ik}|
+
|A_{4ik}-\widehat{\chi}_{ik}^S|.
\end{equation}

For \(
g_{S,ik}(t)
=
\{S_i(t)-\mu_S(t)\}\varphi_k^S(t)
\), Assumption~(A1) gives \(
\|g_{S,ik}'\|_2
\le
C(1+\|(\varphi_k^S)'\|_2),
\) uniformly in $i$ and $k$. Hence, by (A3), \(
|A_{1ik}-A_{2ik}|
\lesssim
m^{-1}(1+\|(\varphi_k^S)'\|_2)\). Similarly, by the eigenequation, (A1), and (R2), \(
\max_{1\le k\le K_1}\|(\varphi_k^S)'\|_2
=O(K_1^{\alpha_1}).
\) Consequently,
\begin{equation}
\label{eq:A1A2_bound}
\max_{1\le k\le K_1}|A_{1ik}-A_{2ik}|
\lesssim \frac{K_1^{\alpha_1}}{m},
\end{equation}
uniformly over $1\le i\le n$.

Using similar argument showing \eqref{eq:uniform_rate_of_empirical_two_norm_of_oracle_eigenfunction} on $\sum_{j=2}^m |\widehat\varphi_k^S(t_{ij})|^2\Delta_{ij}$, then by Lemma~\ref{lem:results_of_L2_norm},
\begin{equation}
\label{eq:A2A3_bound}
\max_{1\le k\le K_1}
|A_{2ik}-A_{3ik}|
=
O_p\left(K_1^{\alpha_1+1}h_{C_S}^2\right).
\end{equation}
Moreover, Lemma~\ref{lem:results_of_L2_norm} gives
\begin{equation}
\label{eq:A3A4_bound}
\max_{1\le k\le K_1}
|A_{3ik}-A_{4ik}|
=
O_p(h_{\mu_S}^2).
\end{equation}
Under (A2.1)--(A2.2), \(
h_{\mu_S}^2
=
o(h_{C_S}^2).    
\) It remains to control the measurement-error projection. Since
$X_i(t_{ij})=S_i(t_{ij})+R_{ij}$, we have \(
A_{4ik}-\widehat{\chi}_{ik}^S
=
-\sum_{j=2}^m
R_{ij}\widehat{\varphi}_k^S(t_{ij})\Delta_{ij}.
\) Write
\[
T_{ik}^{S,(0)}
=
\sum_{j=2}^m
R_{ij}\varphi_k^S(t_{ij})\Delta_{ij},
\qquad
T_{ik}^{S,(1)}
=
\sum_{j=2}^m
R_{ij}
\{\widehat{\varphi}_k^S(t_{ij})-\varphi_k^S(t_{ij})\}
\Delta_{ij} .
\]

Recall that \(
R_{ij}
=
\xi_{ij}
\{\exp(V_i(t_{ij})+W_{ij})\}^{1/2}\), \(\xi_{ij}=\operatorname{sign}(\epsilon_{ij})
\),
where the $\xi_{ij}$ are independent across $j$, independent of all
other random quantities, and satisfy $E(\xi_{ij})=0$. 
Consequently,
any mixed product of the $R_{ij}$'s for which at least one observation
index occurs with odd multiplicity has zero expectation. Thus, in the
eighth-moment expansion of $T_{ik}^{S,(0)}$, only the index configurations
$8$, $6+2$, $4+4$, $4+2+2$, and $2+2+2+2$ remain.
By (A1), (A3) and (A4), these configurations contribute
$O(m^{-7})$, $O(m^{-6})$, $O(m^{-6})$, $O(m^{-5})$,
and $O(m^{-4})$, respectively. Consequently,
\begin{equation}
\label{eq:uniform k rate of expectation of fourth moment of T_ik^S(0)}
\max_{1\le i\le n}\max_{1\le k\le K_1}
E\left[\bigl|T_{ik}^{S,(0)}\bigr|^8\right]
=
O(m^{-4}).
\end{equation}
Hence, for each fixed $i$, Markov's inequality and the union bound yield
\begin{equation}
\label{eq:rate_of_max_Tik_S}
\max_{1\le k\le K_1}
\left|T_{ik}^{S,(0)}\right|
=
O_p\left(\frac{K_1^{1/8}}{\sqrt m}\right).
\end{equation}

For the eigenfunction-estimation component,
\[
\left|T_{ik}^{S,(1)}\right|
\le
\left(\sum_{j=2}^m\{\widehat{\varphi}_k^S(t_{ij})-\varphi_k^S(t_{ij})\}^2\Delta_{ij}\right)^{1/2}
\left(\sum_{j=2}^m |R_{ij}|^2\Delta_{ij}\right)^{1/2} .
\]
Since $R_{ij}$ has bounded second moment uniformly in $i,j$, \(
\sum_{j=2}^m |R_{ij}|^2\Delta_{ij}=O_p(1)
\)
for each fixed $i$, Lemma~\ref{lem:results_of_L2_norm} yields
\begin{equation}
\max_{1\le k\le K_1}
\left|T_{ik}^{S,(1)}\right|
=
O_p\left(K_1^{\alpha_1+1}h_{C_S}^2\right).
\label{eq:rate of max_k Tik^S(1)}
\end{equation}

Combining \eqref{eq:A1A2_bound}--\eqref{eq:rate of max_k Tik^S(1)}, for each fixed $i$,
\[
\boxed{\max_{1\le k\le K_1}
|\Delta_{ik}^S|
=
O_p\left(r_{S,n,K_1}^{\text{ind}}\right),}
\]
where \(
r_{S,n,K_1}^{\text{ind}}
=
K_1^{\alpha_1+1}h_{C_S}^2
\vee
(K_1^{1/8}/\sqrt m).
\)

This proves \textit{(i)} in the part (b) of Theorem 1.

\noindent\textbf{Step 5: Empirical second-moment bound of mean scores.}
We next establish the empirical $L_2$ bound for the mean-score
estimation errors. By \eqref{eq:expansion of error of mean scores}
and $(a_1+\cdots+a_4)^2\le 4\sum_{r=1}^4a_r^2$,
it suffices to control the squared contribution of each term.

Lemma~\ref{lem:results_of_L2_norm},
together with (A1), yield
\begin{align}
\max_{1\le k\le K_1}
\frac1n\sum_{i=1}^n |A_{1ik}-A_{2ik}|^2
&=O_p\left(\frac{K_1^{2\alpha_1}}{m^2}\right),
\notag\\
\max_{1\le k\le K_1}
\frac1n\sum_{i=1}^n |A_{2ik}-A_{3ik}|^2
&=O_p\left(K_1^{2\alpha_1+2}h_{C_S}^4\right),
\notag\\
\max_{1\le k\le K_1}
\frac1n\sum_{i=1}^n |A_{3ik}-A_{4ik}|^2
&=O_p\left(h_{\mu_S}^4\right).
\notag
\end{align}

For $T_{ik}^{S,(0)}$, the same moment argument used for
$T_{ik}^{(0)}$ in the variance-process part gives, uniformly in $k$, \(
E\left[|T_{ik}^{S,(0)}|^2\right]=O(m^{-1})\) and \(
E\left[|T_{ik}^{S,(0)}|^4\right]=O(m^{-2})
\). Since the subjects are independent, Chebyshev's inequality,
a union bound over $1\le k\le K_1$, and (A5) yield
\begin{equation*}
\max_{1\le k\le K_1}
\frac1n\sum_{i=1}^n |T_{ik}^{S,(0)}|^2
=O_p(m^{-1}).
\end{equation*}
Moreover, by the same argument used for $T_{ik}^{(1)}$,
\begin{equation*}
\max_{1\le k\le K_1}
\frac1n\sum_{i=1}^n |T_{ik}^{S,(1)}|^2
=O_p\left(K_1^{2\alpha_1+2}h_{C_S}^4\right),
\end{equation*}
where we use
\(
n^{-1}\sum_{i=1}^n
\left(\sum_{j=2}^m |R_{ij}|^2\Delta_{ij}\right)
=O_p(1).
\) Finally, under $n\asymp m^2$ and (A2.1), \(
h_{\mu_S}^4=o(m^{-1}).\)
Combining the preceding bounds gives
\begin{equation}
\label{eq:uniform k rate of L2 norm of mean score error}
\boxed{\max_{1\le k\le K_1}
\frac1n\sum_{i=1}^n|\Delta_{ik}^S|^2
=
O_p\left(
K_1^{2\alpha_1+2}h_{C_S}^4
\vee m^{-1}
\right).}
\end{equation}

This proves the first assertion in \textit{(ii)} in the part (b) of Theorem 1. Moreover, by (A1) and \eqref{eq:uniform k rate of L2 norm of mean score error}, we also get
\[
\boxed{\max_{1\le k\le K_1}\frac{1}{n}\sum_{i=1}^n|\widehat\chi_{ik}^S|^2=O_p(1).}
\]

This proves the second assertion in \textit{(ii)} in the part (b) of Theorem 1.

\noindent\textbf{Step 6: Uniform magnitude of the estimated mean scores.}
By (A1) and $\|\varphi_k^S\|_2=1$,
\begin{equation}
\label{eq:max true score of mean process is O(1)}
\max_{1\le i\le n}
\max_{1\le k\le K_1}
|\chi_{ik}^S|
=
O(1).
\end{equation}
From \eqref{eq:uniform k rate of expectation of fourth moment of T_ik^S(0)}, the union bound over $(i,k)$, $n\asymp m^2$, and
Markov's inequality,
\begin{equation}
\label{eq:rate_of_max_ik__of_Tik_S}
\max_{\substack{1\le i\le n\\1\le k\le K_1}}
|T_{ik}^{S,(0)}|
=
O_p\left(\frac{K_1^{1/8}}{m^{1/4}}\right)=o_p(1).
\end{equation}
Moreover, \(
\max_{1\le i\le n}
\sum_{j=2}^m|R_{ij}|^2\Delta_{ij}
=
O_p(1).
\) This can be shown by considering the decomposition
\[\sum_j |R_{ij}|^2\Delta_{ij} = \sum_j E\left\{|R_{ij}|^2|V_i\right\}\Delta_{ij} + \sum_j \left\{|R_{ij}|^2 - E\left\{|R_{ij}|^2|V_i\right\}\right\}\Delta_{ij},\]
the first term is clearly $O(1)$ uniformly over $i$ by (A1). Define $U_{ij} = |R_{ij}|^2 - E\{|R_{ij}|^2 | V_i\}$, then combining (A1), (A4), $W_{ij}\perp V_i$ and $W_{ij}\perp W_{i\ell}$ for all $j\neq\ell$, we can get 
\[
E\left[\left(\sum_j U_{ij}\Delta_{ij}\right)^4 \middle | V_i\right] = O\left(m^{-2}\right), 
\]
uniformly over $i$. Applying the tower rule and a union bound over $1\le i\le n$, together with $n\asymp m^2$, gives the desired result. Therefore
\begin{equation}
\label{eq:uniform ik rate of T_ik^S(1)}
\max_{\substack{1\le i\le n\\1\le k\le K_1}}
|T_{ik}^{S,(1)}|
=
O_p\left(K_1^{\alpha_1+1}h_{C_S}^2\right).
\end{equation}
The remaining terms in \eqref{eq:expansion of error of mean scores} are $O_p(1)$ uniformly over $i$ and
$k$. It follows from \eqref{eq:max true score of mean process is O(1)}--\eqref{eq:uniform ik rate of T_ik^S(1)} that
\[
\boxed{\max_{\substack{1\le i\le n\\1\le k\le K_1}}
|\widehat{\chi}_{ik}^S|
=
O_p(1).}
\]

This proves \textit{(iii)} in the part (b) of Theorem 1 and completes the proof.
\end{proof}
\vspace{0.75em}

We next establish an auxiliary result needed for the regression analysis.

\begin{lemma}[Stability of the empirical Gram matrix]
\label{lem:gram_mat}
Let \(
\boldsymbol{\chi}_i^S
=
(\chi_{i1}^S,\ldots,\chi_{iK_1}^S)^\top\), \(
\boldsymbol{\chi}_i^V
=
(\chi_{i1}^V,\ldots,\chi_{iK_2}^V)^\top,
\) and define their estimated counterparts by
\(
\widehat{\boldsymbol{\chi}}_i^S
=
(\widehat{\chi}_{i1}^S,\ldots,\widehat{\chi}_{iK_1}^S)^\top\),
\(\widehat{\boldsymbol{\chi}}_i^V
=
(\widehat{\chi}_{i1}^V,\ldots,\widehat{\chi}_{iK_2}^V)^\top,
\) then define
\[
\boldsymbol{\chi}_i
=
\begin{pmatrix}
1 \\
\boldsymbol{\chi}_i^S \\
\boldsymbol{\chi}_i^V \\
\boldsymbol{\chi}_i^S \otimes \boldsymbol{\chi}_i^V
\end{pmatrix},
\qquad
\widehat{\boldsymbol{\chi}}_i
=
\begin{pmatrix}
1 \\
\widehat{\boldsymbol{\chi}}_i^S \\
\widehat{\boldsymbol{\chi}}_i^V \\
\widehat{\boldsymbol{\chi}}_i^S
\otimes
\widehat{\boldsymbol{\chi}}_i^V
\end{pmatrix},
\]
and \(
\boldsymbol\Lambda_K
=
E\left[
\boldsymbol{\chi}_i
\boldsymbol{\chi}_i^\top
\right]\),
\(
\widehat{\boldsymbol\Lambda}_K
=
n^{-1}
\sum_{i=1}^n
\widehat{\boldsymbol{\chi}}_i
\widehat{\boldsymbol{\chi}}_i^\top.
\)
Under the conditions of Theorem~1 and (S4),
\[
\left\|
\widehat{\boldsymbol\Lambda}_K-\boldsymbol\Lambda_K
\right\|_{\mathrm{op}}
=
O_p\left\{
\sqrt{K_1K_2}
\left(
r_{S,n,K_1}\vee r_{V,n,K_2}
\right)
\right\},
\]
and, with probability tending to one, \(\widehat{\boldsymbol\Lambda}_K\) is nonsingular, and
\[
\lambda_{\min}(\widehat{\boldsymbol\Lambda}_K)
=
\Omega_p(\underline{\lambda}_K)\quad\text{ and }\quad
\left\|
\widehat{\boldsymbol\Lambda}_K^{-1}
\right\|_{\mathrm{op}}
=
O_p(\underline{\lambda}_K^{-1}).
\]
\end{lemma}

\begin{proof}
Define \(\mathbf{d}_i=\widehat{\boldsymbol{\chi}}_i - \boldsymbol{\chi}_i\), $\Delta_{ik}^S=\widehat\chi_{ik}^S - \chi_{ik}^S$ and $\Delta_{ik}^V=\widehat{\chi}_{ik}^V-\chi_{ik}^V$, then
\[
\begin{aligned}
\frac{1}{n}\sum_{i=1}^n\|\mathbf d_i\|_2^2
={}&
\sum_{k=1}^{K_1}
\frac{1}{n}\sum_{i=1}^n|\Delta_{ik}^S|^2
+
\sum_{\ell=1}^{K_2}
\frac{1}{n}\sum_{i=1}^n|\Delta_{i\ell}^V|^2\\
&+
\sum_{k=1}^{K_1}\sum_{\ell=1}^{K_2}
\frac{1}{n}\sum_{i=1}^n
\left|
\widehat{\chi}_{ik}^S\widehat{\chi}_{i\ell}^V
-
\chi_{ik}^S\chi_{i\ell}^V
\right|^2.
\end{aligned}
\]
Using \(
\widehat{\chi}_{ik}^S\widehat{\chi}_{i\ell}^V
-
\chi_{ik}^S\chi_{i\ell}^V
=
\Delta_{ik}^S{\chi}_{i\ell}^V
+
\widehat\chi_{ik}^S\Delta_{i\ell}^V,
\) together with Theorem~1, gives
\begin{equation}
\label{eq:rate_of_two_norm_of_scores_vector}
\frac{1}{n}\sum_{i=1}^n\|\mathbf d_i\|_2^2
=
O_p\left[
K_1K_2
\left(
r_{S,n,K_1}^2
\vee
r_{V,n,K_2}^2
\right)
\right].
\end{equation}

Define $\boldsymbol\Lambda_K^o = n^{-1}\sum_{i=1}^n\boldsymbol{\chi}_i\boldsymbol{\chi}_i^\top$, then
\[
\begin{aligned}
\|\widehat{\boldsymbol\Lambda}_K-\boldsymbol\Lambda_K^o\|_\text{op}=&\left\|\frac{1}{n}\sum_{i=1}^n(\boldsymbol{\widehat\chi}_i\mathbf{d}_i^\top + \mathbf{d}_i\boldsymbol{\chi}_i^\top)\right\|_\text{op}\\
\le&\frac{1}{n}\sum_{i=1}^n\|\mathbf{d}_i\|_2(\|\boldsymbol{\widehat\chi}_i\|_2+\|\boldsymbol{\chi}_i\|_2)\\
\le&\left(\frac{1}{n}\sum_{i=1}^n(\|\boldsymbol{\chi}_i\|+\|\boldsymbol{\widehat{\chi}_i\|})^2\right)^{1/2} \left(\frac{1}{n}\sum_{i=1}^n \|\mathbf{d}_i\|^2_2\right)^{1/2},
\end{aligned}
\]
where the first inequality holds by triangle inequality and the operator norm of rank one matrix, and the second inequality holds by Cauchy-Schwarz inequality.

Furthermore, by (A1) and Bessel's inequality,
\[
\begin{aligned}
\|\boldsymbol{\chi}_i\|_2^2=&1+\|\boldsymbol{\chi}_i^S\|_2^2 + \|\boldsymbol{\chi}_i^V\|_2^2+\|\boldsymbol{\chi_i}^S\|_2^2\|\boldsymbol{\chi}_i^V\|_2^2\\
\le&1+\|S_i-\mu_S\|_2^2+\|V_i-\mu_V\|_2^2+\|S_i-\mu_S\|_2^2\|V_i-\mu_V\|_2^2\le C,
\end{aligned}
\]
for some positive constant $C$ that does not depend on $i$. Hence, $n^{-1}\sum_{i=1}^n\|\boldsymbol{\chi}_i\|_2^2=O(1)$. Consequently, $n^{-1}\sum_{i=1}^n\|\boldsymbol{\widehat\chi}_i\|_2^2=O(1)$ by \eqref{eq:rate_of_two_norm_of_scores_vector}. Therefore, we get
\[
\|\widehat{\boldsymbol\Lambda}_K - \boldsymbol\Lambda_K^o\|_\text{op}=O_p\left[\sqrt{K_1K_2}(r_{S,n,K_1}\vee r_{V,n,K_2})\right]
\]

Next, by i.i.d. assumption between subjects and (A1), we have
\[
\begin{aligned}
E\left[\|\boldsymbol\Lambda_K^o - \boldsymbol\Lambda_K\|_\text{F}^2\right]\le&\frac{1}{n}E\left[\|\boldsymbol{\chi}_i\boldsymbol{\chi}_i^\top-E[\boldsymbol{\chi}_i\boldsymbol\chi_i^\top]\|_\text{F}^2\right]\\
    \le&\frac{1}{n}E\left[\|\boldsymbol{\chi}_i\boldsymbol{\chi}_i^\top\|_\text{F}^2\right]=\frac{1}{n}E[\|\boldsymbol{\chi}_i\|_2^4] = O(n^{-1}).
\end{aligned}
\]
Therefore, \(\|\boldsymbol\Lambda_K^o - \boldsymbol\Lambda\|_\text{op}\le\|\boldsymbol\Lambda_K^o - \boldsymbol\Lambda \|_\text{F}=O_p(n^{-1/2})\).

Since $n^{-1/2} = o(r_{S,n,K_1})$, we get
\[
\|\widehat{\boldsymbol\Lambda}_K - \boldsymbol\Lambda_K\|_\text{op}=O_p\left(\sqrt{K_1K_2}(r_{S,n,K_1}\vee r_{V,n,K_2})\right).
\]

It remains to establish the stability of the smallest eigenvalue.
By Weyl's inequality,
\[
\lambda_{\min}(\widehat{\boldsymbol\Lambda}_K)
\ge
\lambda_{\min}(\boldsymbol\Lambda_K)
-
\left\|
\widehat{\boldsymbol\Lambda}_K-\boldsymbol\Lambda_K
\right\|_{\mathrm{op}}.
\]
By the assumed lower bound on the population Gram matrix and the
operator-norm bound established above,
\[
\begin{aligned}
\lambda_{\min}(\widehat{\boldsymbol\Lambda}_K)
&\ge
c_\Lambda\underline{\lambda}_K
-
O_p\left\{
\sqrt{K_1K_2}
\left(
r_{S,n,K_1}\vee r_{V,n,K_2}
\right)
\right\} \\
&=
c_\Lambda\underline{\lambda}_K
-
o_p(\underline{\lambda}_K) \\
&=
\underline{\lambda}_K
\{c_\Lambda-o_p(1)\},
\end{aligned}
\]
for some constant $c_{\Lambda} > 0$. Consequently, \(
P\left(
\lambda_{\min}(\widehat{\boldsymbol\Lambda}_K)
\ge
{c_\Lambda}\underline{\lambda}_K/2
\right)
\longrightarrow 1,
\) and hence \(
\lambda_{\min}(\widehat{\boldsymbol\Lambda}_K)
=
\Omega_p(\underline{\lambda}_K).
\) In particular, $\widehat{\boldsymbol\Lambda}_K$ is nonsingular with probability
tending to one. Since $\widehat{\boldsymbol\Lambda}_K$ is symmetric and positive
semidefinite,
\[
\left\|
\widehat{\boldsymbol\Lambda}_K^{-1}
\right\|_{\mathrm{op}}
=
\frac{1}
{\lambda_{\min}(\widehat{\boldsymbol\Lambda}_K)}
=
O_p(\underline{\lambda}_K^{-1}).
\]
This completes the proof.
\end{proof}

\vspace{1em}

\begin{proof}[\hspace{-1.25em}Proof of Theorem 2]
Recall that for subject $i$,
\[
\boldsymbol{\chi}_i^S
=
(\chi_{i1}^S,\ldots,\chi_{iK_1}^S)^\top,
\qquad
\boldsymbol{\chi}_i^V
=
(\chi_{i1}^V,\ldots,\chi_{iK_2}^V)^\top,
\]
define the retained true and estimated design vectors by
\[
\mathbf z_i
=
\begin{pmatrix}
\boldsymbol{\chi}_i^S\\
\boldsymbol{\chi}_i^V\\
\boldsymbol{\chi}_i^S\otimes\boldsymbol{\chi}_i^V
\end{pmatrix},
\qquad
\widehat{\mathbf z}_i
=
\begin{pmatrix}
\widehat{\boldsymbol{\chi}}_i^S\\
\widehat{\boldsymbol{\chi}}_i^V\\
\widehat{\boldsymbol{\chi}}_i^S
\otimes
\widehat{\boldsymbol{\chi}}_i^V
\end{pmatrix},
\]
and augment them with the intercept as
\[
\mathbf z_{+,i}
=
\begin{pmatrix}
1\\
\mathbf z_i
\end{pmatrix},
\qquad
\widehat{\mathbf z}_{+,i}
=
\begin{pmatrix}
1\\
\widehat{\mathbf z}_i
\end{pmatrix}.
\]
Let $\mathbf{Z}$ and $\widehat{\mathbf Z}$ denote the $n\times d_K$ design matrices
whose $i$th rows are $\mathbf z_i^\top$ and
$\widehat{\mathbf z}_i^\top$, respectively, and let $\mathbf{Z}_+$ and
$\widehat {\mathbf Z}_+$ denote their intercept-augmented counterparts.
Thus, \(
\widehat {\mathbf Z}_+=(\mathbf 1_n,\mathbf Z)\) and \( \widehat {\mathbf Z}_+=(\mathbf 1_n,\widehat {\mathbf Z}),
\) where $d_K=K_1+K_2+K_1K_2$,
and the empirical Gram matrix is \(
\widehat{\boldsymbol{\Lambda}}_K
=
n^{-1}
\widehat{\mathbf Z}_+^\top\widehat{\mathbf Z}_+.
\) Since \(\widehat{\boldsymbol{\Lambda}}_K\) is positive definie with propability tending to 1 by Lemma~\ref{lem:gram_mat}, we can define the corresponding $\widehat{\boldsymbol{\Lambda}}_K^{-1/2}$. Write the scalar response model as
\[
\mathbf Y
=
\mathbf Z_+\boldsymbol{\theta}_+
+
\boldsymbol{\varepsilon}
+
\mathbf R_K,
\]
where $\mathbf {Y}$ is the vector of scalar response, \(\boldsymbol{\theta}_+ =(\alpha, \boldsymbol b, \boldsymbol c, \operatorname{vec}(\boldsymbol{D})^\top)^\top \) collects the regression cofficients as defined in Section~\ref{sec: consistency properties} of the main text, $\boldsymbol{\varepsilon}=(\varepsilon_1,...,\varepsilon_n)^\top$ is the vector of regression error, and $\mathbf R_K=(R_{1,K},\ldots,R_{n,K})^\top$ collects the
truncation remainder arising from the omitted functional principal components, where for each subject $i$,
\begin{equation}
\label{eq:decomposition of Rik}
\begin{aligned}
R_{i,K}
={}&
\sum_{k>K_1} b_k\chi_{ik}^S
+
\sum_{\ell>K_2} c_\ell\chi_{i\ell}^V+
\sum_{k>K_1}\sum_{\ell=1}^{\infty}
d_{k\ell}\chi_{ik}^S\chi_{i\ell}^V
+
\sum_{k=1}^{K_1}\sum_{\ell>K_2}
d_{k\ell}\chi_{ik}^S\chi_{i\ell}^V.
\end{aligned}
\end{equation}
Since \(
\widehat{\boldsymbol{\theta}}_+
=
\widehat{\boldsymbol{\Lambda}}_K^{-1}
\left(
n^{-1}
\widehat{\mathbf Z}_+^\top\mathbf Y
\right),
\) substitution of the preceding model gives
\[
\widehat{\boldsymbol{\theta}}_+
-\boldsymbol{\theta}_+
={}
\widehat{\boldsymbol{\Lambda}}_K^{-1}
\left\{
\frac{1}{n}
\widehat{\mathbf Z}_+^\top
(\mathbf Z_+-\widehat{\mathbf Z}_+)
\boldsymbol{\theta}_+
\right\}+
\widehat{\boldsymbol{\Lambda}}_K^{-1}
\left(
\frac{1}{n}
\widehat{\mathbf Z}_+^\top
\boldsymbol{\varepsilon}
\right)+
\widehat{\boldsymbol{\Lambda}}_K^{-1}
\left(
\frac{1}{n}
\widehat{\mathbf Z}_+^\top
\mathbf R_K
\right).
\]
Accordingly, define
\[
T_{1n}
=
\left\|
\widehat{\boldsymbol{\Lambda}}_K^{-1/2}\frac{1}{n}
\widehat{\mathbf Z}_+^\top
(\mathbf Z_+-\widehat{\mathbf Z}_+)
\boldsymbol{\theta}_+
\right\|_2,\quad
T_{2n}
=
\left\|\widehat{\boldsymbol{\Lambda}}_K^{-1/2}
\frac{1}{n}
\widehat{\mathbf Z}_+^\top
\boldsymbol{\varepsilon}
\right\|_2,
\quad
T_{3n}
=
\left\|
\widehat{\boldsymbol{\Lambda}}_K^{-1/2}\frac{1}{n}
\widehat{\mathbf Z}_+^\top
\mathbf R_K
\right\|_2.
\]
It follows that
\begin{equation*}
\left\|\widehat{\boldsymbol{\theta}}_+
-\boldsymbol{\theta}_+\right\|_2
\le
\left\|\widehat{\boldsymbol{\Lambda}}_K^{-1/2}\right\|_{\mathrm{op}}
(T_{1n}+T_{2n}+T_{3n}).
\end{equation*}

By Lemma~\ref{lem:gram_mat} and the symmetry of the empirical gram matrix, \(
\|\widehat{\boldsymbol{\Lambda}}_K^{-1/2}\|_\text{op}=\sqrt{\|\widehat{\boldsymbol{\Lambda}}_K^{-1}\|_{\mathrm{op}}}\) 
\(=
O_p(\underline{\lambda}_K^{-1/2}).
\) We first consider $T_{1n}$. Since the intercept coordinate is estimated
without design error, we have\(
(\mathbf Z_+-\widehat{\mathbf Z}_+)
\boldsymbol{\theta}_+
=
(\mathbf Z-\widehat{\mathbf Z})
\boldsymbol{\theta}.
\) Define $
u_i
=
(\mathbf z_i-\widehat{\mathbf z}_i)^\top
\boldsymbol{\theta}.
$ and $\boldsymbol{u}=(u_1,...,u_n)^\top$.
Then we have,
\[
T_{1n}^2=
\left\|
\mathbf{\widehat \Lambda}_K^{-1/2}\frac{1}{n}
\widehat{\mathbf{Z}}_+^\top\boldsymbol{u}
\right\|_2^2\le\frac{1}{n}\boldsymbol{u}^\top P\boldsymbol{u}\le
\frac{1}{n}\sum_{i=1}^n u_i^2
,
\]
where $P = \widehat{\mathbf{Z}}_+(\widehat{\mathbf{Z}}_+^\top\widehat{\mathbf{Z}}_+)^{-1}\widehat{\mathbf{Z}}_+^\top$ is an $n \times n$ projection matrix. Consequently,
\begin{equation*}
T_{1n}
=
O_p\left[
\left(
\frac{1}{n}\sum_{i=1}^n u_i^2
\right)^{1/2}
\right].
\end{equation*}

Using \(
\Delta_{ik}^S
=
\widehat{\chi}_{ik}^S-\chi_{ik}^S
\text{ and }
\Delta_{i\ell}^V
=
\widehat{\chi}_{i\ell}^V-\chi_{i\ell}^V,
\)
we may write \(
u_i=-(A_i+B_i+C_i),
\)
where
\[
A_i
=
\sum_{k=1}^{K_1}b_k\Delta_{ik}^S,
\quad
B_i
=
\sum_{\ell=1}^{K_2}c_\ell\Delta_{i\ell}^V, \quad C_i=\sum_{k=1}^{K_1}\sum_{\ell=1}^{K_2}d_{k\ell}
\left(\Delta_{ik}^S\chi_{i\ell}^V
+\widehat{\chi}_{ik}^S\Delta_{i\ell}^V\right).
\]
Therefore, \(
u_i^2
\le
3(A_i^2+B_i^2+C_i^2).
\) By the weighted Cauchy--Schwarz inequality,
\[
A_i^2
\le
\left(\sum_{k=1}^{K_1}|b_k|\right)
\left(
\sum_{k=1}^{K_1}
|b_k||\Delta_{ik}^S|^2
\right),
\]
and hence, under Theorem~1 and the coefficient-decay condition (S2),
\[
\frac{1}{n}\sum_{i=1}^n A_i^2
=
O_p(r_{S,n,K_1}^2).
\]
Similarly,
\[
\frac{1}{n}\sum_{i=1}^n B_i^2
=
O_p(r_{V,n,K_2}^2).
\]

For the interaction term, write \(
C_i=C_{1i}+C_{2i},
\) where
\[
C_{1i}
=
\sum_{k=1}^{K_1}\sum_{\ell=1}^{K_2}
d_{k\ell}
\Delta_{ik}^S{\chi}_{i\ell}^V,
\quad
C_{2i}
=
\sum_{k=1}^{K_1}\sum_{\ell=1}^{K_2}
d_{k\ell}
\widehat\chi_{ik}^S\Delta_{i\ell}^V.
\]
Again by weighted Cauchy--Schwarz, Theorem 1 and (S3),
\[
\frac{1}{n}\sum_{i=1}^n C_{1i}^2
=
O_p\left(
r_{S,n,K_1}^2
\right).
\]
Since the estimated mean-process scores are uniformly bounded under
Theorem~1,
\[
\frac{1}{n}\sum_{i=1}^n C_{2i}^2
=
O_p(r_{V,n,K_2}^2).
\]
It follows that
\[
\frac{1}{n}\sum_{i=1}^n u_i^2
=
O_p\left(
r_{S,n,K_1}^2
\vee
r_{V,n,K_2}^2
\right),
\]
and therefore \(
T_{1n}
= 
O_p\left(
r_{S,n,K_1}
\vee
r_{V,n,K_2}
\right).
\) We next consider the regression-noise term. let $\mathcal F$ denote the sigma-field generated by the functional
data used to construct the estimated scores. Conditional on $\mathcal F$, since the regression error is independent with the functional data, the regression errors remain
independent with \(
E(\varepsilon\mid\mathcal F)=\mathbf{0}\) and \(E(\varepsilon\varepsilon^\top\mid\mathcal F)=\sigma_\varepsilon^2I_n,
\) where $\mathbf{0}$ is the $n$-dimensional zero vector and $I_n$ is the $n\times n$ identity matrix. Then note that
\[
E\left[\|\widehat{\boldsymbol{\Lambda}}^{-1/2}n^{-1}\widehat{\mathbf{Z}}^\top\varepsilon\|_2^2 \mid \mathcal F\right]=\frac{1}{n}\operatorname{tr}(\sigma^2_\varepsilon P) = \frac{\sigma_\varepsilon^2 d_K}{n},
\]
where $d_K=1+K_1+K_2+K_1K_2=O(K_1K_2)$. Therefore, by the conditional Markov's inequality, we get
\[
T_{2n} = O_p\left(\sqrt{\frac{d_K}{n}}\right)
=O_p\left(\sqrt\frac{K_1K_2}{n}\right)\]

It remains to control the truncation term \(T_{3n}\). By the identity of a projection matrix and operator-norm inequality,
\[
\begin{aligned}
T_{3n}^2\le
\frac{1}{n}\|R_K\|_2^2.
\end{aligned}
\]
It therefore remains to bound the mean squared truncation
remainder. Recall the formula of $R_{i,K}$ in \eqref{eq:decomposition of Rik}, using $(a_1+\cdots+a_4)^2
\leq 4\sum_{j=1}^4a_j^2$, it suffices to bound the four
terms separately. For the two main-effect truncation terms, (R2), (S2) and
the orthogonality of the FPC scores yield
\[
\frac{1}{n}\sum_{i=1}^n
\left(
\sum_{k>K_1}b_k\chi_{ik}^S
\right)^2
=
O_p\left(
K_1^{1-2\eta_1-\alpha_1}
\right)\text{ and }\frac{1}{n}\sum_{i=1}^n
\left(
\sum_{\ell>K_2}c_\ell\chi_{i\ell}^V
\right)^2
=
O_p\left(
K_2^{1-2\eta_2-\alpha_2}
\right),
\]
by Markov's inequality. For the interaction terms, repeated applications of
Cauchy--Schwarz and (S1) give \(
\left|
E\left(
\chi_{ik}^S\chi_{ik'}^S
\chi_{i\ell}^V\chi_{i\ell'}^V
\right)
\right|
\lesssim
\left(
\rho_k^S\rho_{k'}^S
\rho_\ell^V\rho_{\ell'}^V
\right)^{1/2}
\). Together with \(
|d_{k\ell}|
\lesssim
k^{-\eta_3}\ell^{-\eta_4},
\)
this implies
\[
\begin{aligned}
E\left[
\left(
\sum_{k>K_1}\sum_{\ell=1}^{\infty}
d_{k\ell}\chi_{ik}^S\chi_{i\ell}^V
\right)^2
\right]
&\lesssim
\left(
\sum_{k>K_1}
k^{-\eta_3-\alpha_1/2}
\right)^2
\left(
\sum_{\ell=1}^{\infty}
\ell^{-\eta_4-\alpha_2/2}
\right)^2=
O\left(
K_1^{2-2\eta_3-\alpha_1}
\right).
\end{aligned}
\]
Hence, by Markov's inequality,
\[
\frac{1}{n}\sum_{i=1}^n
\left(
\sum_{k>K_1}\sum_{\ell=1}^{\infty}
d_{k\ell}\chi_{ik}^S\chi_{i\ell}^V
\right)^2
=
O_p\left(
K_1^{2-2\eta_3-\alpha_1}
\right).
\]
Similarly,
\[
\frac{1}{n}\sum_{i=1}^n
\left(
\sum_{k=1}^{K_1}\sum_{\ell>K_2}
d_{k\ell}\chi_{ik}^S\chi_{i\ell}^V
\right)^2
=
O_p\left(
K_2^{2-2\eta_4-\alpha_2}
\right).
\]
Therefore,
\[
\frac{1}{n}\|R_K\|_2^2
=
O_p\left(
K_1^{\{(2-2\eta_3)\vee (1-2\eta_1)\}-\alpha_1}
\vee
K_2^{\{(2-2\eta_4)\vee (1-2\eta_2)\}-\alpha_2}
\right).
\]
Consequently,
\[
T_{3n}
=
O_p\left(
K_1^{\{(1-\eta_3)\vee (1/2-\eta_1)\}-\alpha_1/2}
\vee
K_2^{\{(1-\eta_4)\vee (1/2-\eta_2)\}-\alpha_2/2}
\right).
\]
Under the stated growth conditions, the regression-noise term
$T_{2n}=o_p(r_{S,n,K_1}\vee r_{V,n,K_2})$  Hence,
\[
\begin{aligned}
T_{1n}+T_{2n}+T_{3n}
=
O_p\Bigl(&
r_{S,n,K_1}
\vee
r_{V,n,K_2}
\\
&\vee
K_1^{\{(1-\eta_3)\vee(1/2-\eta_1)\}-\alpha_1/2}
\\
&\vee
K_2^{\{(1-\eta_4)\vee(1/2-\eta_2)\}-\alpha_2/2}
\Bigr).
\end{aligned}
\]
Since \(
\|\widehat{\boldsymbol \Lambda}_K^{-1/2}\|_{\mathrm{op}}
=
O_p(\lambda_K^{-1/2}),
\)
using the notation defined in Section~\ref{sec: consistency properties} of the main text, we obtain \(
\|\widehat\theta_+-\theta_+\|_2
=
O_p\left(
A_{n,K}
\right).
\)

We now translate the coefficient-vector bound into the corresponding
$L^2$ convergence rates for the coefficient functions and the interaction
surface. We first consider $\beta_1$. Recall that \(
\beta_1(s)
=
\sum_{k=1}^{\infty} b_k\varphi_k^S(s)\) and \(
\widehat{\beta}_1(s)
=
\sum_{k=1}^{K_1}\widehat b_k\widehat\varphi_k^S(s)
\). Then
\[
\widehat{\beta}_1-\beta_1
=
\sum_{k=1}^{K_1}
(\widehat b_k-b_k)\widehat\varphi_k^S
+
\sum_{k=1}^{K_1}
b_k(\widehat\varphi_k^S-\varphi_k^S)
-
\sum_{k>K_1}b_k\varphi_k^S.
\]
Hence,
\[
\begin{aligned}
\|\widehat{\beta}_1-\beta_1\|_2^2
\lesssim{}&
\left\|
\sum_{k=1}^{K_1}
(\widehat b_k-b_k)\widehat\varphi_k^S
\right\|_2^2+
\left\|
\sum_{k=1}^{K_1}
b_k(\widehat\varphi_k^S-\varphi_k^S)
\right\|_2^2
+
\left\|
\sum_{k>K_1}b_k\varphi_k^S
\right\|_2^2.
\end{aligned}
\]
Since the estimated eigenfunctions are orthonormal,
\[
\left\|
\sum_{k=1}^{K_1}
(\widehat b_k-b_k)\widehat\varphi_k^S
\right\|_2^2
\le
\|\widehat{\boldsymbol\theta}_+
-\boldsymbol\theta_+\|_2^2
=
O_p(A_{n,K}^2).
\]
For the eigenfunction-estimation term, the weighted
Cauchy--Schwarz inequality gives
\[
\left\|
\sum_{k=1}^{K_1}
b_k(\widehat\varphi_k^S-\varphi_k^S)
\right\|_2^2
\le
\left(\sum_{k=1}^{K_1}|b_k|\right)
\left\{
\sum_{k=1}^{K_1}
|b_k|
\|\widehat\varphi_k^S-\varphi_k^S\|_2^2
\right\}.
\]
By (S2),
$\sum_{k=1}^{K_1}|b_k|=O(1)$. Together with the uniform
eigenfunction bound \(
\max_{1\le k\le K_1}
\|\widehat\varphi_k^S-\varphi_k^S\|_2
=
O_p\left(K_1^{\alpha_1+1}h_{C_S}^2\right),
\)
this yields
\[
\left\|
\sum_{k=1}^{K_1}
b_k(\widehat\varphi_k^S-\varphi_k^S)
\right\|_2^2
=
O_p\left(K_1^{2\alpha_1+2}h_{C_S}^4\right)
=
O_p(A_{n,K}^2),
\]
where the last relation follows from the definition of
$r_{S,n,K_1}$ and $A_{n,K}$.

Finally, by orthonormality and (S2),
\[
\left\|
\sum_{k>K_1}b_k\varphi_k^S
\right\|_2^2
=
\sum_{k>K_1}b_k^2
\lesssim
\sum_{k>K_1}k^{-2\eta_1}
=
O(K_1^{1-2\eta_1}).
\]
Therefore, \(
\|\widehat{\beta}_1-\beta_1\|_2^2
=
O_p\left(
A_{n,K}^2
\vee K_1^{1-2\eta_1}
\right).
\) The argument for $\widehat{\beta}_2$ is analogous. In particular,
using \(
\max_{1\le \ell\le K_2}
\|\widehat\varphi_\ell^V-\varphi_\ell^V\|_2
=
O_p\left(K_2^{\alpha_2+1}
\theta_{m,2}
\right)
\), together with (S2) and the definition of $r_{V,n,K_2}$, we obtain
\[
\left\|
\sum_{\ell=1}^{K_2}
c_\ell(\widehat\varphi_\ell^V-\varphi_\ell^V)
\right\|_2^2
=
O_p\left(K_2^{2\alpha_2+2}
\theta_{m,2}^2
\right)
=
O_p(A_{n,K}^2),
\]
whereas
\[
\left\|
\sum_{\ell>K_2}c_\ell\varphi_\ell^V
\right\|_2^2
=
O(K_2^{1-2\eta_2}).
\]
Consequently, \(
\|\widehat{\beta}_2-\beta_2\|_2^2
=
O_p\left(
A_{n,K}^2
\vee K_2^{1-2\eta_2}
\right).
\)

It remains to consider the interaction surface. Write
\[
\gamma(s,t)
=
\sum_{k=1}^{\infty}\sum_{\ell=1}^{\infty}
d_{k\ell}\varphi_k^S(s)\varphi_\ell^V(t),
\]
and
\[
\widehat\gamma(s,t)
=
\sum_{k=1}^{K_1}\sum_{\ell=1}^{K_2}
\widehat d_{k\ell}
\widehat\varphi_k^S(s)\widehat\varphi_\ell^V(t).
\]
Decompose \(
\widehat\gamma-\gamma
=
G_{1n}+G_{2n}+G_{3n},
\) where
\[
G_{1n}
=
\sum_{k=1}^{K_1}\sum_{\ell=1}^{K_2}
(\widehat d_{k\ell}-d_{k\ell})
\widehat\varphi_k^S\otimes\widehat\varphi_\ell^V,
\]
\[
G_{2n}
=
\sum_{k=1}^{K_1}\sum_{\ell=1}^{K_2}
d_{k\ell}
\left(
\widehat\varphi_k^S\otimes\widehat\varphi_\ell^V
-
\varphi_k^S\otimes\varphi_\ell^V
\right),
\]
and $G_{3n}$ denotes the part of the true expansion outside
$\{1,\ldots,K_1\}\times\{1,\ldots,K_2\}$.

Since
$\{\widehat\varphi_k^S\otimes\widehat\varphi_\ell^V\}$
is an orthonormal system in $L^2(\mathcal T^2)$,
\[
\|G_{1n}\|_2^2
=
\sum_{k=1}^{K_1}\sum_{\ell=1}^{K_2}
(\widehat d_{k\ell}-d_{k\ell})^2
\le
\|\widehat{\boldsymbol\theta}_+
-\boldsymbol\theta_+\|_2^2
=
O_p(A_{n,K}^2).
\]

For $G_{2n}$, use \(
\widehat\varphi_k^S\otimes\widehat\varphi_\ell^V
-
\varphi_k^S\otimes\varphi_\ell^V
=
(\widehat\varphi_k^S-\varphi_k^S)
\otimes\widehat\varphi_\ell^V
+
\varphi_k^S\otimes
(\widehat\varphi_\ell^V-\varphi_\ell^V).
\)
By (S3),
\[
\sum_{k=1}^{K_1}\sum_{\ell=1}^{K_2}|d_{k\ell}|
\lesssim
\left(\sum_{k=1}^{K_1}k^{-\eta_3}\right)
\left(\sum_{\ell=1}^{K_2}\ell^{-\eta_4}\right)
=
O(1).
\]
Weighted Cauchy--Schwarz and the
uniform eigenfunction bounds therefore imply
\[
\|G_{2n}\|_2^2
=
O_p\left\{
\left(K_1^{\alpha_1+1}h_{C_S}^2\right)^2
\vee
\left(K_2^{\alpha_2+1}
\theta_{m,2}\right)^2
\right\}
=
O_p(A_{n,K}^2).
\]

Finally, by orthonormality of
$\{\varphi_k^S\otimes\varphi_\ell^V\}$,
\[
\begin{aligned}
\|G_{3n}\|_2^2
={}&
\sum_{k>K_1}\sum_{\ell=1}^{\infty}d_{k\ell}^2
+
\sum_{k=1}^{K_1}\sum_{\ell>K_2}d_{k\ell}^2.
\end{aligned}
\]
Using (S3),
\[
\sum_{k>K_1}\sum_{\ell=1}^{\infty}d_{k\ell}^2
=
O(K_1^{1-2\eta_3}),
\qquad
\sum_{k=1}^{K_1}\sum_{\ell>K_2}d_{k\ell}^2
=
O(K_2^{1-2\eta_4}).
\]
Hence \(
\|G_{3n}\|_2^2
=
O\left(
K_1^{1-2\eta_3}
\vee
K_2^{1-2\eta_4}
\right).
\)
Combining the preceding bounds gives
\[
\|\widehat\gamma-\gamma\|_2^2
=
O_p\left(
A_{n,K}^2
\vee
K_1^{1-2\eta_3}
\vee
K_2^{1-2\eta_4}
\right),
\]
which completes the proof.
\end{proof}

\vspace{0.5em}

\begin{proof}[\hspace{-1.25em}Proof of Lemma 1]

We first consider the non-oracle mean perturbation.
By the same local linear arguments as in Step 1 of the
proof of Proposition 1, the smoothing weights
$w_{k\ell}^{\mu}(t)$ satisfy
\[
\sup_{t\in\mathcal T}\sum_{k,\ell}
|w_{k\ell}^{\mu}(t)|=O(1),
\qquad
\sup_{t\in\mathcal T}\max_{k,\ell}
|w_{k\ell}^{\mu}(t)|
=O\left(\frac{1}{nmh_{\mu_V}}\right).
\]
Moreover, by the compact support of the kernel and (A3),
\[
\sup_{1\le i\le n}
\sum_{j=2}^m
|w_{k\ell}^{\mu}(t_{ij})|\Delta_{ij}
=O\left(\frac{1}{nm}\right),
\]
uniformly in $k,\ell$.

Since \(
\widehat\mu_V(t)-\widetilde\mu_V(t)
=
\sum_{k=1}^n\sum_{\ell=1}^m
w_{k\ell}^{\mu}(t)
(\widehat Z_{k\ell}-Z_{k\ell}),
\) the Cauchy--Schwarz inequality yields
\[
\begin{aligned}
&\sup_{1\le i\le n}
\sum_{j=2}^m
|\widehat\mu_V(t_{ij})
-\widetilde\mu_V(t_{ij})|^2\Delta_{ij}\\
&\quad\le
C\sup_{1\le i\le n}
\sum_{k=1}^n\sum_{\ell=1}^m
|\widehat Z_{k\ell}-Z_{k\ell}|^2
\sum_{j=2}^m
|w_{k\ell}^{\mu}(t_{ij})|\Delta_{ij}\\
&\quad\le
\frac{C}{nm}
\sum_{k=1}^n\sum_{\ell=1}^m
|\widehat Z_{k\ell}-Z_{k\ell}|^2
=O_p(\theta_{m,2}^2),
\end{aligned}
\]
where the last step follows from Proposition 1.

We next consider the covariance function. Let
\[
u_{ij}(s)=\frac{t_{ij}-s}{h_{C_V}},
\qquad
v_{i\ell}(t)=\frac{t_{i\ell}-t}{h_{C_V}},
\qquad
\mathbf{x}_{ij\ell}(s,t)
=
\begin{pmatrix}
1\\ u_{ij}(s)\\ v_{i\ell}(t)
\end{pmatrix},
\]
and write \(
K_{ij\ell}(s,t)
=
\kappa_2\{u_{ij}(s),v_{i\ell}(t)\},
\)
where $\kappa_2(\cdot,\cdot)$ is a bivariate kernel that satisfies (B2). The two-dimensional local linear estimator admits the representation
\[
\widehat C_V(s,t)
=
\sum_{i=1}^n\sum_{j\ne\ell}
w_{ij\ell}(s,t)\widehat C_{ij\ell},
\]
where
\[
w_{ij\ell}(s,t)
=
K_{ij\ell}(s,t)
\mathbf e_1^\top M(s,t)^{-1}
\mathbf{x}_{ij\ell}(s,t),
\]
\[
M(s,t)
=
\sum_{i=1}^n\sum_{j\ne\ell}
K_{ij\ell}(s,t)
\mathbf{x}_{ij\ell}(s,t)
\mathbf{x}_{ij\ell}(s,t)^\top,
\]
and
\[
\widehat C_{ij\ell}
=
\{\widehat Z_{ij}-\widehat\mu_V(t_{ij})\}
\{\widehat Z_{i\ell}-\widehat\mu_V(t_{i\ell})\}.
\]
The oracle quantities $\widetilde C_V$ and
$\widetilde C_{ij\ell}$ are defined analogously.

By (A3), the compact support and continuity of $\kappa_2$,
and the positivity of $\kappa_2$ on a nondegenerate
two-dimensional neighborhood, standard local linear design
arguments yield, uniformly in $(s,t)\in\mathcal T^2$, \(
\lambda_{\min}\{M(s,t)\}
\gtrsim nm^2h_{C_V}^2.
\) Indeed, three sufficiently small rectangles with
noncollinear centers can be selected within the kernel-positive
region, each containing at least $cm^2h_{C_V}^2$
off-diagonal observation pairs per subject. Consequently,
\[
\sup_{s,t}\max_{i,j\ne\ell}|w_{ij\ell}(s,t)|
=O\left(\frac{1}{nm^2h_{C_V}^2}\right),
\]
and
\[
\sup_{s,t}\sum_{i,j\ne\ell}|w_{ij\ell}(s,t)|
=O(1),
\qquad
\sup_{i,j\ne\ell}
\iint_{\mathcal T^2}|w_{ij\ell}(s,t)|\,ds\,dt
=O\left(\frac{1}{nm^2}\right).
\]

By Proposition 1 and the mean-function weight bounds
established above,
\[
\sup_{i,j}E\left[|\widehat Z_{ij}-Z_{ij}|^q\right]
=O(\theta_{m,2}^q),
\qquad
\sup_{t\in\mathcal T}
E\left[|\widehat\mu_V(t)-\widetilde\mu_V(t)|^q\right]
=O(\theta_{m,2}^q),
\quad q=2,4.
\]
Expanding $\widehat C_{ij\ell}-\widetilde C_{ij\ell}$
and applying the Cauchy--Schwarz inequality, together with
the uniformly bounded fourth moments of the oracle
centered observations, gives
\[
\sup_{i,j\ne\ell}
E\left[|\widehat C_{ij\ell}-\widetilde C_{ij\ell}|^2\right]
=O(\theta_{m,2}^2).
\]
It follows that
\[
\begin{aligned}
E\left[\|\widehat C_V-\widetilde C_V\|_2^2\right]
&\lesssim
\sum_{i=1}^n\sum_{j\ne\ell}
E\left[|\widehat C_{ij\ell}-\widetilde C_{ij\ell}|^2\right]
\iint_{\mathcal T^2}|w_{ij\ell}(s,t)|\,ds\,dt=O(\theta_{m,2}^2).
\end{aligned}
\]
Hence, by Markov's inequality, \(
\|\widehat C_V-\widetilde C_V\|_2
=O_p(\theta_{m,2}).
\)
These proves $(i)$. We next consider the oracle mean estimator. Let \(
U_{k\ell}=Z_{k\ell}-\mu_V(t_{k\ell}).
\) By the local linear reproduction properties,
\(
\widetilde\mu_V(t)-\mu_V(t)
=
B_\mu(t)+R_\mu(t),
\)
where
\(
B_\mu(t)
=
\sum_{k,\ell}w_{k\ell}^{\mu}(t)
\{\mu_V(t_{k\ell})-\mu_V(t)\}\) and \(R_\mu(t)
=
\sum_{k,\ell}w_{k\ell}^{\mu}(t)U_{k\ell}\). Taylor's expansion and the boundedness of
$\mu_V''$ yield
\(
\sup_{t\in\mathcal T}|B_\mu(t)|
=O(h_{\mu_V}^2).
\) Moreover, the local linear weight bounds imply \(
\sup_{k,t}\sum_{\ell=1}^m
|w_{k\ell}^{\mu}(t)|=O(n^{-1}).
\) Since observations from different subjects are
independent and the centered observations have
uniformly bounded second moments,
\[
\begin{aligned}
E\left[|R_\mu(t)|^2\right]
&=
\sum_{k=1}^n
E\left[\left|
\sum_{\ell=1}^m
w_{k\ell}^{\mu}(t)U_{k\ell}
\right|^2\right]\le
C\sum_{k=1}^n
\left(
\sum_{\ell=1}^m
|w_{k\ell}^{\mu}(t)|
\right)^2
=O(n^{-1}),
\end{aligned}
\]
uniformly in $t\in\mathcal T$.
Consequently,
\[
\begin{aligned}
\sup_{1\le i\le n}
E\left[\|\widetilde\mu_V-\mu_V\|_{i,m}^2\right]
&\le
C h_{\mu_V}^4
+C\sup_{1\le i\le n}
\sum_{j=2}^m
E\left[|R_\mu(t_{ij})|^2\right]\Delta_{ij}=O(h_{\mu_V}^4+n^{-1}).
\end{aligned}
\]

We next consider the oracle covariance estimator. Let \(C_{ij\ell}^{\circ}=U_{ij}U_{i\ell},
\)
and define
\[
C_V^\circ(s,t)
=
\sum_{i=1}^n\sum_{j\ne\ell}
w_{ij\ell}(s,t)C_{ij\ell}^{\circ}.
\]
We decompose \(
\widetilde C_V-C_V
=
(\widetilde C_V-C_V^\circ)
+
(C_V^\circ-C_V).
\) For the second term, write
\[
C_V^\circ(s,t)-C_V(s,t)
=
B_G(s,t)+R_G(s,t),
\]
where
\[
\begin{aligned}
B_G(s,t)
&=
\sum_{i,j\ne\ell}w_{ij\ell}(s,t)
\{C_V(t_{ij},t_{i\ell})-C_V(s,t)\},\\
R_G(s,t)
&=
\sum_{i,j\ne\ell}w_{ij\ell}(s,t)
\{C_{ij\ell}^{\circ}
-C_V(t_{ij},t_{i\ell})\}.
\end{aligned}
\]
By the second-order Taylor expansion, the local linear
reproduction properties, and the boundedness of the
second derivatives of $C_V$,
\[
\sup_{(s,t)\in\mathcal T^2}|B_G(s,t)|
\le
Ch_{C_V}^2
\sup_{s,t}\sum_{i,j\ne\ell}|w_{ij\ell}(s,t)|
=
O(h_{C_V}^2).
\]

Moreover, observations from different subjects are
independent. The uniformly bounded second moments
of $C_{ij\ell}^{\circ}$ and the subject-level weight bound
\[
\sup_{i,s,t}\sum_{j\ne\ell}
|w_{ij\ell}(s,t)|=O(n^{-1})
\]
yield
\[
\begin{aligned}
E\left[|R_G(s,t)|^2\right]
&=
\sum_{i=1}^n
E\left[\left|
\sum_{j\ne\ell}w_{ij\ell}(s,t)
\{C_{ij\ell}^{\circ}
-C_V(t_{ij},t_{i\ell})\}
\right|\right]^2\\
&\le
C\sum_{i=1}^n
\left(
\sum_{j\ne\ell}|w_{ij\ell}(s,t)|
\right)^2
=O(n^{-1}),
\end{aligned}
\]
uniformly in $(s,t)\in\mathcal T^2$.
Consequently, \(
\|C_V^\circ-C_V\|_2
=
O_p(h_{C_V}^2+n^{-1/2}).
\)

It remains to control the error due to estimating
the mean function. Write \(
d_{ij}
=
\widetilde\mu_V(t_{ij})-\mu_V(t_{ij}).
\) Then \(
\widetilde C_{ij\ell}-C_{ij\ell}^{\circ}
=
-d_{ij}U_{i\ell}
-d_{i\ell}U_{ij}
+d_{ij}d_{i\ell}.
\) Recall the oracle mean decomposition
$d(t)=B_\mu(t)+R_\mu(t)$, where
$\sup_t|B_\mu(t)|=O(h_{\mu_V}^2)$.
By subject-level independence, the uniformly bounded
fourth moments of $U_{ij}$, and
$\sup_{k,t}\sum_\ell|w_{k\ell}^{\mu}(t)|=O(n^{-1})$,
\[
\sup_t E\left[|R_\mu(t)|^4\right]=O(n^{-2}).
\]
Hence,
\[
\sup_{t\in\mathcal T}E\left[|d(t)|^4\right]
=
O(h_{\mu_V}^8+n^{-2}).
\]

By the Cauchy--Schwarz inequality and the uniformly
bounded fourth moments of $U_{ij}$,
\[
\begin{aligned}
&E\left[|\widetilde C_{ij\ell}
-C_{ij\ell}^{\circ}|^2\right]\le
C\left(
E[d_{ij}^2U_{i\ell}^2]
+E[d_{i\ell}^2U_{ij}^2]
+E[d_{ij}^2d_{i\ell}^2]
\right)=
O(h_{\mu_V}^4+n^{-1}),
\end{aligned}
\]
uniformly in $i,j,\ell$. Applying the weighted Cauchy--Schwarz inequality
and the integrated smoothing-weight bound gives
\[
\begin{aligned}
\|\widetilde C_V-C_V^\circ\|_2^2
&\le
C\sum_{i=1}^n\sum_{j\ne\ell}
|\widetilde C_{ij\ell}-C_{ij\ell}^{\circ}|^2
\int_{\mathcal T^2}|w_{ij\ell}(s,t)|\,ds\,dt\\
&\le
\frac{C}{nm^2}
\sum_{i=1}^n\sum_{j\ne\ell}
|\widetilde C_{ij\ell}-C_{ij\ell}^{\circ}|^2.
\end{aligned}
\]
Taking expectations yields \(
E\left[\|\widetilde C_V-C_V^\circ\|_2^2\right]
=
O(h_{\mu_V}^4+n^{-1}).
\) Therefore \(
\|\widetilde C_V-C_V^\circ\|_2
=
O_p(h_{\mu_V}^2+n^{-1/2}).
\) Combining the two bounds, we conclude that
\[
\|\widetilde C_V-C_V\|_2
=
O_p(h_{\mu_V}^2+h_{C_V}^2+n^{-1/2}).
\]
These proves $(ii)$. We next establish the eigenfunction estimation rates in (iii).
Let \(
\delta_k^V
=
\min\{\rho_{k-1}^V-\rho_k^V,
       \rho_k^V-\rho_{k+1}^V\}.
\) By (R2), $\delta_k^V\gtrsim k^{-(\alpha_2+1)}$.
The standard eigenfunction perturbation inequality,
together with the covariance estimation rates in $(i)$
and $(ii)$, yields \(
\|\widetilde{\varphi}_k^V-\varphi_k^V\|_2
=
O_p\left(
k^{\alpha_2+1}(h_{\mu_V}^2+h_{C_V}^2+n^{-1/2})
\right),
\)
and \(
\|\widehat{\varphi}_k^V-\widetilde{\varphi}_k^V\|_2
=
O_p(k^{\alpha_2+1}\theta_{m,2}).
\) Here, Weyl's inequality and the prescribed growth
conditions ensure that the estimated eigengaps
remain of order $k^{-(\alpha_2+1)}$ uniformly
over $k\le K_2$, with probability tending to one.

To transfer these continuous $L^2$ rates to the
observation grids, we first control the derivatives
of the estimated covariance surfaces.
Recall that
\[
w_{ij\ell}(s,t)
=
K_{ij\ell}(s,t)
\mathbf e_1^\top M(s,t)^{-1}x_{ij\ell}(s,t).
\]
Since the kernel is Lipschitz continuous,
the smoothing weights admit weak partial
derivatives. By the local linear design bounds
established above and \(
\partial_s M^{-1}
=
-M^{-1}(\partial_s M)M^{-1},
\) we obtain
\[
\operatorname*{ess\,sup}_{s,t}
\max_{i,j\ne\ell}
|\partial_s w_{ij\ell}(s,t)|
\le
\frac{C}{nm^2h_{C_V}^3}.
\]
The compact support of the kernel further gives
\[
\operatorname*{ess\,sup}_{s,t}
\sum_{i,j\ne\ell}
|\partial_s w_{ij\ell}(s,t)|
\le \frac{C}{h_{C_V}},
\]
and
\[
\max_{i,j\ne\ell}
\iint_{\mathcal T^2}
|\partial_s w_{ij\ell}(s,t)|\,ds\,dt
\le
\frac{C}{nm^2h_{C_V}}.
\]

Since the raw covariance products do not depend
on $(s,t)$, the weighted Cauchy--Schwarz
inequality yields
\[
\begin{aligned}
\|\partial_1\widehat C_V\|_2^2
&\le
\frac{C}{nm^2h_{C_V}^2}
\sum_{i=1}^n\sum_{j\ne\ell}
|\widehat C_{ij\ell}|^2.
\end{aligned}
\]
Under the uniform second-moment bounds for
the covariance products, Markov's inequality gives \(
\|\partial_1\widehat C_V\|_2
=O_p(h_{C_V}^{-1}).
\) The same argument applies to the oracle
covariance estimator, giving \(
\|\partial_1\widetilde C_V\|_2
=O_p(h_{C_V}^{-1}).
\)

The Lipschitz continuity of the covariance
estimators also ensures that their eigenfunctions
associated with nonzero eigenvalues admit
Lipschitz continuous representatives.
Differentiating the eigenequations in the weak
sense gives
\[
\widehat\rho_k^V
(\widehat{\varphi}_k^V)'(s)
=
\int_{\mathcal T}
\partial_1\widehat C_V(s,t)
\widehat{\varphi}_k^V(t)\,dt,
\]
and an analogous identity for
$\widetilde{\varphi}_k^V$.
Hence, by the Cauchy--Schwarz inequality,
\[
\|(\widehat{\varphi}_k^V)'\|_2
\le
\frac{\|\partial_1\widehat C_V\|_2}
{|\widehat\rho_k^V|},
\qquad
\|(\widetilde{\varphi}_k^V)'\|_2
\le
\frac{\|\partial_1\widetilde C_V\|_2}
{|\widetilde\rho_k^V|}.
\]
By (R2), Weyl's inequality, and the growth
conditions,
\[
\max_{k\le K_2}
\left\{
\frac{1}{|\widehat\rho_k^V|}
+
\frac{1}{|\widetilde\rho_k^V|}
\right\}
=O_p(K_2^{\alpha_2}).
\]
Consequently,
\[
\max_{k\le K_2}
\left(
\|(\widehat{\varphi}_k^V)'\|_2
+
\|(\widetilde{\varphi}_k^V)'\|_2
\right)
=
O_p\left(\frac{K_2^{\alpha_2}}{h_{C_V}}\right).
\]
Under the uniform boundedness of the second-order
partial derivatives of $G_V$, differentiating the
population eigenequation yields
\[
\max_{1\le k\le K_2}\|(\varphi_k^V)'\|_2
\le
\|\partial_1C_V\|_2
\max_{1\le k\le K_2}(\rho_k^V)^{-1}
=O(K_2^{\alpha_2}).
\]

Finally, for any absolutely continuous function \(f\) on \(\mathcal T\) with \(f'\in L^2(\mathcal T)\), the fundamental theorem of calculus and the Cauchy–Schwarz inequality imply, under (A3),
\[
\|f\|_{i,m}
\le
\|f\|_2+\frac{C}{m}\|f'\|_2,
\]
uniformly in $i$.
Applying this inequality to
$f=\widetilde{\varphi}_k^V-\varphi_k^V$
and
$f=\widehat{\varphi}_k^V-
\widetilde{\varphi}_k^V$, respectively, yields
\[
\max_{\substack{1\le k\le K_2\\1\le i\le n}}
\|\widetilde{\varphi}_k^V-\varphi_k^V\|_{i,m}
=
O_p\left(
K_2^{\alpha_2+1}
(h_{\mu_V}^2 + h_{C_V}^2 + n^{-1/2})
+
\frac{K_2^{\alpha_2}}{mh_{C_V}}
\right),
\]
and
\[
\max_{\substack{1\le k\le K_2\\1\le i\le n}}
\|\widehat{\varphi}_k^V-
\widetilde{\varphi}_k^V\|_{i,m}
=
O_p\left(
K_2^{\alpha_2+1}\theta_{m,2}
+
\frac{K_2^{\alpha_2}}{mh_{C_V}}
\right).
\]
Under (A2) and (A5), the
discretization terms are absorbed into the
respective continuous estimation rates, giving the desired result.

The corresponding mean and covariance estimation bounds
for the mean process follow by the same arguments as
those for the oracle variance process, since the
off-diagonal covariance products of the observed
process are unbiased for $C_S$ under equation~\eqref{equa: expec and cov of X} in the main text.

\end{proof}

\section{Additional simulation results}
Unless otherwise specified, the simulation settings and notation
follow those introduced in Section~\ref{subsec: simulation setting} of the main text.

\subsection{Estimation and prediction under alternative simulation settings}

This section presents additional simulation results examining the robustness
of the findings in the main simulation study to alternative observation-design
and dependence settings. We first replace the jittered regular observation
design by a random observation design while retaining independence between the
latent mean and variance processes. We then retain the jittered regular design
but allow the two latent processes to be positively dependent, with
$\rho_{SV}=0.5$. All other simulation settings are the same as those described
in Section~\ref{subsec: simulation setting} of the main text.

Table~\ref{tab:random_response_relRMSE} reports the response-prediction results under the random observation
design. Compared with the jittered regular design used in the main simulation
study, the relRMSEs are generally somewhat larger, while the main qualitative
patterns remain unchanged. In particular, incorporating the variance process
improves prediction when it contributes additional response information, and
the full model provides further improvement when an interaction effect is
present. Thus, the predictive advantages observed in the main simulation study
remain evident under the random observation design.

Table~\ref{tab:dependent_response_relRMSE} reports the corresponding results when the latent mean and variance
processes are positively dependent, with $\rho_{SV}=0.5$. The main qualitative
conclusions again remain unchanged under this dependence structure. The
additive model improves upon the mean-only model when the variance process
contributes to the response, while the full model provides further gains when
an interaction effect is present. The relRMSEs are also generally lower than
those under the corresponding independent-process setting in the main
simulation study, particularly for the full model. Since introducing dependence
changes the joint distribution of the functional predictors and response,
however, these differences in error magnitude should not be interpreted as the
effect of dependence alone. Overall, the results indicate that the predictive
performance of the proposed approach remains robust when the latent mean and
variance processes are correlated.

Before discussing the coefficient-estimation results, we note that the relRMISE is normalized by the \(L^2\)-norm of the true coefficient function or interaction surface. In particular, when an effect size is 0.25, the normalization factor is one-fourth of that under unit effect size, so the relRMISE is four times the corresponding RMISE normalized by the unit-effect coefficient norm. This scaling should be kept in mind when interpreting Tables~\ref{tab:simulation}–\ref{tab:dependent_simulation}, as larger relative errors do not necessarily indicate larger absolute estimation errors.

Table~\ref{tab:simulation} reports the coefficient-estimation results under the jittered regular design for two basis settings. Overall, the relRMISEs exhibit similar qualitative patterns across the two settings. When an interaction effect is present, the full model generally provides more accurate estimates of \(\beta_1\) than the additive model, while the estimation errors for \(\beta_2\) tend to be larger when the variance process contributes less strongly to the response. The results are also broadly comparable across the FVE thresholds, although some differences arise in the estimation of the interaction coefficient. These findings suggest that the main conclusions regarding coefficient estimation are relatively insensitive to the choice of basis setting.

Tables~\ref{tab:random_simulation} and~\ref{tab:dependent_simulation} present the corresponding coefficient-estimation results.
Under the random observation design (Table~\ref{tab:random_simulation}), the relRIMSEs are broadly
comparable to those in Table~\ref{tab:simulation}, with modest differences in
estimation accuracy. Under the dependent-process setting with
$\rho_{SV}=0.5$ (Table~\ref{tab:dependent_simulation}), the full-model coefficient estimates remain
similarly accurate. In contrast, substantially larger errors for
$\widehat{\beta}_1$ occur in some reduced-model configurations, particularly
for the mean-only model when the variance process contributes to the response.
This illustrates the potential consequence of omitting a relevant functional
process when the latent mean and variance processes are correlated. Overall,
the additional results support the robustness of the proposed full model to
both alternative observation designs and dependence between the latent
processes.


\begin{table}[!h]
\centering
\caption{Relative root mean squared errors (relRMSEs) of the fitted
responses under two basis settings for the random observation design.
The percentages (80\%, 85\%, and 90\%) correspond to the fraction of
variance explained (FVE) thresholds used to determine the numbers of
retained functional principal components. The mean-only model includes only the mean process, the additive model
includes both the mean and variance processes, and the full model further
includes their interaction. The model and basis labels in the left block
apply to the corresponding rows in both blocks.}
\label{tab:random_response_relRMSE}

\small
\setlength{\tabcolsep}{5.5pt}
\renewcommand{\arraystretch}{1}

\begin{tabular}{@{}
c l c c c c
@{\hspace{18pt}}
c c c c
@{}}

\toprule

$(\delta_S,\delta_V,\delta_I)$
& Model
& Basis
& 80\%
& 85\%
& 90\%
&
$(\delta_S,\delta_V,\delta_I)$
& 80\%
& 85\%
& 90\% \\

\cmidrule(r){1-6}
\cmidrule(l){7-10}


\multirow{6}{*}{$(1,1,1)$}
& \multirow{2}{*}{Mean-only}
& 1 & 0.817 & 0.817 & 0.814
&
\multirow{6}{*}{$(1,0.25,1)$}
& 0.717 & 0.717 & 0.713 \\

&
& 2 & 0.814 & 0.814 & 0.812
&
& 0.712 & 0.712 & 0.709 \\

\cmidrule(r){2-6}

&
\multirow{2}{*}{Additive}
& 1 & 0.603 & 0.601 & 0.597
&
& 0.692 & 0.691 & 0.686 \\

&
& 2 & 0.598 & 0.596 & 0.594
&
& 0.686 & 0.685 & 0.682 \\

\cmidrule(r){2-6}

&
\multirow{2}{*}{Full}
& 1 & 0.431 & 0.424 & 0.404
&
& 0.467 & 0.459 & 0.433 \\

&
& 2 & 0.400 & 0.392 & 0.384
&
& 0.426 & 0.416 & 0.406 \\


\midrule

\multirow{6}{*}{$(1,0,1)$}
& \multirow{2}{*}{Mean-only}
& 1 & 0.707 & 0.707 & 0.703
&
\multirow{6}{*}{$(0.25,1,1)$}
& 0.978 & 0.978 & 0.976 \\

&
& 2 & 0.701 & 0.701 & 0.698
&
& 0.978 & 0.978 & 0.976 \\

\cmidrule(r){2-6}

&
\multirow{2}{*}{Additive}
& 1 & 0.700 & 0.699 & 0.694
&
& 0.713 & 0.711 & 0.708 \\

&
& 2 & 0.694 & 0.692 & 0.690
&
& 0.712 & 0.710 & 0.708 \\

\cmidrule(r){2-6}

&
\multirow{2}{*}{Full}
& 1 & 0.470 & 0.462 & 0.436
&
& 0.472 & 0.464 & 0.445 \\

&
& 2 & 0.428 & 0.418 & 0.408
&
& 0.448 & 0.437 & 0.429 \\


\midrule

\multirow{6}{*}{$(1,0,0)$}
& \multirow{2}{*}{Mean-only}
& 1 & 0.376 & 0.376 & 0.370
&
\multirow{6}{*}{$(1,1,0)$}
& 0.740 & 0.740 & 0.737 \\

&
& 2 & 0.359 & 0.359 & 0.355
&
& 0.736 & 0.736 & 0.733 \\

\cmidrule(r){2-6}

&
\multirow{2}{*}{Additive}
& 1 & 0.374 & 0.372 & 0.366
&
& 0.362 & 0.360 & 0.355 \\

&
& 2 & 0.357 & 0.356 & 0.352
&
& 0.350 & 0.348 & 0.346 \\

\cmidrule(r){2-6}

&
\multirow{2}{*}{Full}
& 1 & 0.363 & 0.360 & 0.349
&
& 0.355 & 0.351 & 0.343 \\

&
& 2 & 0.350 & 0.346 & 0.339
&
& 0.345 & 0.341 & 0.335 \\

\bottomrule

\end{tabular}
\end{table}


\begin{table}[!t]
\centering
\caption{Relative root mean squared errors (relRMSEs) under the dependent-process
setting with $\rho_{SV}=0.5$. The observation times follow the jittered
regular design used in the main simulation study. All other simulation
settings and table entries are defined as in Table~\ref{tab:random_response_relRMSE}.}
\label{tab:dependent_response_relRMSE}

\small
\setlength{\tabcolsep}{5.5pt}
\renewcommand{\arraystretch}{1}

\begin{tabular}{@{}
c l c c c c
@{\hspace{18pt}}
c c c c
@{}}

\toprule

$(\delta_S,\delta_V,\delta_I)$
& Model
& Basis
& 80\%
& 85\%
& 90\%
&
$(\delta_S,\delta_V,\delta_I)$
& 80\%
& 85\%
& 90\% \\

\cmidrule(r){1-6}
\cmidrule(l){7-10}


\multirow{6}{*}{$(1,1,1)$}
& \multirow{2}{*}{Mean-only}
& 1 & 0.677 & 0.677 & 0.673
&
\multirow{6}{*}{$(1,0.25,1)$}
& 0.686 & 0.686 & 0.682 \\

&
& 2 & 0.672 & 0.671 & 0.668
&
& 0.681 & 0.680 & 0.678 \\

\cmidrule(r){2-6}

&
\multirow{2}{*}{Additive}
& 1 & 0.488 & 0.486 & 0.483
&
& 0.649 & 0.647 & 0.643 \\

&
& 2 & 0.490 & 0.487 & 0.485
&
& 0.647 & 0.645 & 0.643 \\

\cmidrule(r){2-6}

&
\multirow{2}{*}{Full}
& 1 & 0.313 & 0.305 & 0.290
&
& 0.390 & 0.380 & 0.361 \\

&
& 2 & 0.292 & 0.283 & 0.276
&
& 0.355 & 0.346 & 0.337 \\


\midrule

\multirow{6}{*}{$(1,0,1)$}
& \multirow{2}{*}{Mean-only}
& 1 & 0.718 & 0.717 & 0.714
&
\multirow{6}{*}{$(0.25,1,1)$}
& 0.872 & 0.872 & 0.868 \\

&
& 2 & 0.713 & 0.713 & 0.710
&
& 0.870 & 0.869 & 0.867 \\

\cmidrule(r){2-6}

&
\multirow{2}{*}{Additive}
& 1 & 0.708 & 0.706 & 0.701
&
& 0.625 & 0.624 & 0.621 \\

&
& 2 & 0.705 & 0.703 & 0.700
&
& 0.632 & 0.630 & 0.628 \\

\cmidrule(r){2-6}

&
\multirow{2}{*}{Full}
& 1 & 0.425 & 0.414 & 0.394
&
& 0.365 & 0.356 & 0.341 \\

&
& 2 & 0.385 & 0.376 & 0.366
&
& 0.350 & 0.339 & 0.332 \\


\midrule

\multirow{6}{*}{$(1,0,0)$}
& \multirow{2}{*}{Mean-only}
& 1 & 0.355 & 0.355 & 0.349
&
\multirow{6}{*}{$(1,1,0)$}
& 0.554 & 0.553 & 0.549 \\

&
& 2 & 0.342 & 0.341 & 0.337
&
& 0.546 & 0.546 & 0.543 \\

\cmidrule(r){2-6}

&
\multirow{2}{*}{Additive}
& 1 & 0.352 & 0.351 & 0.345
&
& 0.282 & 0.279 & 0.275 \\

&
& 2 & 0.339 & 0.337 & 0.333
&
& 0.274 & 0.269 & 0.266 \\

\cmidrule(r){2-6}

&
\multirow{2}{*}{Full}
& 1 & 0.344 & 0.341 & 0.331
&
& 0.272 & 0.268 & 0.262 \\

&
& 2 & 0.333 & 0.328 & 0.322
&
& 0.266 & 0.259 & 0.255 \\

\bottomrule

\end{tabular}
\end{table}

\begin{table}[!t]
\centering
\caption{Relative root mean integrated squared errors (relRIMSEs) of the
estimated coefficient functions under two basis settings for the jitter design and the same FVE thresholds as in
Table~\ref{tab:random_response_relRMSE}. The model, coefficient, and basis
labels in the left block apply to the corresponding rows in both blocks.}
\label{tab:simulation}

\small
\setlength{\tabcolsep}{4.5pt}
\renewcommand{\arraystretch}{1}

\begin{tabular}{@{}
c l c c c c c
@{\hspace{18pt}}
c c c c
@{}}

\toprule

$(\delta_S,\delta_V,\delta_I)$
& Model
& Coef.
& Basis
& 80\%
& 85\%
& 90\%
&
$(\delta_S,\delta_V,\delta_I)$
& 80\%
& 85\%
& 90\% \\

\cmidrule(r){1-7}
\cmidrule(l){8-11}


\multirow{12}{*}{$(1,1,1)$}
& \multirow{2}{*}{Mean-only}
& \multirow{2}{*}{$\widehat{\beta}_1$}
& 1 & 0.252 & 0.256 & 0.297
&
\multirow{12}{*}{$(1,0.25,1)$}
& 0.216 & 0.218 & 0.228 \\

&
&
& 2 & 0.253 & 0.256 & 0.295
&
& 0.217 & 0.217 & 0.222 \\

\cmidrule(r){2-7}

&
\multirow{4}{*}{Additive}
& \multirow{2}{*}{$\widehat{\beta}_1$}
& 1 & 0.216 & 0.217 & 0.231
&
& 0.213 & 0.215 & 0.224 \\

&
&
& 2 & 0.215 & 0.215 & 0.223
&
& 0.214 & 0.214 & 0.219 \\

\cmidrule(r){3-7}

&
& \multirow{2}{*}{$\widehat{\beta}_2$}
& 1 & 0.233 & 0.254 & 0.263
&
& 0.695 & 0.870 & 0.939 \\

&
&
& 2 & 0.237 & 0.243 & 0.244
&
& 0.696 & 0.878 & 0.894 \\

\cmidrule(r){2-7}

&
\multirow{6}{*}{Full}
& \multirow{2}{*}{$\widehat{\beta}_1$}
& 1 & 0.189 & 0.187 & 0.172
&
& 0.185 & 0.184 & 0.161 \\

&
&
& 2 & 0.183 & 0.183 & 0.166
&
& 0.180 & 0.179 & 0.158 \\

\cmidrule(r){3-7}

&
& \multirow{2}{*}{$\widehat{\beta}_2$}
& 1 & 0.202 & 0.197 & 0.194
&
& 0.515 & 0.592 & 0.620 \\

&
&
& 2 & 0.201 & 0.173 & 0.169
&
& 0.479 & 0.551 & 0.554 \\

\cmidrule(r){3-7}

&
& \multirow{2}{*}{$\widehat{\gamma}$}
& 1 & 0.522 & 0.486 & 0.529
&
& 0.519 & 0.480 & 0.505 \\

&
&
& 2 & 0.483 & 0.428 & 0.445
&
& 0.479 & 0.420 & 0.426 \\


\midrule

\multirow{12}{*}{$(1,0,1)$}
& \multirow{2}{*}{Mean-only}
& \multirow{2}{*}{$\widehat{\beta}_1$}
& 1 & 0.213 & 0.214 & 0.221
&
\multirow{12}{*}{$(0.25,1,1)$}
& 0.817 & 0.847 & 1.134 \\

&
&
& 2 & 0.213 & 0.213 & 0.216
&
& 0.824 & 0.842 & 1.117 \\

\cmidrule(r){2-7}

&
\multirow{4}{*}{Additive}
& \multirow{2}{*}{$\widehat{\beta}_1$}
& 1 & 0.213 & 0.215 & 0.224
&
& 0.626 & 0.642 & 0.842 \\

&
&
& 2 & 0.214 & 0.214 & 0.219
&
& 0.628 & 0.637 & 0.808 \\

\cmidrule(r){3-7}

&
& \multirow{2}{*}{$\widehat{\beta}_2$}
& 1 & 0.191 & 0.242 & 0.262
&
& 0.227 & 0.245 & 0.255 \\

&
&
& 2 & 0.191 & 0.246 & 0.251
&
& 0.233 & 0.237 & 0.238 \\

\cmidrule(r){2-7}

&
\multirow{6}{*}{Full}
& \multirow{2}{*}{$\widehat{\beta}_1$}
& 1 & 0.185 & 0.184 & 0.160
&
& 0.455 & 0.459 & 0.565 \\

&
&
& 2 & 0.181 & 0.179 & 0.158
&
& 0.433 & 0.452 & 0.537 \\

\cmidrule(r){3-7}

&
& \multirow{2}{*}{$\widehat{\beta}_2$}
& 1 & 0.137 & 0.163 & 0.172
&
& 0.193 & 0.184 & 0.178 \\

&
&
& 2 & 0.127 & 0.153 & 0.155
&
& 0.195 & 0.162 & 0.157 \\

\cmidrule(r){3-7}

&
& \multirow{2}{*}{$\widehat{\gamma}$}
& 1 & 0.519 & 0.480 & 0.505
&
& 0.501 & 0.453 & 0.458 \\

&
&
& 2 & 0.479 & 0.421 & 0.427
&
& 0.471 & 0.403 & 0.397 \\


\midrule

\multirow{12}{*}{$(1,0,0)$}
& \multirow{2}{*}{Mean-only}
& \multirow{2}{*}{$\widehat{\beta}_1$}
& 1 & 0.161 & 0.159 & 0.117
&
\multirow{12}{*}{$(1,1,0)$}
& 0.205 & 0.205 & 0.223 \\

&
&
& 2 & 0.160 & 0.158 & 0.118
&
& 0.205 & 0.208 & 0.231 \\

\cmidrule(r){2-7}

&
\multirow{4}{*}{Additive}
& \multirow{2}{*}{$\widehat{\beta}_1$}
& 1 & 0.161 & 0.159 & 0.117
&
& 0.165 & 0.163 & 0.130 \\

&
&
& 2 & 0.160 & 0.158 & 0.118
&
& 0.162 & 0.160 & 0.127 \\

\cmidrule(r){3-7}

&
& \multirow{2}{*}{$\widehat{\beta}_2$}
& 1 & 0.070 & 0.091 & 0.094
&
& 0.171 & 0.156 & 0.146 \\

&
&
& 2 & 0.066 & 0.088 & 0.089
&
& 0.177 & 0.135 & 0.129 \\

\cmidrule(r){2-7}

&
\multirow{6}{*}{Full}
& \multirow{2}{*}{$\widehat{\beta}_1$}
& 1 & 0.158 & 0.156 & 0.112
&
& 0.163 & 0.160 & 0.127 \\

&
&
& 2 & 0.158 & 0.156 & 0.115
&
& 0.161 & 0.159 & 0.125 \\

\cmidrule(r){3-7}

&
& \multirow{2}{*}{$\widehat{\beta}_2$}
& 1 & 0.068 & 0.090 & 0.096
&
& 0.172 & 0.157 & 0.148 \\

&
&
& 2 & 0.065 & 0.088 & 0.089
&
& 0.177 & 0.136 & 0.130 \\

\cmidrule(r){3-7}

&
& \multirow{2}{*}{$\widehat{\gamma}$}
& 1 & 0.148 & 0.194 & 0.302
&
& 0.169 & 0.228 & 0.364 \\

&
&
& 2 & 0.133 & 0.192 & 0.270
&
& 0.159 & 0.225 & 0.323 \\

\bottomrule

\end{tabular}
\end{table}


\begin{table}[!t]
\centering
\caption{Relative root mean integrated squared errors (relRIMSEs) of the
estimated coefficient functions under two basis settings for the random
observation design and the same FVE thresholds as in
Table~\ref{tab:random_response_relRMSE}. The model, coefficient, and basis
labels in the left block apply to the corresponding rows in both blocks.}
\label{tab:random_simulation}

\small
\setlength{\tabcolsep}{4.5pt}
\renewcommand{\arraystretch}{1}

\begin{tabular}{@{}
c l c c c c c
@{\hspace{16pt}}
c c c c
@{}}

\toprule

$(\delta_S,\delta_V,\delta_I)$
& Model
& Coef.
& Basis
& 80\%
& 85\%
& 90\%
&
$(\delta_S,\delta_V,\delta_I)$
& 80\%
& 85\%
& 90\% \\

\cmidrule(r){1-7}
\cmidrule(l){8-11}


\multirow{12}{*}{$(1,1,1)$}
& \multirow{2}{*}{Mean-only}
& \multirow{2}{*}{$\widehat{\beta}_1$}
& 1 & 0.258 & 0.258 & 0.297
&
\multirow{12}{*}{$(1,0.25,1)$}
& 0.225 & 0.224 & 0.236 \\

&
&
& 2 & 0.255 & 0.256 & 0.294
&
& 0.222 & 0.223 & 0.229 \\

\cmidrule(r){2-7}

&
\multirow{4}{*}{Additive}
& \multirow{2}{*}{$\widehat{\beta}_1$}
& 1 & 0.227 & 0.227 & 0.243
&
& 0.223 & 0.222 & 0.234 \\

&
&
& 2 & 0.223 & 0.223 & 0.230
&
& 0.220 & 0.221 & 0.225 \\

\cmidrule(r){3-7}

&
& \multirow{2}{*}{$\widehat{\beta}_2$}
& 1 & 0.233 & 0.250 & 0.260
&
& 0.694 & 0.847 & 0.918 \\

&
&
& 2 & 0.244 & 0.251 & 0.252
&
& 0.721 & 0.888 & 0.903 \\

\cmidrule(r){2-7}

&
\multirow{6}{*}{Full}
& \multirow{2}{*}{$\widehat{\beta}_1$}
& 1 & 0.204 & 0.203 & 0.191
&
& 0.199 & 0.198 & 0.178 \\

&
&
& 2 & 0.194 & 0.193 & 0.181
&
& 0.190 & 0.189 & 0.173 \\

\cmidrule(r){3-7}

&
& \multirow{2}{*}{$\widehat{\beta}_2$}
& 1 & 0.211 & 0.209 & 0.206
&
& 0.561 & 0.646 & 0.661 \\

&
&
& 2 & 0.212 & 0.195 & 0.192
&
& 0.535 & 0.617 & 0.615 \\

\cmidrule(r){3-7}

&
& \multirow{2}{*}{$\widehat{\gamma}$}
& 1 & 0.548 & 0.524 & 0.574
&
& 0.541 & 0.514 & 0.547 \\

&
&
& 2 & 0.499 & 0.468 & 0.506
&
& 0.492 & 0.454 & 0.475 \\


\midrule

\multirow{12}{*}{$(1,0,1)$}
& \multirow{2}{*}{Mean-only}
& \multirow{2}{*}{$\widehat{\beta}_1$}
& 1 & 0.222 & 0.221 & 0.231
&
\multirow{12}{*}{$(0.25,1,1)$}
& 0.815 & 0.825 & 1.105 \\

&
&
& 2 & 0.220 & 0.221 & 0.224
&
& 0.811 & 0.829 & 1.096 \\

\cmidrule(r){2-7}

&
\multirow{4}{*}{Additive}
& \multirow{2}{*}{$\widehat{\beta}_1$}
& 1 & 0.223 & 0.222 & 0.233
&
& 0.638 & 0.646 & 0.855 \\

&
&
& 2 & 0.220 & 0.222 & 0.226
&
& 0.631 & 0.648 & 0.805 \\

\cmidrule(r){3-7}

&
& \multirow{2}{*}{$\widehat{\beta}_2$}
& 1 & 0.190 & 0.236 & 0.257
&
& 0.226 & 0.241 & 0.250 \\

&
&
& 2 & 0.197 & 0.248 & 0.253
&
& 0.238 & 0.244 & 0.245 \\

\cmidrule(r){2-7}

&
\multirow{6}{*}{Full}
& \multirow{2}{*}{$\widehat{\beta}_1$}
& 1 & 0.199 & 0.197 & 0.177
&
& 0.499 & 0.502 & 0.619 \\

&
&
& 2 & 0.190 & 0.189 & 0.174
&
& 0.473 & 0.483 & 0.570 \\

\cmidrule(r){3-7}

&
& \multirow{2}{*}{$\widehat{\beta}_2$}
& 1 & 0.151 & 0.178 & 0.183
&
& 0.199 & 0.192 & 0.189 \\

&
&
& 2 & 0.142 & 0.171 & 0.170
&
& 0.203 & 0.182 & 0.180 \\

\cmidrule(r){3-7}

&
& \multirow{2}{*}{$\widehat{\gamma}$}
& 1 & 0.541 & 0.513 & 0.545
&
& 0.518 & 0.483 & 0.496 \\

&
&
& 2 & 0.492 & 0.454 & 0.474
&
& 0.482 & 0.436 & 0.449 \\


\midrule

\multirow{12}{*}{$(1,0,0)$}
& \multirow{2}{*}{Mean-only}
& \multirow{2}{*}{$\widehat{\beta}_1$}
& 1 & 0.171 & 0.168 & 0.130
&
\multirow{12}{*}{$(1,1,0)$}
& 0.210 & 0.209 & 0.225 \\

&
&
& 2 & 0.169 & 0.165 & 0.133
&
& 0.209 & 0.211 & 0.230 \\

\cmidrule(r){2-7}

&
\multirow{4}{*}{Additive}
& \multirow{2}{*}{$\widehat{\beta}_1$}
& 1 & 0.171 & 0.168 & 0.129
&
& 0.176 & 0.174 & 0.145 \\

&
&
& 2 & 0.169 & 0.165 & 0.133
&
& 0.173 & 0.169 & 0.143 \\

\cmidrule(r){3-7}

&
& \multirow{2}{*}{$\widehat{\beta}_2$}
& 1 & 0.074 & 0.093 & 0.096
&
& 0.175 & 0.159 & 0.151 \\

&
&
& 2 & 0.071 & 0.090 & 0.091
&
& 0.181 & 0.147 & 0.145 \\

\cmidrule(r){2-7}

&
\multirow{6}{*}{Full}
& \multirow{2}{*}{$\widehat{\beta}_1$}
& 1 & 0.166 & 0.164 & 0.120
&
& 0.171 & 0.169 & 0.139 \\

&
&
& 2 & 0.166 & 0.162 & 0.125
&
& 0.170 & 0.167 & 0.138 \\

\cmidrule(r){3-7}

&
& \multirow{2}{*}{$\widehat{\beta}_2$}
& 1 & 0.072 & 0.091 & 0.095
&
& 0.175 & 0.160 & 0.152 \\

&
&
& 2 & 0.070 & 0.091 & 0.092
&
& 0.181 & 0.148 & 0.146 \\

\cmidrule(r){3-7}

&
& \multirow{2}{*}{$\widehat{\gamma}$}
& 1 & 0.164 & 0.203 & 0.306
&
& 0.194 & 0.245 & 0.377 \\

&
&
& 2 & 0.147 & 0.197 & 0.283
&
& 0.183 & 0.248 & 0.357 \\

\bottomrule

\end{tabular}
\end{table}


\begin{table}[!t]
\centering
\caption{Relative root mean integrated squared errors (relRIMSEs) of the
estimated coefficient functions under two basis settings for the jittered
observation design with $\rho_{SV}=0.5$ and the same FVE thresholds as in
Table~\ref{tab:random_response_relRMSE}. The model, coefficient, and
basis labels in the left block apply to the corresponding rows in both
blocks.}
\label{tab:dependent_simulation}

\small
\setlength{\tabcolsep}{4.5pt}
\renewcommand{\arraystretch}{1}

\begin{tabular}{@{}
c l c c c c c
@{\hspace{16pt}}
c c c c
@{}}

\toprule

$(\delta_S,\delta_V,\delta_I)$
& Model
& Coef.
& Basis
& 80\%
& 85\%
& 90\%
&
$(\delta_S,\delta_V,\delta_I)$
& 80\%
& 85\%
& 90\% \\

\cmidrule(r){1-7}
\cmidrule(l){8-11}


\multirow{12}{*}{$(1,1,1)$}
& \multirow{2}{*}{Mean-only}
& \multirow{2}{*}{$\widehat{\beta}_1$}
& 1 & 0.482 & 0.485 & 0.518
&
\multirow{12}{*}{$(1,0.25,1)$}
& 0.225 & 0.228 & 0.249 \\

&
&
& 2 & 0.497 & 0.502 & 0.522
&
& 0.232 & 0.235 & 0.243 \\

\cmidrule(r){2-7}

&
\multirow{4}{*}{Additive}
& \multirow{2}{*}{$\widehat{\beta}_1$}
& 1 & 0.220 & 0.222 & 0.255
&
& 0.234 & 0.237 & 0.266 \\

&
&
& 2 & 0.220 & 0.223 & 0.237
&
& 0.230 & 0.233 & 0.246 \\

\cmidrule(r){3-7}

&
& \multirow{2}{*}{$\widehat{\beta}_2$}
& 1 & 0.285 & 0.294 & 0.322
&
& 0.927 & 1.044 & 1.192 \\

&
&
& 2 & 0.279 & 0.277 & 0.284
&
& 0.874 & 1.023 & 1.056 \\

\cmidrule(r){2-7}

&
\multirow{6}{*}{Full}
& \multirow{2}{*}{$\widehat{\beta}_1$}
& 1 & 0.186 & 0.184 & 0.177
&
& 0.198 & 0.195 & 0.182 \\

&
&
& 2 & 0.184 & 0.182 & 0.161
&
& 0.188 & 0.184 & 0.165 \\

\cmidrule(r){3-7}

&
& \multirow{2}{*}{$\widehat{\beta}_2$}
& 1 & 0.204 & 0.200 & 0.198
&
& 0.546 & 0.643 & 0.674 \\

&
&
& 2 & 0.205 & 0.178 & 0.171
&
& 0.500 & 0.619 & 0.608 \\

\cmidrule(r){3-7}

&
& \multirow{2}{*}{$\widehat{\gamma}$}
& 1 & 0.523 & 0.486 & 0.509
&
& 0.518 & 0.480 & 0.499 \\

&
&
& 2 & 0.480 & 0.444 & 0.452
&
& 0.476 & 0.434 & 0.436 \\


\midrule

\multirow{12}{*}{$(1,0,1)$}
& \multirow{2}{*}{Mean-only}
& \multirow{2}{*}{$\widehat{\beta}_1$}
& 1 & 0.220 & 0.223 & 0.237
&
\multirow{12}{*}{$(0.25,1,1)$}
& 1.988 & 2.003 & 2.170 \\

&
&
& 2 & 0.221 & 0.223 & 0.228
&
& 2.021 & 2.044 & 2.170 \\

\cmidrule(r){2-7}

&
\multirow{4}{*}{Additive}
& \multirow{2}{*}{$\widehat{\beta}_1$}
& 1 & 0.242 & 0.245 & 0.274
&
& 0.675 & 0.695 & 0.959 \\

&
&
& 2 & 0.237 & 0.240 & 0.254
&
& 0.678 & 0.719 & 0.906 \\

\cmidrule(r){3-7}

&
& \multirow{2}{*}{$\widehat{\beta}_2$}
& 1 & 0.254 & 0.290 & 0.334
&
& 0.271 & 0.283 & 0.308 \\

&
&
& 2 & 0.240 & 0.287 & 0.296
&
& 0.272 & 0.268 & 0.278 \\

\cmidrule(r){2-7}

&
\multirow{6}{*}{Full}
& \multirow{2}{*}{$\widehat{\beta}_1$}
& 1 & 0.206 & 0.203 & 0.190
&
& 0.476 & 0.477 & 0.603 \\

&
&
& 2 & 0.194 & 0.190 & 0.173
&
& 0.481 & 0.495 & 0.554 \\

\cmidrule(r){3-7}

&
& \multirow{2}{*}{$\widehat{\beta}_2$}
& 1 & 0.148 & 0.180 & 0.190
&
& 0.192 & 0.181 & 0.179 \\

&
&
& 2 & 0.135 & 0.177 & 0.175
&
& 0.201 & 0.163 & 0.161 \\

\cmidrule(r){3-7}

&
& \multirow{2}{*}{$\widehat{\gamma}$}
& 1 & 0.519 & 0.481 & 0.503
&
& 0.504 & 0.454 & 0.442 \\

&
&
& 2 & 0.477 & 0.435 & 0.438
&
& 0.467 & 0.413 & 0.401 \\


\midrule

\multirow{12}{*}{$(1,0,0)$}
& \multirow{2}{*}{Mean-only}
& \multirow{2}{*}{$\widehat{\beta}_1$}
& 1 & 0.161 & 0.159 & 0.116
&
\multirow{12}{*}{$(1,1,0)$}
& 0.458 & 0.459 & 0.471 \\

&
&
& 2 & 0.162 & 0.157 & 0.116
&
& 0.473 & 0.474 & 0.481 \\

\cmidrule(r){2-7}

&
\multirow{4}{*}{Additive}
& \multirow{2}{*}{$\widehat{\beta}_1$}
& 1 & 0.167 & 0.165 & 0.128
&
& 0.159 & 0.159 & 0.132 \\

&
&
& 2 & 0.167 & 0.162 & 0.127
&
& 0.163 & 0.159 & 0.129 \\

\cmidrule(r){3-7}

&
& \multirow{2}{*}{$\widehat{\beta}_2$}
& 1 & 0.081 & 0.107 & 0.107
&
& 0.172 & 0.158 & 0.149 \\

&
&
& 2 & 0.079 & 0.113 & 0.104
&
& 0.184 & 0.140 & 0.134 \\

\cmidrule(r){2-7}

&
\multirow{6}{*}{Full}
& \multirow{2}{*}{$\widehat{\beta}_1$}
& 1 & 0.164 & 0.161 & 0.121
&
& 0.157 & 0.156 & 0.130 \\

&
&
& 2 & 0.165 & 0.159 & 0.121
&
& 0.162 & 0.159 & 0.128 \\

\cmidrule(r){3-7}

&
& \multirow{2}{*}{$\widehat{\beta}_2$}
& 1 & 0.083 & 0.107 & 0.110
&
& 0.177 & 0.163 & 0.157 \\

&
&
& 2 & 0.080 & 0.113 & 0.104
&
& 0.187 & 0.144 & 0.139 \\

\cmidrule(r){3-7}

&
& \multirow{2}{*}{$\widehat{\gamma}$}
& 1 & 0.139 & 0.190 & 0.287
&
& 0.201 & 0.252 & 0.373 \\

&
&
& 2 & 0.130 & 0.195 & 0.265
&
& 0.188 & 0.263 & 0.348 \\

\bottomrule

\end{tabular}
\end{table}

\clearpage

\subsection{Estimation accuracy under increasing sample size}

This section provides the detailed simulation settings and results
for the increasing-sample-size analysis summarized in the main text. The simulation design is aligned with the assumptions in Section~\ref{appen S1}, particularly those regarding bandwidths. For local linear smoothing, \(b_S(m)\) denotes the common bandwidth used in the local linear smoothing procedure for estimating the individual mean curves \( \widetilde{S}_i(t) \). The bandwidths \(h_{\mu_S}(n)\) and \(h_{\mu_V}(n)\) are used to smooth the pooled trajectories across subjects to estimate the population mean functions. Similarly, \(h_{C_S}(n)\) and \(h_{C_V}(n)\) are used for two-dimensional smoothing of the empirical covariance surfaces. These bandwidths are set as $b_S(m)=r_{b_S}m^{-k_{b_S}}$, $h_{\mu_S}(n)=r_{\mu_S}n^{-k_{\mu_S}}$, and $h_{C_S}(n)=r_{C_S}n^{-k_{C_S}}$ (similarly for $h_{\mu_V}$ and $h_{C_V}$), 
where the constants $(r_{b_S}, r_{\mu_S}, r_{C_S}, r_{\mu_V}, r_{C_V})=(0.5, 0.45, 0.4, 0.45, 0.4)$ and exponents $(k_{b_S}, k_{\mu_S}, k_{C_S}, k_{\mu_V}, k_{C_V})=(1/3, 1/5, 1/7, 1/5, 1/7)$  are chosen within the theoretically admissible ranges implied by the asymptotic conditions.

To reflect the increasing-truncation regime in the asymptotic analysis, the truncation points are allowed to increase slowly with the sample size. Specifically, for $ (n,m)=(100,50),$ $(200,71),(400,100),(800,141),(1600,200)$,  we set $K_1=K_2=1,2,2,3,3$, respectively. Thus, both the number of subjects and the number of observations per subject increase, with \(m\) approximately proportional to \(n^{1/2}\), while the truncation levels increase conservatively.

We use the baseline configuration
$(\delta_S,\delta_V,\delta_I)=(1,1,1)$ and focus on Basis setting 1. To reduce additional variation from the observation design and more clearly examine the estimation-error pattern along the increasing-\((n,m)\) sequence, all subjects are observed on a common equally spaced grid over \(\mathcal T=[0,1]\). All other data-generating mechanisms also
remain unchanged.

\begin{figure}[!t]
\centering
\includegraphics[width=0.45\textwidth]{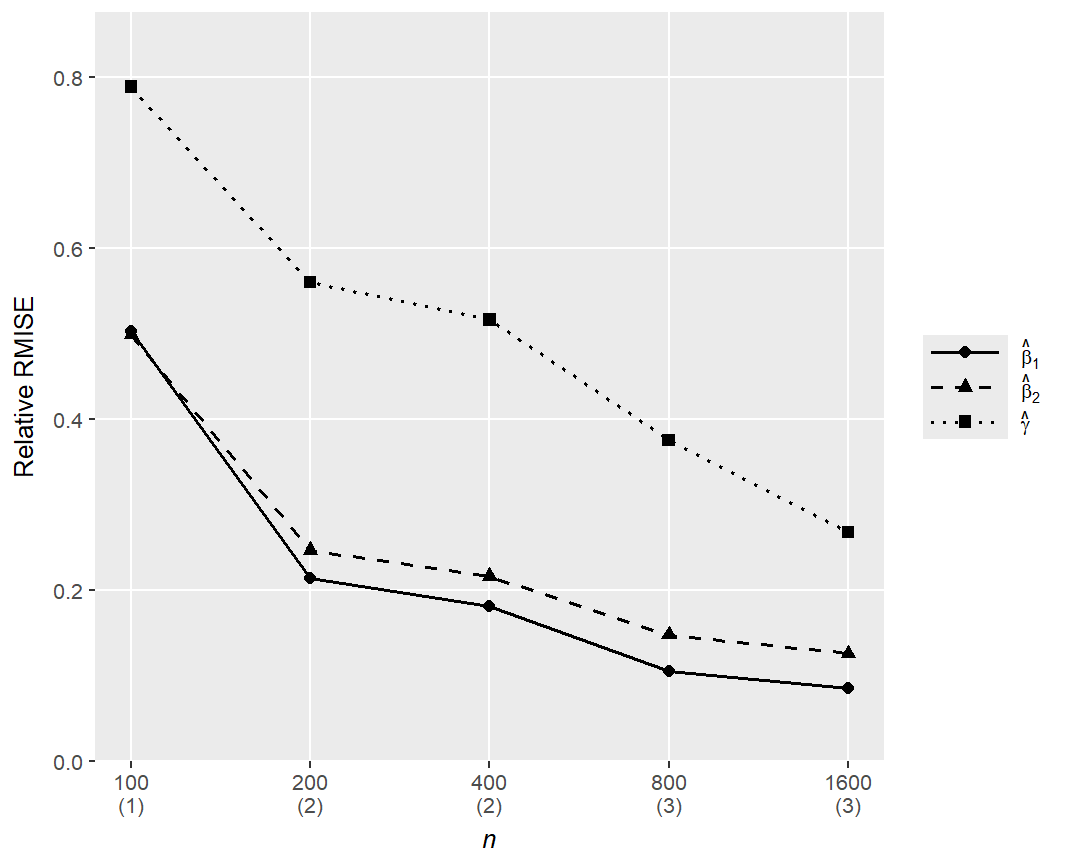}
\caption{Relative root mean integrated squared errors (relative RMISEs)
of $\widehat{\beta}_1$, $\widehat{\beta}_2$, and $\widehat{\gamma}$ along the
increasing-$(n,m)$ sequence. The numbers in parentheses below the sample
sizes indicate the common truncation level $K_1=K_2$.}
\label{fig:consistency}
\end{figure}

Figure~\ref{fig:consistency} displays the relRMISEs of \(\widehat{\beta}_1\), \(\widehat{\beta}_2\), and \(\widehat{\gamma}\) along the increasing-\((n,m)\) sequence. The estimation errors of all three coefficient components decrease steadily as \(n\) and \(m\) increase. The two main-effect coefficient functions are estimated more accurately than the interaction surface throughout, while the relRMISEs of \(\widehat{\gamma}\) nevertheless decreases substantially from approximately \(0.80\) to \(0.27\). These patterns provide numerical evidence consistent with the large-sample results in Section~\ref{sec: consistency properties} of the main text.

\subsection{Finite-sample performance of bootstrap confidence bands}
\label{subsec: simulation_bootstrap}

This section provides the detailed simulation settings and results
for the bootstrap simultaneous confidence bands summarized in
the main text. We consider the full-signal scenario
$(\delta_S,\delta_V,\delta_I)=(1,1,1)$ under Basis setting 2.

The subject-specific observation times are generated using the jittered
regular design described in Section~\ref{subsec: simulation setting} of the main text. Rather than considering a full factorial grid of \((n,m)\) and FVE thresholds, we use a sequence of representative configurations in which the FVE threshold is gradually increased as the sample size and observation frequency increase. This design allows more FPCs to be retained as the amount of information in the data increases, thereby reducing truncation while avoiding an unnecessarily large estimation burden in the smaller-sample settings. Selected configurations also vary the FVE threshold while holding \((n,m)\) fixed, allowing the finite-sample trade-off between truncation and estimation variability to be examined directly.

For each configuration of
$(n,m)$ and the FVE threshold, we generate \(L=100\) independent Monte Carlo
datasets. Within each dataset, 95\% simultaneous confidence bands (\(\alpha=0.05\)) are constructed using \(B=1,000\) subject-level pairs bootstrap replicates. 
We assess the resulting bands using their empirical simultaneous coverage
probabilities (SCP). For $j=1,2$, the simultaneous coverage probability is defined as
\[
\operatorname{SCP}_j
=
\frac{1}{L}
\sum_{r=1}^{L}
\mathbf{1}
\left\{
\beta_j(t_k)
\in
\left[
\ell_j^{(r)}(t_k),
u_j^{(r)}(t_k)
\right]
\text{ for all } k
\right\},
\]
where \(\mathbf 1(\cdot)\) denotes the indicator function and \([\ell_j^{(r)}(t_k),u_j^{(r)}(t_k)]\) denotes the pointwise endpoints of the \(100(1-\alpha)\%\) simultaneous confidence band for \(\beta_j\) in the \(r\)th simulation replicate. The empirical SCP for the interaction surface, \(\mathrm{SCP}_\gamma\), is defined analogously, requiring the entire true surface \(\gamma(s,t)\) to lie within its confidence band over the two-dimensional evaluation grid.

To assess the width of the resulting bands relative to the magnitude of the
underlying effect, we also report the median normalized half-width across
simulation replicates. For each coefficient function or interaction surface,
the band half-width is normalized by the sup norm of the corresponding true
effect. Thus, the coverage probabilities assess calibration of the
simultaneous bands, whereas the normalized half-widths quantify their
relative uncertainty.

Table~\ref{tab:simultaneous_coverage} summarizes the empirical simultaneous
coverage probabilities and median normalized half-widths under the simulation
settings considered. Overall, the simultaneous bands attain coverage reasonably close to the
nominal level under many of the settings considered, although their
finite-sample performance depends on the FVE threshold and the sampling
configuration. The comparisons at fixed \((n,m)\) further reveal a finite-sample trade-off between truncation and estimation variability. For example, when $n=200$ and $m=50$, the
coverage probability for $\beta_1$ increases from 0.68 under an 85\% FVE
threshold to 0.89 under a 90\% threshold. This improvement is accompanied
by an increase in the median normalized half-width from 0.289 to 0.370,
suggesting that retaining additional components reduces truncation bias at
the cost of increased uncertainty.

The variance-process coefficient is less sensitive to this change in
truncation, with coverage probabilities of 0.92 and 0.93 under the two
thresholds in the same setting. For the interaction surface, increasing the
FVE threshold improves coverage from 0.90 to 0.98, while the median
normalized half-width increases from 1.107 to 1.574. More generally, the
interaction surface exhibits substantially larger normalized band widths
than the two main-effect coefficient functions, reflecting the greater
finite-sample uncertainty associated with estimating the two-dimensional
interaction effect.

\begin{table}[!t]
\centering
\caption{Empirical simultaneous coverage probabilities and median normalized
half-widths under different simulation settings.}
\label{tab:simultaneous_coverage}

\small
\setlength{\tabcolsep}{5pt}

\begin{tabular}{ccc cc cc cc}
\toprule
& & &
\multicolumn{2}{c}{$\beta_1$} &
\multicolumn{2}{c}{$\beta_2$} &
\multicolumn{2}{c}{$\gamma$} \\
\cmidrule(lr){4-5}
\cmidrule(lr){6-7}
\cmidrule(lr){8-9}
$n$ & $m$ & FVE
& Cov. & Med. NHW
& Cov. & Med. NHW
& Cov. & Med. NHW \\
\midrule

50 & 30 & 0.80
& 0.92 & 0.724
& 0.94 & 0.946
& 1.00 & 2.576 \\

100 & 30 & 0.80
& 0.90 & 0.547
& 0.95 & 0.707
& 0.96 & 1.829 \\

100 & 30 & 0.85
& 0.92 & 0.575
& 0.97 & 0.737
& 0.98 & 2.340 \\

\addlinespace

100 & 50 & 0.85
& 0.90 & 0.479
& 0.96 & 0.600
& 0.99 & 1.713 \\

100 & 50 & 0.90
& 0.97 & 0.614
& 0.96 & 0.651
& 1.00 & 2.802 \\

\addlinespace

200 & 50 & 0.85
& 0.68 & 0.289
& 0.92 & 0.407
& 0.90 & 1.107 \\

200 & 50 & 0.90
& 0.89 & 0.370
& 0.93 & 0.419
& 0.98 & 1.574 \\

\addlinespace

200 & 100 & 0.90
& 0.79 & 0.277
& 0.97 & 0.316
& 0.95 & 1.149 \\

200 & 100 & 0.95
& 0.97 & 0.310
& 0.97 & 0.413
& 0.99 & 1.944 \\

\bottomrule
\end{tabular}

\vspace{0.5em}

\begin{minipage}{0.94\textwidth}
\footnotesize
\textit{Note.}
Cov.\ denotes the empirical SCP, with nominal
level 0.95. Med.\ NHW denotes the median normalized half-width. At the nominal coverage level, the Monte Carlo standard error is approximately
$\{0.95(1-0.95)/100\}^{1/2}=0.022$.
\end{minipage}

\end{table}

\section{Additional results from real-data analyses}

\subsection{Additional results for the tecator data}

Table~\ref{tab:tecator_fat_fixedK} presents the detailed
leave-one-out cross-validation (LOOCV) prediction results
for the Tecator data, complementing the results summarized
in the main text. The table reports RMSEs and MAEs (in
parentheses) across different truncation levels for the
mean-only, functional quadratic (FQM), additive, and interaction
models, together with the PLS benchmark.

The mean-only and FQM models generally yielded higher RMSEs
than PLS, with FQM outperforming PLS only at $K_S=7$.
In contrast, the additive model achieved lower RMSEs than PLS
in 36 of the 40 truncation configurations. Incorporating
interactions further improved prediction accuracy in 22 of
the 28 configurations with $K_S\leq 7$, although performance
deteriorated at larger truncation levels.

\begin{table}[!t]
\centering
\caption{Leave-one-out prediction RMSE for the Tecator data
with fat content as the response.}
\label{tab:tecator_fat_fixedK}

\small
\setlength{\tabcolsep}{2.5pt}
\renewcommand{\arraystretch}{1.05}

\begin{threeparttable}
\begin{tabular}{@{}c cc cccc cccc@{}}
\toprule
& \multicolumn{2}{c}{Mean-process models}
& \multicolumn{4}{c}{Additive}
& \multicolumn{4}{c}{Interaction} \\
\cmidrule(lr){2-3}
\cmidrule(lr){4-7}
\cmidrule(lr){8-11}

$K_S$
& Mean
& FQM
& $K_V=2$
& $K_V=3$
& $K_V=4$
& $K_V=5$
& $K_V=2$
& $K_V=3$
& $K_V=4$
& $K_V=5$ \\
\midrule

1
& \rmsecell{11.56}{9.41}
& \rmsecell{11.55}{9.36}
& \rmsecell{5.65}{4.12}
& \rmsecell{2.49}{1.96}
& \rmsecell{2.51}{1.98}
& \rmsecell{1.91}{1.39}
& \rmsecell{5.14}{3.91}
& \rmsecell{2.20}{1.74}
& \rmsecell{2.20}{1.74}
& \rmsecell{1.89}{1.37} \\[3pt]

2
& \rmsecell{11.50}{9.27}
& \rmsecell{11.28}{8.88}
& \rmsecell{4.55}{3.40}
& \rmsecell{2.52}{1.98}
& \rmsecell{2.49}{1.95}
& \rmsecell{1.90}{1.39}
& \rmsecell{4.01}{2.95}
& \rmsecell{2.18}{1.70}
& \rmsecell{2.08}{1.63}
& \rmsecell{1.78}{1.29} \\[3pt]

3
& \rmsecell{8.60}{6.67}
& \rmsecell{7.89}{5.85}
& \rmsecell{1.91}{1.47}
& \rmsecell{1.70}{1.32}
& \rmsecell{1.68}{1.28}
& \rmsecell{1.73}{1.27}
& \rmsecell{1.88}{1.47}
& \rmsecell{1.55}{1.18}
& \rmsecell{1.53}{1.15}
& \rmsecell{1.50}{1.13} \\[3pt]

4
& \rmsecell{4.33}{3.60}
& \rmsecell{3.49}{2.66}
& \rmsecell{1.78}{1.40}
& \rmsecell{1.68}{1.32}
& \rmsecell{1.66}{1.29}
& \rmsecell{1.66}{1.29}
& \rmsecell{1.71}{1.36}
& \rmsecell{1.55}{1.25}
& \rmsecell{1.59}{1.24}
& \rmsecell{1.49}{1.14} \\[3pt]

5
& \rmsecell{4.53}{3.79}
& \rmsecell{3.82}{3.01}
& \rmsecell{1.77}{1.38}
& \rmsecell{1.75}{1.39}
& \rmsecell{1.75}{1.37}
& \rmsecell{1.71}{1.35}
& \rmsecell{1.82}{1.45}
& \rmsecell{1.72}{1.41}
& \rmsecell{1.56}{1.26}
& \rmsecell{1.46}{1.19} \\[3pt]

6
& \rmsecell{4.11}{3.43}
& \rmsecell{3.71}{3.25}
& \rmsecell{1.69}{1.26}
& \rmsecell{1.67}{1.27}
& \rmsecell{1.66}{1.25}
& \rmsecell{1.64}{1.23}
& \rmsecell{1.94}{1.53}
& \rmsecell{1.72}{1.36}
& \rmsecell{1.82}{1.46}
& \rmsecell{1.74}{1.46} \\[3pt]

7
& \rmsecell{\textbf{3.22}}{\textbf{2.63}}
& \rmsecell{\textbf{2.44}}{\textbf{1.95}}
& \rmsecell{1.65}{1.24}
& \rmsecell{1.64}{1.23}
& \rmsecell{1.64}{1.23}
& \rmsecell{1.66}{1.27}
& \rmsecell{1.67}{1.30}
& \rmsecell{1.40}{1.09}
& \rmsecell{1.40}{1.03}
& \rmsecell{\textbf{1.32}}{\textbf{1.00}} \\[3pt]

8
& \rmsecell{3.41}{2.71}
& \rmsecell{6.25}{4.61}
& \rmsecell{1.77}{1.34}
& \rmsecell{1.77}{1.36}
& \rmsecell{1.68}{1.29}
& \rmsecell{1.62}{1.26}
& \rmsecell{3.71}{3.05}
& \rmsecell{3.38}{2.78}
& \rmsecell{2.73}{2.24}
& \rmsecell{2.56}{2.12} \\[3pt]

9
& \rmsecell{3.79}{2.96}
& \rmsecell{7.29}{3.67}
& \rmsecell{1.67}{1.21}
& \rmsecell{\textbf{1.60}}{\textbf{1.18}}
& \rmsecell{1.64}{1.20}
& \rmsecell{1.62}{1.20}
& \rmsecell{4.38}{3.53}
& \rmsecell{4.96}{3.76}
& \rmsecell{4.57}{3.62}
& \rmsecell{3.96}{3.25} \\[3pt]

10
& \rmsecell{7.72}{4.69}
& \rmsecell{90.33}{56.46}
& \rmsecell{3.37}{2.70}
& \rmsecell{2.70}{2.05}
& \rmsecell{2.63}{2.01}
& \rmsecell{3.43}{2.48}
& \rmsecell{9.26}{5.95}
& \rmsecell{6.65}{5.14}
& \rmsecell{6.82}{5.08}
& \rmsecell{8.44}{5.89} \\[3pt]

\midrule
PLS
& \multicolumn{10}{c}{\rmsecell{2.54}{1.84}} \\
\bottomrule

\end{tabular}
\end{threeparttable}
\end{table}

\subsection{Additional results for the R-peak-to-R-peak interval data}

\medskip
\noindent\textbf{Scalar summary of variability.}
To assess whether modeling subject-specific variability as a
functional process provides additional predictive information
beyond a scalar summary, we consider a scalar variability
measure motivated by the detrended standard deviation of RR
intervals (SDRR) used in \citealp{mongin2022decrease}.
In that study, the RR interval series was detrended to remove
low-frequency trends, and the standard deviation of the
residuals was computed as a global measure of short-term
variability.

In our analysis, we adopt a simplified approach by applying
local smoothing to each RR interval trajectory to obtain a
detrended signal and then computing the standard deviation
of the residuals as a scalar variability measure.
This measure is included as an additional covariate alongside
the FPC scores of the mean process, allowing comparison with
the proposed functional mean--variance regression model.

\medskip
\noindent\textbf{Exploratory analysis of mean--variance
interactions.}
As noted in the main text, the full interaction model did not
consistently improve prediction accuracy for the RR interval data,
although this does not necessarily imply that mean--variance
interactions are uninformative. We therefore conducted an
exploratory score-level analysis to investigate whether a smaller
subset of main effects and interactions could provide useful
predictive information.

Based on the full-data FPCA representation under the 95\% FVE criterion, which retained three mean components and four variance components, we considered the seven corresponding main-effect scores together with their 12 pairwise mean--variance interactions, yielding 19 candidate predictors in total. We then performed an exhaustive search over regression specifications formed from these candidate predictors, excluding specifications for which ordinary least squares estimation failed due to rank deficiency. No heredity constraint was imposed, allowing an interaction to enter the model without requiring its corresponding main effects to be included. This choice was motivated by the possibility that an interaction may contain predictive information even when one of its corresponding main effects contributes little marginal predictive information.

\begin{figure}[!t]
\centering

\begin{subfigure}[t]{0.43\textwidth}
    \centering
    \includegraphics[
        width=\textwidth,
        trim=0 0 7.5 0,
        clip
    ]{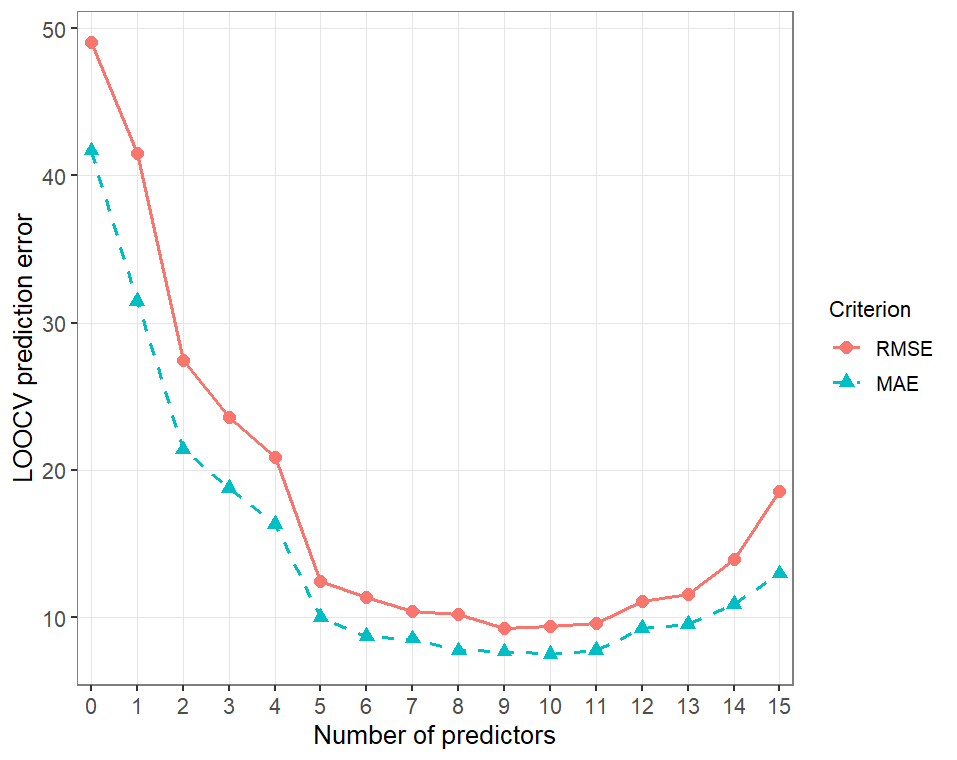}
    \caption{}
    \label{fig:rr_model_size}
\end{subfigure}
\hspace{0.3em}
\begin{subfigure}[t]{0.43\textwidth}
    \centering
    \includegraphics[
        width=\textwidth,
        trim=7.5 0 0 0,
        clip
    ]{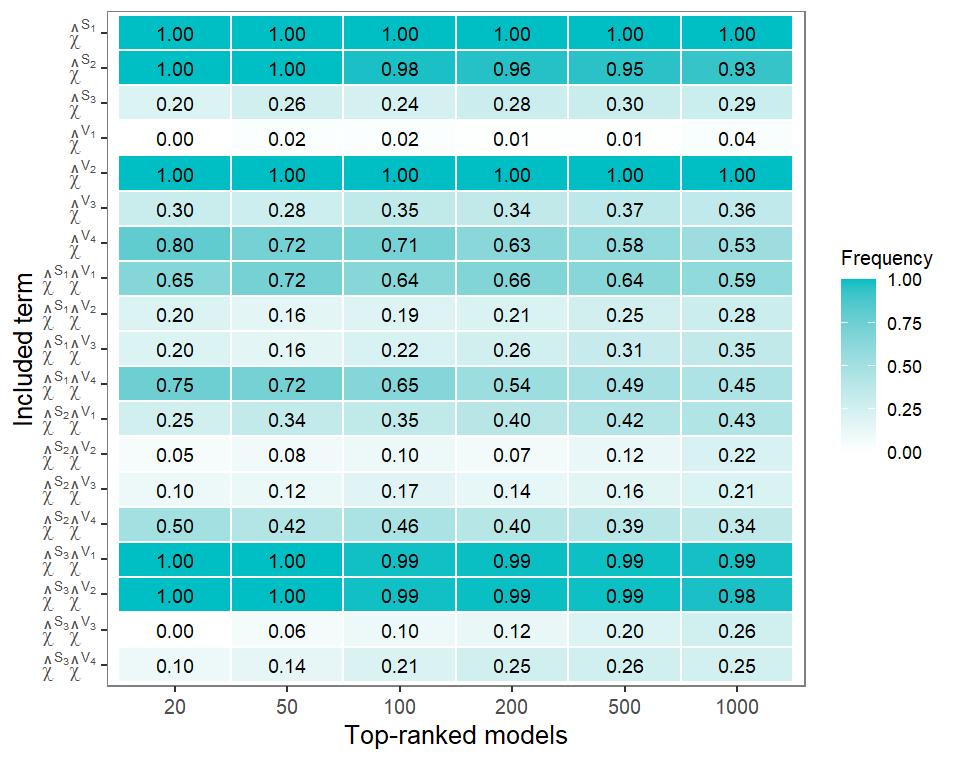}
    \caption{}
    \label{fig:rr_inclusion_heatmap}
\end{subfigure}

\caption{
Exploratory exhaustive-search results for the RR interval data.
Panel (a) shows the minimum LOOCV RMSE and MAE achieved at each model size.
Panel (b) shows the inclusion frequencies of the score-level main effects and interactions among the top 20, 50, 100, 200, 500, and 1000 models ranked by LOOCV RMSE.
}
\label{fig:rr_exhaustive_search}

\end{figure}

Predictive performance was evaluated by LOOCV using both RMSE and MAE. Table~\ref{tab:rr_exhaustive} summarizes the benchmark models and the best-performing specifications identified by the exhaustive search. The best-performing unrestricted specification achieved a markedly smaller LOOCV error than the additive benchmarks. Figure~\ref{fig:rr_exhaustive_search} provides a broader view of the predictive patterns across the exhaustive search. Panel (a) shows the minimum LOOCV prediction error attained at each model size. Both criteria improve substantially as the number of predictors increases from the smallest specifications, with the minimum RMSE attained at nine predictors and the minimum MAE at ten predictors. Beyond this range, the best achievable prediction error begins to increase, indicating that the strong predictive performance is associated with moderately sized specifications rather than simply with the inclusion of more predictors.

\begin{table}[!t]
\centering
\caption{LOOCV prediction errors for the RR interval data
under exploratory exhaustive model search
(95\% FVE; $K_1=3$, $K_2=4$).}
\label{tab:rr_exhaustive}

\small
\setlength{\tabcolsep}{3pt}
\renewcommand{\arraystretch}{1}

\begin{tabular}{@{}p{0.19\textwidth} p{0.57\textwidth} cc@{}}
\toprule
Model & Included terms & RMSE & MAE \\
\midrule

Mean only
& $\widehat{\chi}^{S}_{1},\widehat{\chi}^{S}_{2},
   \widehat{\chi}^{S}_{3}$
& 38.17
& 29.18 \\

Full additive
& $\widehat{\chi}^{S}_{1},\widehat{\chi}^{S}_{2},
   \widehat{\chi}^{S}_{3},
   \widehat{\chi}^{V}_{1},\widehat{\chi}^{V}_{2},
   \widehat{\chi}^{V}_{3},\widehat{\chi}^{V}_{4}$
& 33.30
& 28.19 \\

Best additive
& $\widehat{\chi}^{S}_{1},
   \widehat{\chi}^{V}_{1},\widehat{\chi}^{V}_{2}$
& 25.08
& 19.68 \\

\midrule

Best (RMSE)
& $\widehat{\chi}^{S}_{1},\widehat{\chi}^{S}_{2},
   \widehat{\chi}^{V}_{2},\widehat{\chi}^{V}_{4},
   \widehat{\chi}^{S}_{1}\widehat{\chi}^{V}_{1},
   \widehat{\chi}^{S}_{1}\widehat{\chi}^{V}_{4},
   \widehat{\chi}^{S}_{2}\widehat{\chi}^{V}_{4},
   \widehat{\chi}^{S}_{3}\widehat{\chi}^{V}_{1},
   \widehat{\chi}^{S}_{3}\widehat{\chi}^{V}_{2}$
& \textbf{9.26}
& 7.68 \\

Best (MAE)
& $\widehat{\chi}^{S}_{1},\widehat{\chi}^{S}_{2},
   \widehat{\chi}^{V}_{2},\widehat{\chi}^{V}_{4},
   \widehat{\chi}^{S}_{1}\widehat{\chi}^{V}_{1},
   \widehat{\chi}^{S}_{1}\widehat{\chi}^{V}_{4},
   \widehat{\chi}^{S}_{2}\widehat{\chi}^{V}_{2},
   \widehat{\chi}^{S}_{2}\widehat{\chi}^{V}_{4},
   \widehat{\chi}^{S}_{3}\widehat{\chi}^{V}_{1},
   \widehat{\chi}^{S}_{3}\widehat{\chi}^{V}_{2}$
& 9.39
& \textbf{7.49} \\

\bottomrule
\end{tabular}
\end{table}

Panel (b) further examines which score-level terms recur among the best-performing models ranked by LOOCV RMSE. Several terms appear consistently across the different ranking thresholds. In particular, $\widehat{\chi}^{S}_1$ and $\widehat{\chi}^{V}_2$ are included in all of the top 20, 50, 100, 200, 500, and 1000 models. More notably, the interactions $\widehat{\chi}^{S}_3\widehat{\chi}^{V}_1$ and $\widehat{\chi}^{S}_3\widehat{\chi}^{V}_2$ also appear in nearly all of these high-performing specifications, with inclusion frequencies remaining above 0.98 even among the top 1000 models. In contrast, the corresponding main effect $\widehat{\chi}^{S}_3$ is included in only about 20--30\% of these models, while $\widehat{\chi}^{V}_1$ is rarely included as a main effect. This contrast suggests that, for these data, some mean--variance interactions may carry predictive information that is not well represented by the marginal contributions of their constituent scores. It also illustrates why imposing a heredity constraint could possibly exclude interaction structures that repeatedly occur among the better-performing models. Given the small sample size and the extensive model search, however, these inclusion patterns should be interpreted as exploratory rather than as evidence of formal variable-selection stability.

The results reported in Table~\ref{tab:rr_exhaustive} and Figure~\ref{fig:rr_exhaustive_search} are based on a preliminary smoothing bandwidth equal to 10\% of each subject's observed time-domain range. Although the absolute LOOCV error values were sensitive to the choice of this bandwidth, the best-performing unrestricted models consistently achieved substantially lower prediction errors than the best additive models across the bandwidth choices considered. Moreover, the overall relationship between prediction error and model size and the inclusion-frequency patterns among the top-ranked models remained qualitatively stable across these choices. Taken together, these exploratory results suggest that the predictive role of subject-specific variability may extend beyond an additive contribution and may depend on particular features of the underlying mean dynamics, while also motivating further investigation of interaction structures that need not satisfy conventional heredity constraints.


\end{document}